\documentclass[12pt]{article}
\usepackage{latexsym}
\usepackage{amsmath,amsfonts}
\usepackage{times}
\allowdisplaybreaks[4]

\catcode`\@=11

\newcount\hour
\newcount\minute
\newtoks\amorpm \hour=\time\divide\hour by 60\minute
=\time{\multiply\hour by 60 \global\advance\minute by-\hour}
\edef\standardtime{{\ifnum\hour<12 \global\amorpm={am}%
        \else\global\amorpm={pm}\advance\hour by-12 \fi
        \ifnum\hour=0 \hour=12 \fi
        \number\hour:\ifnum\minute<10
        0\fi\number\minute\the\amorpm}}
\edef\militarytime{\number\hour:\ifnum\minute<10 0\fi\number\minute}

\def\draftlabel#1{{\@bsphack\if@filesw {\let\thepage\relax
   \xdef\@gtempa{\write\@auxout{\string
      \newlabel{#1}{{\@currentlabel}{\thepage}}}}}\@gtempa
   \if@nobreak \ifvmode\nobreak\fi\fi\fi\@esphack}
        \gdef\@eqnlabel{#1}}
\def\@eqnlabel{}
\def\@vacuum{}
\def\marginnote#1{}
\def\draftmarginnote#1{\marginpar{\raggedright\scriptsize\tt#1}}
\def\draft{
        \pagestyle{plain}
        \overfullrule=2pt
        \oddsidemargin -.5truein
        \def\@oddhead{\sl \phantom{\today\quad\militarytime} \hfil
        \smash{\Large\sl DRAFT} \hfil \today\quad\militarytime}
        \let\@evenhead\@oddhead
        \let\label=\draftlabel
        \let\marginnote=\draftmarginnote
        \def\ps@empty{\let\@mkboth\@gobbletwo
        \def\@oddfoot{\hfil \smash{\Large\sl DRAFT} \hfil}
        \let\@evenfoot\@oddhead}
        \def\@eqnnum{(\theequation)\rlap{\kern\marginparsep\tt\@eqnlabel}%
        \global\let\@eqnlabel\@vacuum}  }

\newcommand{\rf}[1]{(\ref{#1})}
\renewcommand{\theequation}{\thesection.\arabic{equation}}
\renewcommand{\thefootnote}{\fnsymbol{footnote}}
\newcommand{\newsection}{   
\setcounter{equation}{0}\section}

\def\appendix#1{\addtocounter{section}{1}\setcounter{equation}{0}
\renewcommand{\thesection}{\Alph{section}}
\section*{Appendix \thesection\protect\indent \parbox[t]{11.15cm}{#1}}
\addcontentsline{toc}{section}{Appendix \thesection\ \ \ #1}}

\def\be{\begin{equation}}
\def\ee{\end{equation}}
\def\beq{\begin{eqnarray}}
\def\eeq{\end{eqnarray}}

\def\parline{\,\partial\kern -0.55em /\,\,}

\def\half{{\frac{1}{2}}}

\def\AA{{\cal A}}

\def\LL{{\cal L}}

\def\VV{{\cal V}}

\def\Wk{|W\rangle}
\def\Phik{|\Phi\rangle}
\def\Xik{|\Xi\rangle}
\def\phik{|\phi\rangle}
\def\phibr{\langle\phi|}

\def\Cb{\bar{C}}
\def\Vb{\bar{V}}

\def\eb{\bar{e}}
\def\mb{\bar{m}}
\def\rb{\bar{r}}

\def\xik{|\xi\rangle}

\def\smG{{\scriptscriptstyle G}}

\def\smzero{{\scriptscriptstyle (0)}}
\def\smone{{\scriptscriptstyle (1)}}
\def\smtwo{{\scriptscriptstyle (2)}}

\def\smn{{\scriptscriptstyle (n)}}

\def\oplussm{{\scriptscriptstyle \oplus}}
\def\ominussm{{\scriptscriptstyle \ominus}}
\def\smminone{{\scriptscriptstyle (-1)}}

\def\smG{{\scriptscriptstyle G}}

\def\smzero{{\scriptscriptstyle (0)}}
\def\smone{{\scriptscriptstyle (1)}}
\def\smtwo{{\scriptscriptstyle (2)}}

\def\oplussm{{\scriptscriptstyle \oplus}}
\def\ominussm{{\scriptscriptstyle \ominus}}

\def\smponetwo{{\scriptscriptstyle [1,2]}}

\def\Cwt{\widetilde{C}}

\def\Vwt{\widetilde{V}}

\def\bwt{\widetilde{b}}
\def\ewt{\widetilde{e}}

\def\mwt{\widetilde{m}}
\def\rwt{\widetilde{r}}

\def\mwt{\widetilde{m}}

\def\kwh{{\widehat{k}}}

\def\alpar{\alpha\partial}
\def\albpar{\bar\alpha\partial}

\def\tr{{\rm tr}}
\def\sh{{\rm sh}}
\def\Irm{{\rm I}}
\def\IIrm{{\rm II}}

\def\ibf{{\bf i}}
\def\iibf{{\bf ii}}
\def\iiibf{{\bf iii}}
\def\ivbf{{\bf iv}}
\def\vbf{{\bf v}}
\def\nbf{{\bf n}}

\def\ebwt{\widetilde{\bar e}}
\def\mbwt{\widetilde{\bar m}}
\def\rbwt{\widetilde{\bar r}}

\def\mubf{{\boldsymbol{\mu}}}

\begin{document}


\begin{flushright}
FIAN-TD-2007-11 \hspace{1.7cm}{}~\\
arXiv: 0709.4392 [hep-th] \hspace{0.5cm}{}~\\
Modified, April 2012\hspace{1.4cm}{}~
\end{flushright}

\vspace{1cm}

\begin{center}

{\Large \bf Ordinary-derivative formulation of conformal

\bigskip totally symmetric arbitrary spin bosonic fields}

\vspace{2.5cm}

R.R. Metsaev\footnote{ E-mail: metsaev@lpi.ru }

\vspace{1cm}

{\it Department of Theoretical Physics, P.N. Lebedev Physical
Institute,
\\ Leninsky prospect 53,  Moscow 119991, Russia }

\vspace{3.5cm}

{\bf Abstract}

\end{center}

Conformal totally symmetric arbitrary spin bosonic fields in flat
space-time of even dimension greater than or equal to four are
studied. Second-derivative (ordinary-derivative) formulation for
such fields is developed. We obtain gauge invariant Lagrangian and
the corresponding gauge transformations. Gauge symmetries are
realized by involving the Stueckelberg and auxiliary fields.
Realization of global conformal boost symmetries on conformal gauge
fields is obtained. Modified de Donder gauge condition and de
Donder-Stueckelberg gauge condition are introduced. Using the de
Donder-Stueckelberg gauge frame, equivalence of the
ordinary-derivative and higher-derivative approaches is
demonstrated. On-shell degrees of freedom of the arbitrary spin
conformal field are analyzed. Ordinary-derivative light-cone gauge
Lagrangian of conformal fields is also presented. Interrelations
between the ordinary-derivative gauge invariant formulation of
conformal fields and the gauge invariant formulation of massive
fields are discussed.

\newpage
\renewcommand{\thefootnote}{\arabic{footnote}}
\setcounter{footnote}{0}

\newsection{\large Introduction}

The present paper is a sequel to our paper \cite{Metsaev:2007fq},
where the ordinary-derivative formulation of conformal fields was
developed. Commonly used Lagrangian formulations of most conformal
fields involve higher derivatives (for review, see
Ref.\cite{Fradkin:1985am}). In Ref.\cite{Metsaev:2007fq}, we
developed the ordinary-derivative gauge invariant and Lagrangian
formulation for free low-spin conformal fields. This is to say that
our Lagrangians for free bosonic conformal fields do not involve
higher than second order terms in derivatives, while our Lagrangians
for free fermionic conformal fields do not involve higher than first
order terms in derivatives. In the present paper, we generalize
results in Ref.\cite{Metsaev:2007fq} to the case of totally
symmetric arbitrary spin bosonic conformal fields.

A long term motivation for our study of conformal fields comes from
the following potentially important applications.

In Ref.\cite{Fradkin:1990ps}, it has been conjectured that string
theory, theory of massless higher-spin fields in AdS space, and
theory of conformal fields, though different, eventually may turn
out to be different phases of one and the same unified field theory
with new forces mediated by higher-spin gauge fields. According to
this conjecture, in the ultra high-energy domain, dynamics of the
unified theory is governed by theory of conformal higher-spin fields
generalizing Weyl gravity, i.e., in the ultra high-energy domain,
microscopic degrees of freedom of the unified theory are described
by conformal low- and higher-spin gauge fields. Spontaneous breaking
of conformal symmetry leads to the theory of massless higher-spin
AdS fields \cite{Vasiliev:1990en} generalizing AdS supergravity. One
expects that, in the AdS phase, symmetries of the unified theory are
realized as infinite-dimensional gauge symmetries of massless low-
and higher-spin fields. Further, spontaneous breaking of higher-spin
gauge symmetries of massless higher-spin AdS fields theory leads to
the string theory in flat space. We believe that use of the
ordinary-derivative formulation of conformal fields might be helpful
for studying the conjecture in Ref.\cite{Fradkin:1990ps}. This is to
say that Lagrangian of free string field theory and Lagrangian of
free massless higher-spin AdS fields theory do not involve higher
derivatives, i.e., these Lagrangians take ordinary-derivative form.
Therefore, the ordinary-derivative formulation of conformal fields
theory seems to be most suitable for the investigation of
possible interrelations between theory of conformal fields, theory
of massless higher-spin AdS fields, and string theory.%
\footnote{ Recent interesting discussion of conformal symmetries of
massless higher-spin $AdS_4$ field theory may be found in
Ref.\cite{Vasiliev:2007yc}.}

The second application is to the various conjectured dualities
\cite{Maldacena:1997re} between  supesymmetric Yang-Mills theory,
string theory and theory of massless higher-spin AdS fields. In the
framework of AdS/CFT correspondence, the conformal fields manifest
themselves in a intriguing way. This is to say that conformal fields
appear as boundary values of non-normalazible solution to equations
of motion for massless AdS fields (see, e.g.,
Refs.\cite{Balasubramanian:1998sn}-\cite{Metsaev:2005ws}%
\footnote{In the earlier literature, discussion of conformal field
dualities may be found in Ref.\cite{Petkou:1994ad}.}).
Therefore the action for massless AdS field, when it is evaluated on
solution of the Dirichlet problem (effective action), leads to the
action of conformal field.%
\footnote{ The computation of effective action for free massless
spin-2 field in $AdS_5$ space was carried out in
Ref.\cite{Liu:1998bu}, while the computation of effective action for
free arbitrary spin-$s$ field, $s\geq 2$, in $AdS_{d+1}$ space was
carried out in Ref.\cite{Metsaev:2009ym}.}
As a side remark we also note that conformal symmetries manifest
themselves in the tensionless limit of string theory (see, e.g.,
Ref.\cite{Isberg:1992ia}). In view of these interesting
interrelations between string theory, theory of massless higher-spin
AdS fields, and theory of conformal fields we think that the
ordinary-derivative formulation of conformal fields might be useful
to understand string/gauge theory dualities better. This is to say
that bulk equations of motion of free string field theory and theory
of free massless higher-spin AdS fields do not involve higher
derivatives. We believe therefore that the ordinary-derivative
approach to theory of conformal fields should be most suitable for
the investigation of bulk/boundary correspondence.

Higher-derivative Lagrangian formulation of the totally symmetric
arbitrary spin bosonic conformal fields in 4-dimensional space and
in $d$-dimensional space, $d\geq 4$, was developed in the respective
Ref.\cite{Fradkin:1985am} and Ref.\cite{Segal:2002gd}. Alternative
higher-derivative description of the conformal fields obtained via
AdS/CFT duality may be found in Ref.\cite{Metsaev:2009ym}. The
purpose of this paper is to develop ordinary-derivative gauge
invariant and Lagrangian formulation for free conformal fields. In
this paper, we discuss the bosonic totally symmetric arbitrary spin
conformal fields in space-time of even dimension $d \geq 4$. Our
approach is summarized as follows.

\noindent \ibf) By introducing additional field degrees of freedom,
we extend the space of fields entering the higher-derivative theory.
Some of the additional fields turn out to be Stueckelberg fields,
while the remaining additional fields turn out to be auxiliary
fields.

\noindent \iibf) Our ordinary-derivative Lagrangian does not contain
higher than second order terms in derivatives. Two-derivative
contributions to the Lagrangian take the form of the standard
Klein-Gordon, Maxwell, Einstein-Hilbert, and Fronsdal kinetic terms
of the respective spin-0, spin-1, spin-2, and spin-$s$, $s>2$,
bosonic fields.

\noindent \iiibf) All vector and tensor fields are
supplemented by appropriate gauge symmetries.%
\footnote{ To realize those additional gauge symmetries we adopt the
approach in Refs.\cite{Zinoviev:2001dt,Metsaev:2006zy} which turns
out to be most useful for our purposes.}
Gauge transformations of conformal fields do not involve higher than
first order terms in derivatives. One-derivative contributions to
the gauge transformations take the form of the standard gradient
gauge transformations of the vector and tensor fields.

\noindent \ivbf) The gauge symmetries of our Lagrangian make it
possible to match  our approach with  the higher-derivative one,
i.e., by gauging away the Stueckelberg fields and by solving
equations of motion for the auxiliary fields, we obtain the
higher-derivative formulation of conformal fields. This implies that
our approach retains propagating D.o.F of the higher-derivative
conformal fields theory, i.e., our approach is equivalent to the
higher-derivative one, at least at the classical level.

The rest of the paper  is organized as follows.

In Sec. \ref{prelim},  we summarize the notation used in this paper.

Section \ref{on-shell} is devoted to the discussion of on-shell
D.o.F appearing in higher-derivative formulation of arbitrary
spin-$s$ conformal field propagating in $d$-dimensional  space.
Because approach in Ref.\cite{Metsaev:2009ym} turns out to be
convenient for this purpose we start with brief review of the gauge
invariant higher-derivative formulation of conformal fields
developed in Ref.\cite{Metsaev:2009ym}. Also, for the reader
convenience, we discuss how our higher-derivative approach in
Ref.\cite{Metsaev:2009ym} is related to the one in
Refs.\cite{Fradkin:1985am,Segal:2002gd}. After this, we describe our
results for total number of on-shell D.o.F for the arbitrary spin
conformal field and decomposition of the on-shell D.o.F into irreps
of the $so(d-2)$ algebra. Light-cone gauge higher-derivative and
ordinary-derivative Lagrangians are also presented.

In Sec.\ref{lagran}, we develop the ordinary-derivative formulation
for arbitrary spin conformal field. We start with the discussion of
field content entering our approach. After this we present our
result for gauge invariant Lagrangian. We discuss various
representations for the Lagrangian. Also, we describe two new gauge
conditions for conformal fields which we refer to as modified de
Donder gauge and de Donder-Stueckelberg gauge. In Sec.\ref{gaugsym}
we discuss gauge symmetries of our ordinary-derivative Lagrangian,
while in Sec.\ref{realiz} we discuss realization of conformal
algebra symmetries on the space of gauge fields entering our
approach.

In Sec.\ref{sec-05onter}, we discuss the interrelations between the
higher-derivative and ordinary-derivative approaches to conformal
fields. We demonstrate that our ordinary-derivative approach in
Sec.\ref{lagran} is equivalent to the higher-derivative approach in
Sec.\ref{subsec-301}.

In Sec. \ref{interel}, we discuss the interrelations between the
ordinary-derivative description of conformal field and gauge
invariant description of massive field. In due course we also
discuss our new representation for gauge invariant Lagrangian of
arbitrary spin massive field by using modified de Donder
divergencies.

Section \ref{conlcus} is devoted to the discussion of directions for
future research.

Technical details are collected in Appendices. In Appendix A, we
outline the procedure of derivation of on-shell D.o.F. appearing in
the higher-derivative formulation of arbitrary spin conformal field
in Ref.\cite{Segal:2002gd}. In Appendix B, we present details of the
derivation of ordinary-derivative gauge invariant Lagrangian and
gauge transformations for the arbitrary spin conformal field. In
Appendices C,D, we discuss some technical details of the
interrelations between the ordinary-derivative and higher-derivative
approaches to conformal fields.

\newsection{\large Preliminaries}\label{prelim}

\subsection{Notation}

Our conventions are as follows. $x^a$ denotes coordinates in
$d$-dimensional flat space-time, while $\partial_a$ denotes
derivatives with respect to $x^a$, $\partial_a \equiv \partial /
\partial x^a$. Vector indices of the Lorentz algebra $so(d-1,1)$ take
the values $a,b,c,e=0,1,\ldots ,d-1$. We use mostly positive flat
metric tensor $\eta^{ab}$. To simplify our expressions we drop
$\eta_{ab}$ in scalar products, i.e., we use $X^aY^a \equiv
\eta_{ab}X^a Y^b$.

A set of the creation operators $\alpha^a$, $\zeta$,
$\upsilon^\oplussm$, $\upsilon^\ominussm$ and the respective set of
annihilation operators $\bar{\alpha}^a$, $\bar{\zeta}$,
$\bar\upsilon^\ominussm$, $\bar\upsilon^\oplussm$ will be referred
to as oscillators in what
follows.%
\footnote{ As in Ref.\cite{Lopatin:1987hz}, we use oscillators to
handle many indices appearing for tensor fields. The oscillator
algebra can also be reformulated as an algebra acting on the
symmetric-spinor bundle on the manifold $M$ \cite{Hallowell:2005np}.
The oscillators $\zeta$, $\bar\zeta$ appearing in gauge invariant
formulation of massive fields arise naturally by a dimensional
reduction \cite{Biswas:2002nk,Hallowell:2005np} from flat space. We
expect that `conformal' oscillators $\upsilon^\oplussm$,
$\upsilon^\ominussm$, $\bar\upsilon^\oplussm$,
$\bar\upsilon^\ominussm$ also allow certain interpretation via
dimensional reduction. Extensive discussion of oscillator
formulation may be found in Ref.\cite{Bekaert:2006ix}.}
Commutation relations, the vacuum, and hermitian conjugation rules
are defined as
\beq
&&{} [\bar{\alpha}^a,\alpha^b]=\eta^{ab}\,, \qquad
[\bar\zeta,\zeta]=1\,, \qquad [\bar{\upsilon}^\oplussm,\,
\upsilon^\ominussm ]=1\,, \qquad\quad [\bar{\upsilon}^\ominussm,\,
\upsilon^\oplussm]=1\,,
\\
&& \bar\alpha^a |0\rangle = 0\,,\qquad\quad  \bar\zeta|0\rangle =
0\,,\qquad\quad \bar\upsilon^\oplussm |0\rangle = 0\,,\qquad\quad
\quad \bar\upsilon^\ominussm |0\rangle = 0\,,
\\
&& \alpha^{a\dagger} = \bar\alpha^a\,, \qquad \zeta^\dagger =
\bar\zeta\,,
\qquad
\upsilon^{\oplussm\dagger} = \bar\upsilon^\oplussm\,,\qquad
\upsilon^{\ominussm \dagger} = \bar\upsilon^\ominus \,.
\eeq
The oscillators $\alpha^a$, $\bar\alpha^a$ and $\zeta$, $\bar\zeta$,
$\upsilon^\oplussm$, $\upsilon^\ominussm$, $\bar\upsilon^\oplussm$,
$\bar\upsilon^\ominussm$ transform in the respective vector and
scalar representations of the Lorentz algebra $so(d-1,1)$. We assume
the following hermitian conjugation rule for the derivatives
$\partial^{a\dagger} = - \partial^a$. We use operators constructed
out of the derivatives, coordinates and oscillators,
\beq
\label{manold-18032012-01} && \hspace{-0.5cm} \Box \equiv
\partial^a\partial^a\,,\qquad\quad  \ \ x\partial \equiv
x^a\partial^a\,,\qquad\ \   x^2\equiv x^ax^a\,,
\\
\label{manold-18032012-02} &&  \hspace{-0.5cm} \alpha\partial \equiv
\alpha^a\partial^a\,,\qquad\quad \bar\alpha\partial \equiv
\bar\alpha^a\partial^a\,,\qquad \ \ \alpha^2 \equiv
\alpha^a\alpha^a\,, \qquad\quad   \bar\alpha^2 \equiv
\bar\alpha^a\bar\alpha^a\,,\qquad
\\
\label{manold-31102011-05} &&  \hspace{-0.5cm} N_\alpha \equiv
\alpha^a \bar\alpha^a \,,\qquad \ \ \ \ N_\zeta \equiv \zeta
\bar\zeta \,, \qquad \quad \ \ \kwh \equiv N_\alpha +
\frac{d-6}{2}\,,\qquad \qquad \
\\
\label{manold-31102011-06} &&  \hspace{-0.5cm} N_{\upsilon^\oplussm}
\equiv \upsilon^\oplussm \bar\upsilon^\ominussm\,, \qquad \ \
N_{\upsilon^\ominussm} \equiv \upsilon^\ominussm
\bar\upsilon^\oplussm\,, \qquad   N_\upsilon \equiv
N_{\upsilon^\oplussm} + N_{\upsilon^\ominussm} \,, \qquad \Delta'
\equiv N_{\upsilon^\oplussm} - N_{\upsilon^\ominussm} \,,\qquad\quad
\\
\label{manold-18032012-03} &&  \hspace{-0.5cm} \Vwt^a \equiv
\alpha^a - \alpha^2 \frac{1}{2N_\alpha + d-2}\bar\alpha^a \,,
\\
\label{manold-18032012-04} && \hspace{-0.5cm} V^a \equiv \alpha^a -
\alpha^2 \frac{1}{2N_\alpha + d}\bar\alpha^a \,, \hspace{1.8cm} V
\equiv V^a\partial^a\,,\qquad
\\
\label{manold-18032012-05} &&  \hspace{-0.5cm} \Vb_\perp^a \equiv
\bar\alpha^a - \half \alpha^a \bar\alpha^2\,, \hspace{3cm} \Vb_\perp
\equiv \Vb_\perp^a \partial^a\,,
\\
\label{manold-18032012-06} &&  \hspace{-0.5cm} \mubf \equiv 1
-\frac{1}{4}\alpha^2\bar\alpha^2\,, \hspace{3.4cm} \Pi^\smponetwo
\equiv 1 - \alpha^2 \frac{1}{2(2N_\alpha +d)}\bar\alpha^2\,.
\eeq

Throughout the paper the notation $k' \in [n]_2$ implies that $k'
=-n,-n+2,-n+4,\ldots,n-4, n-2,n$:
\be
k' \in [n]_2 \quad \Longrightarrow \quad k'
=-n,-n+2,-n+4,\ldots,n-4, n-2,n\,.
\ee

We adopt the following conventions for the light-cone frame. The
space-time coordinates are decomposed as $x^a= x^+, x^-,x^i$, where
the coordinates $x^\pm$ are defined as $x^\pm = (x^{d-1} \pm
x^0)/\sqrt{2}$ and $x^+$ is taken to be a light-cone time. Vector
indices of the $so(d-2)$ algebra take values $i,j =1,\ldots, d-2$.
We use the following conventions for the derivatives: $
\partial^i=\partial_i\equiv\partial/\partial x^i$, $\partial^\pm=\partial_\mp \equiv
\partial/\partial x^\mp$.

\subsection{Global conformal symmetries }

The conformal algebra $so(d,2)$ of $d$-dimensional space-time taken
to be in basis of the Lorentz algebra $so(d-1,1)$ consists of
translation generators $P^a$, dilatation generator $D$, conformal
boost generators $K^a$, and generators $J^{ab}$ which span
$so(d-1,1)$ Lorentz algebra. We assume the following normalization
for commutators of the conformal algebra:
\beq
\label{ppkk}
&& {}[D,P^a]=-P^a\,, \hspace{2cm}  {}[P^a,J^{bc}]=\eta^{ab}P^c
-\eta^{ac}P^b \,,
\\
&& [D,K^a]=K^a\,, \hspace{2.2cm} [K^a,J^{bc}]=\eta^{ab}K^c -
\eta^{ac}K^b\,,
\\
\label{pkjj} && \hspace{2.5cm} {}[P^a,K^b]=\eta^{ab}D-J^{ab}\,,
\\
&& \hspace{2.5cm} [J^{ab},J^{ce}]=\eta^{bc}J^{ae}+3\hbox{ terms} \,.
\eeq

Let $\phik$ denotes field propagating in flat space-time of
dimension $d\geq 4$. Let Lagrangian for the free field $\phik$ be
conformal invariant. This implies, that Lagrangian is invariant with
respect to transformation (invariance of the Lagrangian is assumed
to be up to total derivatives)
\be \label{man-14022012-01} \delta_{\hat{G}} \phik  = \hat{G} \phik
\,, \ee
where realization of the conformal algebra generators $\hat{G}$ in
terms of differential operators takes the form
\beq
\label{conalggenlis01} && P^a = \partial^a \,,
\\
\label{JIJdef}
\label{conalggenlis02} && J^{ab} = x^a\partial^b -  x^b\partial^a +
M^{ab}\,,
\\
\label{conalggenlis03} && D = x\partial  + \Delta\,,
\\
\label{conalggenlis04} && K^a = K_{\Delta,M}^a + R^a\,,
\\
\label{conalggenlis04n1} && \hspace{1cm} K_{\Delta,M}^a \equiv
-\frac{1}{2}x^2\partial^a + x^a D + M^{ab}x^b\,.
\eeq
In \rf{conalggenlis02}-\rf{conalggenlis04n1}, $\Delta$ is operator
of conformal dimension, $M^{ab}$ is spin operator of the Lorentz
algebra,
\be  [M^{ab},M^{ce}]=\eta^{bc}M^{ae}+3\hbox{ terms} \,, \ee
and $R^a$ is operator depending on derivatives with respect to
space-time coordinates and not depending on space-time coordinates
$x^a$, $[P^a,R^b]=0$.%
\footnote{For the case of conformal currents and shadow fields
studied in Refs.\cite{Metsaev:2008fs,Metsaev:2010zu}, the operator
$R^a$ does not dependent on the derivatives.}
The spin operator of the Lorentz algebra is well known for arbitrary
spin conformal field. In higher-derivative formulations of conformal
fields, the operator $R^a$ is often equal to zero, while in the
ordinary-derivative approach, we develop in this paper, the operator
$R^a$ is non-trivial. This implies that complete description of
conformal fields in the ordinary-derivative approach requires
finding not only gauge invariant Lagrangian but also the operator
$R^a$ as well. It turns out that requiring the Lagrangian to be
invariant under gauge transformations and conformal algebra
transformations we determine both the Lagrangian and the operator
$R^a$.

\newsection{ \large On-shell degrees of freedom of conformal
field in higher -derivative approach}\label{on-shell}

As is well known, higher-derivative theory can be cast into
ordinary-derivative form by using the Ostrogradsky method. We note
however that the conventional use of this method leads to field
content involving only auxiliary fields. In
Ref.\cite{Metsaev:2007fq}, we proposed method which leads
automatically to the field content involving both auxiliary and
Stueckelberg fields. Our method for finding field content that we
use for building the ordinary-derivative gauge invariant Lagrangian
involves the following two steps.

\noindent {\bf i}) Starting with higher-derivative formulation of
conformal field, we use the light-cone gauge to classify on-shell
D.o.F of the conformal field according to irreps of the $so(d-2)$
algebra.

\noindent {\bf ii}) We replace the $so(d-2)$ algebra irreps entering
the on-shell D.o.F of the conformal field by the corresponding
representations of the Lorentz algebra $so(d-1,1)$.

For arbitrary spin values, $s\geq 1$, and arbitrary dimension of
space, $d\geq 4$, on-shell D.o.F of the spin-$s$ conformal field in
$d$-dimensional space have not been discussed so far in the
literature. Therefore, in this section, we present our result for
on-shell D.o.F of the totally symmetric arbitrary spin-$s$ conformal
field in $d$-dimensional space, $d\geq 4$. Since our method for
counting on-shell D.o.F is based on the use of higher-derivative
formulation, we begin with recalling of higher-derivative
formulations of conformal fields available in the literature.

\subsection{ Higher-derivative formulation of arbitrary spin
conformal field} \label{subsec-301}

At present time, two higher-derivative formulations of conformal
totally symmetric fields are available in the literature. These
formulations were developed in
Refs.\cite{Fradkin:1985am,Segal:2002gd} and
Ref.\cite{Metsaev:2009ym}. Because formulation developed in
Ref.\cite{Metsaev:2009ym} turns out to be more convenient for the
discussion of on-shell D.o.F we begin with the review of our results
in Ref.\cite{Metsaev:2009ym}. In section \ref{subsubsec-02}, we
describe how formulation in Refs.\cite{Fradkin:1985am,Segal:2002gd}
is obtained from our higher-derivative approach in
Ref.\cite{Metsaev:2009ym}. In section \ref{subsubsec-03}, we discuss
higher-derivative light-cone gauge Lagrangian which provides easy
and quick access to on-shell D.o.F. of conformal fields. Outline of
the derivation of on-shell D.o.F by using approach in
Refs.\cite{Fradkin:1985am,Segal:2002gd} may be found in Appendix A.

\subsubsection{ Higher-derivative Lagrangian with Stueckelberg
and auxiliary fields} \label{subsubsec-301}

In Ref.\cite{Metsaev:2009ym}, to discuss higher-derivative gauge
invariant formulation of the totally symmetric arbitrary spin-$s$
conformal field in flat space of even dimension $d\geq 4$ we use the
following set of scalar, vector, and tensor fields of the Lorentz
algebra $so(d-1,1)$:
\be \label{manold-06032012-02}
\phi^{a_1\ldots a_{s'}}\,, \hspace{2cm} s'=0,1,\ldots,s\,.
\ee
We note that\\
\ibf) In \rf{manold-06032012-02}, the fields $\phi$ and $\phi^a$ are
the respective scalar and vector fields of the Lorentz algebra,
while the field $\phi^{a_1\ldots a_{s'}}$, $s'>1$, is rank-$s'$
totally symmetric tensor field of the Lorentz algebra $so(d-1,1)$.
The tensor fields $\phi^{a_1\ldots a_{s'}}$ with $s'\geq 4
$ satisfy the double-tracelessness constraint,%
\be \phi^{aabba_5\ldots a_{s'}}=0\,, \qquad s'\geq 4\,. \ee
\iibf) Conformal dimension of the field $\phi^{a_1\ldots a_{s'}}$ is
given by
\be
\Delta(\phi^{a_1\ldots a_{s'}}) = 2-s'\,.
\ee
\iiibf) Fields \rf{manold-06032012-02} are subject to differential
constraints. To discuss the differential constraints we use the
oscillators $\alpha^a$, $\zeta$ and collect fields
\rf{manold-06032012-02} into ket-vector $\phik$ defined as
\be \label{manold-06032012-03}
|\phi\rangle \equiv \sum_{s'=0}^s
\frac{\zeta^{s-s'}}{\sqrt{(s-s')!}} |\phi^{s'}\rangle \,,
\qquad\quad |\phi^{s'}\rangle \equiv \frac{1}{s'!} \alpha^{a_1}
\ldots  \alpha^{a_{s'}} \phi^{a_1\ldots a_{s'}} |0\rangle\,.
\ee
In terms of ket-vector $\phik$ \rf{manold-06032012-03}, the
differential constraints can be presented as
\beq
\label{manold-06032012-04} && \hspace{-1cm} \Cb_\sh \phik  =  0 \,,
\\
\label{manold-06032012-04n1} && \Cb_\sh  \equiv  \albpar - \half
\alpar \bar\alpha^2  - \eb_{1,\sh} \Pi^\smponetwo + \half e_{1,\sh}
\bar\alpha^2\,,
\\
\label{manold-07032012-01} && e_{1,\sh} \equiv  \zeta \ewt_1 \Box\,,
\qquad
\eb_{1,\sh} \equiv  - \ewt_1 \bar\zeta\,, \qquad \ewt_1 \equiv
\Bigl(\frac{2s+d-4-N_\zeta}{2s+d-4-2N_\zeta}\Bigr)^{1/2}\,,
\eeq
where operator $\Pi^\smponetwo$ is defined in
\rf{manold-18032012-06}.

Lagrangian for the spin-$s$ conformal field found in
Ref.\cite{Metsaev:2009ym} takes the form
\be \label{manold-06032012-05}
\LL = \half \phibr \mubf \Box^{\kwh +1}\phik \,, \qquad \mubf \equiv
1 -\frac{1}{4}\alpha^2\bar\alpha^2\,, \qquad \kwh \equiv N_\alpha +
\frac{d-6}{2}\,,
\ee
where operator $N_\alpha$ is defined in \rf{manold-31102011-05}.
Note that throughout this paper bra-vectors are defined according
the rule $\phibr \equiv (\phik)^\dagger$.

To illustrate the structure of the Lagrangian we note that, in terms
of the tensor fields $\phi^{a_1\ldots a_{s'}}$, Lagrangian
\rf{manold-06032012-05} takes the form
\beq
\label{manold27032012-01} && \LL = \sum_{s'=0}^s \LL_{s'}\,,
\\
&& \LL_{s'} = \frac{1}{2 s'!}\Bigl( \phi^{a_1\ldots a_{s'}}
\Box^{k_{s'}+1} \phi^{a_1\ldots a_{s'}}
- \frac{s'(s'-1)}{4} \phi^{aaa_3\ldots a_{s'}} \Box^{k_{s'}+1}
\phi^{bba_3\ldots a_{s'}}\Bigr)\,,\qquad
\\
\label{oldman-06032012-01} && \hspace{1cm} k_{s'} \equiv
s'+\frac{d-6}{2}\,.
\eeq

To discuss realization of gauge symmetries in our higher-derivative
approach we introduce the following set of gauge transformation
parameters:
\be \label{manold-06032012-06}
\xi^{a_1\ldots a_{s'}}\,,\hspace{1.5cm} s'=0,1,\ldots,s-1\,.
\ee
We note that, in \rf{manold-06032012-06}, the gauge transformation
parameters $\xi$ and $\xi^a$ are the respective scalar and vector
fields of the Lorentz algebra, while the gauge transformation
parameter $\xi^{a_1\ldots a_{s'}}$, $s' \geq 2$, is rank-$s'$
totally symmetric traceless tensor field of the Lorentz algebra
$so(d-1,1)$,
\be \xi^{aaa_3\ldots a_{s'}}=0\,, \qquad s'\geq 2\,. \ee
Now, as usually, we collect gauge transformation parameters
\rf{manold-06032012-06} into ket-vector $\xik$ defined by
\be  \label{manold-06032012-06n1}
\xik \equiv \sum_{s'=0}^{s-1}
\frac{\zeta^{s-1-s'}}{\sqrt{(s-1-s')!}} |\xi^{s'}\rangle \,,
\qquad\quad |\xi^{s'}\rangle \equiv \frac{1}{s'!} \alpha^{a_1}
\ldots  \alpha^{a_{s'}} \xi^{a_1\ldots a_{s'}} |0\rangle\,.
\ee

Gauge transformations can entirely be written in terms of the
ket-vectors $\phik$ and $\xik$. This is to say that gauge
transformations take the form
\be  \label{manold-06032012-18}
\delta |\phi\rangle  =   G_\sh |\xi\rangle \,, \qquad G_\sh \equiv
\alpar - e_{1,\sh} - \alpha^2\frac{1}{2N_\alpha + d- 2}
\eb_{1,\sh}\,,
\ee
where operators $e_{1,\sh}$, $\eb_{1,\sh}$ are given in
\rf{manold-07032012-01}. We make sure that differential constraint
\rf{manold-06032012-04} is invariant under gauge transformations
\rf{manold-06032012-18}. Also we make sure that Lagrangian
\rf{manold-06032012-05} is invariant under gauge transformations
\rf{manold-06032012-18} provided the field $\phik$ satisfies
differential constraint \rf{manold-06032012-04}.

{\bf Higher-derivative Lagrangian for spin-2 field}. For the later
use and illustration purposes let us demonstrate our
higher-derivative approach for the case of spin-2 conformal field.
For this case, the field content involves rank-2 tensor field
$\phi^{ab}$, vector field $\phi^a$, and scalar field $\phi$. In
terms of these fields, Lagrangian \rf{manold-06032012-05} and
differential constraint \rf{manold-06032012-04} take the form
\beq
\label{oldman-21032012-07} && \hspace{-1cm} \LL  =  \frac{1}{4}
\phi^{ab} \Box^{k+1}\phi^{ab} - \frac{1}{8} \phi^{aa}
\Box^{k+1}\phi^{bb} + \half \phi^a \Box^k \phi^a + \half \phi
\Box^{k-1} \phi \,, \quad k \equiv \frac{d-2}{2}\,,\quad
\\
\label{oldman-21032012-08} && \hspace{1cm} \partial^b \phi^{ab} -
\half \partial^a \phi^{bb} + \phi^a = 0 \,,
\\
\label{oldman-21032012-09} && \hspace{1cm} \partial^a \phi^a + \half
\Box \phi^{aa} + u \phi = 0 \,,  \qquad u \equiv
\Bigl(2\frac{d-1}{d-2}\Bigr)^{1/2}\,.
\eeq
Lagrangian \rf{oldman-21032012-07} and constraints
\rf{oldman-21032012-08}, \rf{oldman-21032012-09} are invariant under
the gauge transformations
\beq
&& \delta \phi^{ab} =\partial^a \xi^b +
\partial^b \xi^a + \frac{2}{d-2} \eta^{ab} \xi\,,
\nonumber\\
\label{oldman-21032012-15} && \delta \phi^a = \partial^a \xi - \Box
\xi^a \,,
\\
&& \delta \phi = - u \Box \xi\,,
\nonumber
\eeq
where $\xi$, $\xi^a$ are gauge transformation parameters.

\subsubsection{Higher-derivative Lagrangian in Stueckelberg gauge
frame}\label{subsubsec-02}

Lagrangian of the conformal field in
Refs.\cite{Fradkin:1985am,Segal:2002gd} is obtained from our
Lagrangian \rf{manold-06032012-05} by using a Stueckelberg gauge
frame. Therefore to illustrate how our approach is related to the
one in Refs.\cite{Fradkin:1985am,Segal:2002gd}, we now present
Stueckelberg gauge fixed Lagrangian of the spin-$s$ conformal field.

From gauge transformations \rf{manold-06032012-18}, we see that a
field $\bar\alpha^2\phik$ transforms as Stueckelberg field.
Therefore we can gauge away the field $\bar\alpha^2\phik$ by using
the Stueckelberg gauge condition,
\be  \label{manold-06032012-17}
\bar\alpha^2\phik = 0 \,.
\ee
Using \rf{manold-06032012-17} and differential constraint
\rf{manold-06032012-04}, we find the following relations:
\beq
\label{manold-06032012-07} && \bar\alpha^2 |\phi^{s'}\rangle
=0\,,\hspace{4.1cm} s'=2,3,\ldots,s\,,
\\
\label{manold-06032012-08} && |\phi^{s'}\rangle = X_{s'}
(\albpar)^{s-s'} |\phi^s\rangle\,, \hspace{2cm} s'=0,1,\ldots,s-1\,,
\\
&& X_{s'} \equiv \frac{(-)^{s-s'}}{(s-s')!} \Bigl(
\frac{2^{s-s'}\Gamma(s+s'+d-3)\Gamma(s+\frac{d-2}{2})}{
\Gamma(2s+d-3)\Gamma(s'+\frac{d-2}{2}) }\Bigr)^{\half}\,.
\eeq
Relation \rf{manold-06032012-07} tells us that the tensor fields
$\phi^{a_1\ldots a_{s'}}$, $s'=2,3,\ldots,s$, are traceless, while,
from relation \rf{manold-06032012-08}, we learn that the fields
$\phi^{a_1\ldots a_{s'}}$, $s'=0,1,\ldots,s-1$, can be expressed in
terms of the traceless rank-$s$ tensor field $\phi^{a_1\ldots a_s}$.
Plugging \rf{manold-06032012-08} into \rf{manold-06032012-05} and
ignoring total derivative, we get the Stueckelberg gauge frame
Lagrangian,
\beq \label{manold-06032012-15}
\LL & = & \half \sum_{s'=0}^s
\frac{2^{s-s'}(s'+\frac{d-2}{2})_{s-s'}}{(s-s')! (s+s'+d-3)_{s-s'}}
\langle (\albpar)^{s-s'}\phi^s|\Box^{k_{s'}+1} |(\albpar)^{s-s'}
\phi^s\rangle\,,
\eeq
where we use the notation $(p)_q$ to indicate the Pochhammer symbol,
$(p)_q \equiv \frac{\Gamma(p+q)}{\Gamma(p)}$, while $k_{s'}$ is
defined in \rf{oldman-06032012-01}. To illustrate the structure of
the Lagrangian we note that, in terms of the tensor field
$\phi^{a_1\ldots a_s}$, Lagrangian \rf{manold-06032012-15} takes the
form
\beq \label{manold-06032012-16}
&& \LL  =  \half \sum_{s'=0}^s
\frac{2^{s-s'}(s'+\frac{d-2}{2})_{s-s'}}{s'!(s-s')!
(s+s'+d-3)_{s-s'}} (\partial^{s-s'}\phi)^{a_1\ldots
a_{s'}}\Box^{k_{s'}+1} (\partial^{s-s'}\phi)^{a_1\ldots a_{s'}}\,,
\qquad
\\
&& \hspace{1cm} (\partial^{s-s'}\phi)^{a_1\ldots a_{s'}} \equiv
\partial^{b_1} \ldots \partial^{b_{s-s'}} \phi^{b_1\ldots b_{s-s'}a_1\ldots
a_{s'}}\,.
\eeq
Representation for Lagrangian given in \rf{manold-06032012-16} was
found in Ref.\cite{Metsaev:2009ym}. In earlier literature,
alternative representations for Lagrangian of conformal fields may
be found in Refs.\cite{Fradkin:1985am,Segal:2002gd}. For the readers
convenience, we write down the leading terms in Lagrangian
\rf{manold-06032012-16},
\beq
\LL & = & \frac{1}{2s!} \phi^{a_1\ldots a_s} \Box^{k_s+1}
\phi^{a_1\ldots a_s} +  \frac{1}{2(s-1)!} (\partial\phi)^{a_1\ldots
a_{s-1}}\Box^{k_s} (\partial\phi)^{a_1\ldots a_{s-1}}
\nonumber\\
& +  & \frac{1}{4(s-2)!} \frac{2s+d-6}{2s+d-5}
(\partial^2\phi)^{a_1\ldots a_{s-2}}\Box^{k_s-
1}(\partial^2\phi)^{a_1\ldots a_{s-2}} + \ldots \,.
\eeq

\subsubsection{Higher-derivative Lagrangian in light-cone gauge
frame}\label{subsubsec-03}

Light-cone gauge frame provides easy and quick access to on-shell
D.o.F of arbitrary spin conformal field. Therefore, we now discuss
light-cone gauge-fixed Lagrangian. To this end we impose the
light-cone gauge condition,
\be \label{manold-06032012-09}
\bar\alpha^+ \Pi^\smponetwo \phik =0\,,
\ee
and analyze differential constraint \rf{manold-06032012-04} for the
field $|\phi\rangle$. We find that solution to the differential
constraint can be expressed in terms of the light-cone ket-vector
$|\phi^{\rm l.c.}\rangle$,
\beq
\label{manold-06032012-10} && |\phi\rangle =
\exp\Bigl(-\frac{\alpha^+}{\partial^+}(\bar\alpha^i\partial^i -
\eb_1)\Bigr) |\phi^{\rm l.c.}\rangle\,,
\\
\label{manold-06032012-11} && \bar\alpha^i \bar\alpha^i | \phi^{\rm
l.c.}\rangle = 0 \,,
\\
\label{manold-06032012-12} && |\phi^{\rm l.c.}\rangle \equiv
|\phi\rangle\bigr|_{\alpha^+=\alpha^-=0} \,.
\eeq
From \rf{manold-06032012-12}, we learn that the light-cone
ket-vector $|\phi^{\rm l.c.}\rangle$ is obtained from $\phik$
\rf{manold-06032012-03} by equating $\alpha^+ = \alpha^- = 0$. Thus,
we are left with fields $\phi^{i_1\ldots i_{s'}}$,
$s'=0,1,\ldots,s$, which, for $s \geq 2$, are traceless $so(d-2)$
algebra tensor fields, $\phi^{iii_3\ldots i_{s'}}=0$. These fields
constitute the field content of the light-cone gauge frame. Note
that the fields $\phi^{i_1\ldots i_{s'}}$, $s'=0,1,\ldots,s$, are
not subject to any differential constraints as it should be in the
light-cone gauge frame. Making use of
\rf{manold-06032012-10}-\rf{manold-06032012-12}, we find that gauge
invariant Lagrangian \rf{manold-06032012-05} leads to the following
light-cone gauge Lagrangian:
\be \label{manold-06032012-14}
\LL^{{\rm l.c.}} =\half \langle \phi^{{\rm l.c.}}|\Box^{\kwh+1}
|\phi^{{\rm l.c.}}\rangle\,.
\ee
To illustrate structure of the light-cone gauge Lagrangian we note
that, in terms of the fields $\phi^{i_1\ldots i_{s'}}$, Lagrangian
\rf{manold-06032012-14} takes the form
\beq \label{manold-06032012-23}
\LL^{{\rm l.c.}} = \sum_{s'=0}^s \LL_{s'}^{{\rm l.c.}}\,, \qquad
\LL_{s'}^{{\rm l.c.}} = \frac{1}{2 s'!} \phi^{i_1\ldots i_{s'}}
\Box^{k_{s'}+1} \phi^{i_1\ldots i_{s'}}\,,
\eeq
where $k_{s'}$ is given in \rf{oldman-06032012-01}. In the next
section, we demonstrate how light-cone gauge  Lagrangian
\rf{manold-06032012-23} can be used for counting on-shell D.o.F of
arbitrary spin conformal field.

\subsection{ On-shell degrees of freedom of arbitrary spin-$s$ conformal field}

As we have already said, to discuss on-shell D.o.F of the conformal
field we exploit the light-cone gauge and use fields transforming in
irreps of the $so(d-2)$
algebra.%
\footnote{ Discussion of alternative methods for counting on-shell
D.o.F may be found in Refs.\cite{Lee:1982cp,Buchbinder:1987vp}.}
Namely, we decompose on-shell D.o.F into irreps of the $so(d-2)$
algebra. At the end of this section, we demonstrate that on-shell
D.o.F of the totally symmetric spin-$s$ conformal field in
$d$-dimensional space, $d\geq 4$, are described by the following set
of fields of the $so(d-2)$ algebra:
\beq
\label{manold-11042012-01} && \phi_{k'}^{i_1\ldots i_{s'}}\,,
\hspace{2cm}
s'= \left\{\begin{array}{l}
0,1,\ldots,s;\qquad  \hbox{for }\  d \geq 6;
\\[3pt]
1,2,\ldots,s;\qquad  \hbox{for }d = 4;
\end{array}\right.
\hspace{2cm} k' \in  [k_{s'}]_2\,;\qquad
\\
&& \hspace{3.6cm} k_{s'} \equiv s'+ \frac{d-6}{2}\,,
\eeq
where vector indices of the $so(d-2)$ algebra take values
$i=1,2,\ldots,d-2$. In \rf{manold-11042012-01}, the fields
$\phi_{k'}$ and $\phi_{k'}^i$ are the respective scalar and vector
fields of the $so(d-2)$ algebra, while the field
$\phi_{k'}^{i_1\ldots i_{s'}}$, $s'\geq 2$, is rank-$s'$ totally
symmetric traceless tensor field of the $so(d-2)$ algebra,
\be
\label{lcepsdoutracon01} \phi_{k'}^{iii_3\ldots i_{s'}}=0\,, \qquad
s'\geq 2\,,
\ee
i.e., the tensor field $\phi_{k'}^{i_1\ldots i_{s'}}$ transforms as
irreps of the $so(d-2)$ algebra. Note that the scalar fields
$\phi_{k'}$ enter the field content only when $d\geq 6$. Obviously,
set of fields in \rf{manold-11042012-01} is related to non-unitary
representation of the conformal algebra $so(d,2)$.%
\footnote{ By now, unitary representations of (super)conformal
algebras that are relevant for elementary particles are well
understood (see, e.g., Refs.\cite{Evans}-\cite{Dobrev:1985qv}). In
contrast to this, non-unitary representations deserve to be
understood better.}

Alternatively, for $d\geq 6$, field content \rf{manold-11042012-01}
can be represented as
\beq
\label{lcspin2DoF01} &&\phi_{k'}^{i_1\ldots i_s}\,, \hspace{2.5cm}
k'\in [k_s]_2\,;
\\
\label{lcspin2DoF02} &&\phi_{k'}^{i_1\ldots i_{s-1}}\,,
\hspace{2.2cm} k'\in [k_s-1]_2\,;
\\
&&\ldots \quad \ldots \hspace{2.5cm}  \ldots \quad \ldots
\nonumber\\
&&\ldots \quad \ldots \hspace{2.5cm} \ldots \quad \ldots
\nonumber\\
\label{lcspin2DoF03ex1} && \phi_{k'}^i\,, \hspace{3cm} k'\in
[k_s-s+1]_2\,;
\\
\label{lcspin2DoF03} && \phi_{k'}\,, \hspace{3cm} k'\in [k_s-s]_2\,;
\eeq
\be k_s \equiv s+ \frac{d-6}{2}\,, \ee
while, for $d=4$, field content is given in
\rf{lcspin2DoF01}-\rf{lcspin2DoF03ex1}.

Total number of on-shell D.o.F shown in \rf{manold-11042012-01} is
given by

\be \label{nunus'def01}
\nbf = \half (d-3)(2s+d-2)(2s+d-4) \frac{(s+d-4)!}{s!(d-2)!} \,.
\ee
For the particular values of $s$ and $d$, relation \rf{nunus'def01}
leads to the following expressions for $\nbf$:
\beq
\label{man-18-022012-01} && \nbf\Bigr|_{s- {\rm arbitrary};\,\,
d=4}\, = \, s(s+1) \,,
\\
\label{man-18-022012-02} &&  \nbf\Bigr|_{s=1;\,\, d-{\rm
arbitrary}}\,  = \,  \frac{1}{2}d(d-3)\,,
\\
\label{man-18-022012-03} && \nbf\Bigr|_{s=2;\,\, d-{\rm
arbitrary}}\,  = \, \frac{1}{4}d(d-3)(d+2) \,.
\eeq
Results for $\nbf$ in \rf{man-18-022012-01} and
\rf{man-18-022012-03} were obtained in Ref.\cite{Fradkin:1985am},
while the expression for $\nbf$ in \rf{man-18-022012-02} was
obtained in Ref.\cite{Metsaev:2007fq}. Thus, our result in
\rf{nunus'def01} agrees with the previously reported results related
to the particular values of $s$ and $d$ in
\rf{man-18-022012-01}-\rf{man-18-022012-03} and gives expression for
$\nbf$ in \rf{nunus'def01} corresponding to the arbitrary values of
$s$ and $d$.

To summarize, our result in \rf{lcspin2DoF01}-\rf{lcspin2DoF03}
gives decomposition of the on-shell D.o.F into irreps of $so(d-2)$
algebra, while expression for $\nbf$ \rf{nunus'def01} gives the
total number of on-shell D.o.F appearing in
\rf{lcspin2DoF01}-\rf{lcspin2DoF03}.%
\footnote{ For the case of $s=2$, $d=4$, decomposition of D.o.F into
irreps of the $so(2)$ algebra was carried out in
Ref.\cite{Lee:1982cp}. For the case of $s=2$, $d\geq 4$,
decomposition of D.o.F into irreps of the $so(d-2)$ algebra was
carried out in Ref.\cite{Metsaev:2007fq}.}

For the reader convenience, we now explain how $\nbf$ given in
\rf{nunus'def01} is obtained from decomposition in
\rf{lcspin2DoF01}-\rf{lcspin2DoF03}. By definition, $\nbf$ given in
\rf{nunus'def01} is a sum of tensorial components of fields
\rf{lcspin2DoF01}-\rf{lcspin2DoF03} subject to tracelessness
constraint \rf{lcepsdoutracon01}. This is to say that $\nbf$ can be
represented as
\beq
\label{nunus'def02} && \nbf = \sum_{s'=0}^s \nbf_{s'}\,, \qquad
\nbf_{s'} = \sum_{k'\in [k_{s'}]_2} \nbf(\phi_{k'}^{s'})\,,
\\
\label{nunus'def04} && \nbf(\phi_{k'}^{s'}) =
\frac{(s'+d-5)!}{(d-4)!\,s'!} (2s'+d-4) \,,
\eeq
where $\nbf(\phi_{k'}^{s'})$ stands for D.o.F of rank-$s'$ traceless
tensor field $\phi_{k'}^{i_1\ldots i_{s'}}$. In other words,
$\nbf(\phi_{k'}^{s'})$ is a dimension of the rank-$s'$ traceless
tensor field of the $so(d-2)$ algebra. Taking into account
\rf{nunus'def04} and the fact that there are $k_{s'}+1$ rank-$s'$
traceless tensor fields $\phi_{k'}^{s'}$, we note that relation for
$\nbf_{s'}$ in \rf{nunus'def02} amounts to
\be  \label{nunus'def05}
\nbf_{s'} = \half \frac{(s'+d-5)!}{(d-4)!\, s'!} (2s'+d-4)^2 \,.
\ee
Plugging $\nbf_{s'}$ \rf{nunus'def05} into expression for $\nbf$ in
\rf{nunus'def02} and using the textbook formula,
\be \sum_{s'=0}^s \frac{(s' + t )!}{s'!} = \frac{(s+ t
+1)!}{(t+1)s!}\,,
\ee
we arrive at the expression for $\nbf$ in \rf{nunus'def01}.

Now let us discuss the derivation of on-shell D.o.F given in
\rf{lcspin2DoF01}-\rf{lcspin2DoF03}. To this end we use light-cone
gauge Lagrangian \rf{manold-06032012-23}. Using light-cone gauge
Lagrangian \rf{manold-06032012-23}, we obtain the following
higher-derivative equations of motion:
\be \label{oldman080321012-01}
\Box^{k_{s'}+1}\phi^{i_1\ldots i_{s'}} = 0 \,, \qquad s'=0,1,\ldots
, s\,.\
\ee
Now, all that remains is to note that it is the use of the set of
the fields given in \rf{lcspin2DoF01}-\rf{lcspin2DoF03} that leads
to the ordinary-derivative form of equations
\rf{oldman080321012-01},
\be \label{oldman080321012-02}
\Box\phi_{k'}^{i_1\ldots i_{s'}} - \phi_{k'+2}^{i_1\ldots i_{s'}}  =
0 \,, \qquad s'=0,1,\ldots , s\,, \qquad k'\in [k_{s'}]_2\,,
\ee
where we use the identification $\phi_{-k_{s'}}^{i_1\ldots i_{s'}}
\equiv \phi^{i_1\ldots i_{s'}}$, $s'=0,1,\ldots, s$, and $k_{s'}$ is
defined in \rf{oldman-06032012-01}. Note that, for $d=4$ and $s'=0$,
Eqs.\rf{oldman080321012-01} imply $\phi=0$. Thus we see that scalar
field does not contribute to on-shell D.o.F when $d=4$.

As a side remark we note that, for $d\geq 6$, ordinary-derivative
light-cone gauge Lagrangian that leads to equations
\rf{oldman080321012-02} is given by
\be \label{oldman-14042012-01}
\LL^{{\rm l.c.}} =  \sum_{s'=0}^s \sum_{k' \in [k_{s'}]_2}
\LL_{s',k'}^{{\rm l.c.}}\,,
\qquad\quad
\LL_{s',k'}^{{\rm l.c.}} = \half \phi_{-k'}^{i_1\ldots i_{s'}} \Box
\phi_{k'}^{i_1\ldots i_{s'}} - \half \phi_{-k'}^{i_1\ldots
i_{s'}}\phi_{k'+2}^{i_1\ldots i_{s'}}\,.
\ee
For $d=4$, we should remove contribution of $\LL_{0,k'}^{\rm
l.c.}$-terms to Lagrangian $\LL^{{\rm l.c.}}$
\rf{oldman-14042012-01}.

\newsection{ \large Ordinary-derivative gauge
invariant Lagrangian}\label{lagran}

To discuss ordinary-derivative gauge invariant formulation of the
totally symmetric arbitrary spin-$s$ conformal field in flat space
of even dimension $d\geq 4$ we use the following set of scalar,
vector, and tensor fields of the Lorentz algebra $so(d-1,1)$:
\beq
\label{phiset01} && \phi_{k'}^{a_1\ldots a_{s'}}\,, \hspace{2cm}
s'= \left\{\begin{array}{l}
0,1,\ldots,s;\qquad  \hbox{for }\  d \geq 6;
\\[3pt]
1,2,\ldots,s;\qquad  \hbox{for }d = 4;
\end{array}\right.
\hspace{2cm} k' \in  [k_{s'}]_2\,;\qquad
\\
\label{oldman-21032012-10} && \hspace{3.6cm} k_{s'} \equiv s'+
\frac{d-6}{2}\,.
\eeq
Alternatively, for $d\geq 6$, field content \rf{phiset01} can be
represented as
\beq
\label{covspin2DoF01} &&\phi_{k'}^{a_1\ldots a_s}\,, \hspace{3cm}
k'\in [k_s]_2\,;
\\
\label{covspin2DoF02} &&\phi_{k'}^{a_1\ldots a_{s-1}}\,,
\hspace{2.7cm} k' \in [k_s-1]_2\,;
\\
&&\ldots\ldots \hspace{3.4cm} \ldots \ldots\ldots
\nonumber\\[-10pt]
&&\ldots\ldots \hspace{3.4cm} \ldots \ldots\ldots
\nonumber\\
\label{covspin2DoF03nn01}&& \phi_{k'}^a\,, \hspace{3.8cm} k' \in
[k_s-s+1]_2\,;
\\
\label{covspin2DoF03} && \phi_{k'}\,, \hspace{3.8cm} k' \in
[k_s-s]_2\,;
\\
&& \qquad\qquad \label{olmman-05022012-01} k_s \equiv s+
\frac{d-6}{2}\,,
\eeq
while, for $d=4$, field content \rf{phiset01} is given in
\rf{covspin2DoF01}-\rf{covspin2DoF03nn01}. We note that\\
\ibf) In \rf{phiset01}, the fields $\phi_{k'}$ and $\phi_{k'}^a$ are
the respective scalar and vector fields of the Lorentz algebra,
while the field $\phi_{k'}^{a_1\ldots a_{s'}}$, $s'>1$, is rank-$s'$
totally symmetric tensor field of the Lorentz algebra $so(d-1,1)$.
Note that the scalar fields $\phi_{k'}$ enter the field content only
when $d\geq 6$.
\\
\iibf) The tensor fields $\phi_{k'}^{a_1\ldots a_{s'}}$ with $s'\geq
4
$ satisfy the double-tracelessness constraint%
\be \label{doutracon01} \phi_{k'}^{aabba_5\ldots a_{s'}}=0\,, \qquad
s'\geq 4\,. \ee
\iiibf) The conformal dimension of the field $\phi_{k'}^{a_1\ldots
a_{s'}}$ is given by
\be
\label{condimarbspi01} \Delta(\phi_{k'}^{a_1\ldots a_{s'}}) =
\frac{d-2}{2} + k'\,.
\ee
\noindent \ivbf) Comparison of light-cone gauge fields
\rf{lcspin2DoF01}-\rf{lcspin2DoF03} and Lorentz fields
\rf{covspin2DoF01}-\rf{covspin2DoF03} demonstrates the general rule
we use to obtain the field content of gauge invariant
ordinary-derivative formulation. Namely, all that is required is to
replace the on-shell light-cone gauge fields of the $so(d-2)$
algebra \rf{lcspin2DoF01}-\rf{lcspin2DoF03} by the respective fields
of the Lorentz $so(d-1,1)$ algebra
\rf{covspin2DoF01}-\rf{covspin2DoF03}.%
\footnote{ Such a rule can unambiguously be used when there is
one-to-one mapping between spin labels of the $so(d-2)$ algebra
fields and those of the Lorentz algebra $so(d-1,1)$ fields. In fact,
such  mapping can unambiguously be realized for all fields with the
exception of self-dual fields. Study of self-dual fields in the
framework of ordinary-derivative approach may be found in
Ref.\cite{Metsaev:2008ba}.}

To illustrate the field content given in \rf{phiset01} we use the
shortcut $\phi_{k'}^{s'}$ for the field $\phi_{k'}^{a_1\ldots
a_{s'}}$ and note that, for $d\geq 6$ and arbitrary $s$, fields in
\rf{phiset01} can be presented as
{\small
\beq
&& \hspace{2.5cm} \hbox{Field content for} \ \ d \geq 6\,,\quad s -
\hbox{ arbitrary}
\nonumber\\
&&\hspace{-1cm} \phi_{-k_s}^s \hspace{1cm} \phi_{2-k_s}^s
\hspace{1cm} \ldots \hspace{1cm}\ldots \hspace{1cm} \ldots
\hspace{1cm} \ldots \hspace{1cm}\ldots \hspace{1cm} \phi_{k_s-2}^s
\hspace{1cm}  \phi_{k_s}^s \qquad
\nonumber\\[15pt]
&& \hspace{-0.3cm} \phi_{1-k_s}^{s-1} \hspace{1cm}
\phi_{3-k_s}^{s-1} \hspace{1cm} \ldots \hspace{1cm} \ldots
\hspace{1cm}\ldots \hspace{1cm}\ldots \hspace{1cm}
\phi_{k_s-3}^{s-1} \hspace{1cm} \phi_{k_s-1}^{s-1}
\nonumber\\[14pt]
&& \hspace{0.5cm} \ldots \hspace{1cm}  \ldots \hspace{1cm}  \ldots
\hspace{1cm} \ldots \hspace{1cm} \ldots   \hspace{1cm} \ldots
\hspace{1cm} \ldots \hspace{1cm} \ldots
\\[14pt]
&& \hspace{0.8cm}  \phi_{s-1-k_s}^1 \hspace{1cm} \phi_{s+1-k_s}^1
\hspace{1cm}  \ldots  \hspace{1cm} \ldots \hspace{1cm}
\phi_{k_s-s-1}^1 \hspace{0.8cm} \phi_{k_s-s+1}^1
\nonumber\\[15pt]
&& \hspace{1.8cm} \phi_{s-k_s}^0 \hspace{1.2cm} \phi_{s-k_s+2}^0
\hspace{1.2cm} \ldots \hspace{1cm} \phi_{k_s-s-2}^0 \hspace{1cm}
\phi_{k_s-s}^0
\nonumber
\eeq }

As we have said, the scalar fields do not enter the field content
when $d=4$. This is to say that, for $d=4$ and arbitrary $s$, the
field content in \rf{phiset01} can be represented as
{\small
$$ \hbox{Field content for} \ \ d = 4\,, \qquad s - \hbox{ arbitrary} $$
\be \label{oldman-27022012-01}
\begin{array}{ccccccccc}
\phi_{1-s}^s & & \phi_{3-s}^s & & \ldots & & \phi_{s-3}^s & &
\phi_{s-1}^s
\\[12pt]
& \phi_{2-s}^{s-1} & & \phi_{4-s}^{s-1} & \ldots & \phi_{s-4}^{s-1}
& & \phi_{s-2}^{s-1} &
\\[12pt]
& & \ldots  &   & \ldots & & \ldots  &  &
\\[12pt]
&  & & \phi_{-1}^2  & & \phi_1^2 & & &
\\[12pt]
&  & &  & \phi_0^1& & & &
\end{array}
\ee }

We note that $d=6$ is the lowest space-time dimension when the
scalar fields appear in the field content. Namely, for $d=6$ and
arbitrary $s$, the field content in \rf{phiset01} can be represented
as
{\small
$$ \hbox{Field content for} \ \ d = 6\,, \qquad s - \hbox{ arbitrary}  $$
\be
\begin{array}{ccccccccc}
\phi_{-s}^s & & \phi_{2-s}^s & & \ldots & & \phi_{s-2}^s & &
\phi_s^s
\\[12pt]
& \phi_{1-s}^{s-1} & & \phi_{3-s}^{s-1} & \ldots & \phi_{s-3}^{s-1}
& & \phi_{s-1}^{s-1} &
\\[12pt]
& & \ldots  &   & \ldots & & \ldots  &  &
\\[12pt]
&  & & \phi_{-1}^1  & & \phi_1^1 & & &
\\[12pt]
&  & &  & \phi_0^0& & & &
\end{array}
\ee }

In order to streamline the presentation of ordinary-derivative
formulation we use the oscillators  $\alpha^a$, $\zeta$,
$\upsilon^\oplussm$, $\upsilon^\ominussm$, and collect fields
\rf{phiset01} into ket-vector $\phik$ defined by
\beq
&& \phik \equiv \sum_{s'=0}^s
\frac{\zeta^{s-s'}}{\sqrt{(s-s')!}}|\phi^{s'}\rangle \,,
\hspace{3cm} \hbox{for } \ d\geq 6\,,
\nonumber\\[-10pt]
&& \label{phikdef01}
\\[-10pt]
&& \phik \equiv \sum_{s'=1}^s
\frac{\zeta^{s-s'}}{\sqrt{(s-s')!}}|\phi^{s'}\rangle \,,
\hspace{3cm} \hbox{for } \ d =4 \,,
\nonumber\\
\label{phikdef02} && |\phi^{s'}\rangle \equiv   \sum_{k'\in
[k_{s'}]_2} \frac{1}{s'!(\frac{k_{s'} + k'}{2})!}\alpha^{a_1} \ldots
\alpha^{a_{s'}} (\upsilon^\oplussm)^{^{\frac{k_{s'}+k'}{2}}}
(\upsilon^\ominussm)^{^{\frac{k_{s'} - k'}{2}}} \,
\phi_{k'}^{a_1\ldots a_{s'}}|0\rangle\,.
\eeq
From \rf{phikdef01},\rf{phikdef02},  we see that the ket-vectors
$\phik$, $|\phi^{s'}\rangle$  satisfy the relations
\beq
\label{olmman-05022012-02} && (N_\alpha + N_\zeta - s)\phik = 0 \,,
\hspace{2cm} (N_\zeta + N_\upsilon - k_s) \phik = 0 \,,
\\
\label{olmman-05022012-04} && (N_\alpha - s') |\phi^{s'}\rangle = 0
\,, \hspace{2.6cm}  (N_\upsilon - k_{s'}) |\phi^{s'}\rangle =0 \,,
\eeq
where $k_{s'}$ and $k_s$ are given in
\rf{oldman-21032012-10},\rf{olmman-05022012-01}. Relations
\rf{olmman-05022012-02} tell us that the ket-vector $\phik$ is
degree-$s$ homogeneous polynomial in the oscillators $\alpha^a$,
$\zeta$ and degree-$k_s$ homogeneous polynomial in the oscillators
$\zeta$, $\upsilon^\oplussm$, $\upsilon^\ominussm$, while relations
\rf{olmman-05022012-04} tell us that the ket-vector
$|\phi^{s'}\rangle$ is degree-$s'$ homogeneous polynomial in the
oscillators $\alpha^a$ and degree-$k_{s'}$ homogeneous polynomial in
the oscillators $\upsilon^\oplussm$, $\upsilon^\ominussm$. We note
that, in terms of the ket-vector $\phik$, double-tracelessness
constraint \rf{doutracon01} takes the form%
\footnote{ We adapt the formulation in terms of the double-traceless
gauge fields in Ref.\cite{Fronsdal:1978vb}. To develop the gauge
invariant approach one can use unconstrained gauge fields studied in
Ref.\cite{Francia:2002aa}.}
\be \label{man-15022012-04}
(\bar{\alpha}^2)^2 \phik = 0 \,.
\ee

We now discuss two representations for gauge invariant Lagrangian in
turn.

\noindent {\bf 1st representation for Lagrangian}. Lagrangian we
found takes the form
\be \label{spi2lag01} \LL = \frac{1}{2} \phibr E \phik\,, \ee
where operator $E$ is given by
\beq
\label{Frosecordope01} E  & = & E_\smtwo + E_\smone + E_\smzero\,,
\\
\label{Frosecordope02} && E_\smtwo \equiv \Box -
\alpha\partial\bar\alpha\partial +
\frac{1}{2}(\alpha\partial)^2\bar\alpha^2 + \frac{1}{2} \alpha^2
(\bar\alpha\partial)^2 - \frac{1}{2}\alpha^2 \Box \bar\alpha^2
-\frac{1}{4}\alpha^2\alpha\partial\,\bar\alpha\partial\bar\alpha^2\,,\qquad
\\
&& E_\smone \equiv   e_1 \bar\AA + \eb_1 \AA\,,
\\
&& E_\smzero \equiv m_1 + \alpha^2\bar\alpha^2m_2 + m_3 \bar\alpha^2
+ \mb_3 \alpha^2 \,,
\\
\label{man-13022012-22} && \AA \equiv  \alpar - \alpha^2 \albpar +
\frac{1}{4}\alpha^2\alpar \bar\alpha^2 \,,
\\
\label{man-13022012-23} && \bar\AA \equiv \albpar -
\alpar\bar\alpha^2 + \frac{1}{4}\alpha^2 \albpar \bar\alpha^2 \,,
\\[10pt]
\label{oldman-05022012-10} && \hspace{1cm} e_1 = \zeta \ewt_1
\bar\upsilon^\ominussm\,,\qquad \qquad \eb_1 = - \upsilon^\ominussm
\ewt_1 \bar\zeta\,,
\\
\label{man-14022012-22} && \hspace{1cm} m_1 =
\frac{2s+d-2-N_\zeta}{2s+d-2-2N_\zeta}(N_\zeta-1) \upsilon^\ominussm
\bar\upsilon^\ominussm \,,
\nonumber\\
\label{man-14022012-23} && \hspace{1cm} m_2 = \frac{2(2s+d-2) +
(2s+d - 7)N_\zeta - N_\zeta^2}{4(2s+d-2-2N_\zeta)}
\upsilon^\ominussm \bar\upsilon^\ominussm \,,
\nonumber\\
\label{man-14022012-24} && \hspace{1cm} m_3  = \half \zeta^2 \ewt_1
\ewt_1^\smone \bar\upsilon^\ominussm \bar\upsilon^\ominussm\,,
\qquad \quad
\label{man-14022012-25} \hspace{1cm} \mb_3 =  \half
\upsilon^\ominussm \upsilon^\ominussm \ewt_1 \ewt_1^\smone
\bar\zeta^2\,,
\\
\label{ewtdef01} && \hspace{1cm} \ewt_1 =
\Bigl(\frac{2s+d-4-N_\zeta}{2s+d-4-2N_\zeta}\Bigr)^{1/2}\,, \qquad
\ewt_1^\smone =
\Bigl(\frac{2s+d-5-N_\zeta}{2s+d-6-2N_\zeta}\Bigr)^{1/2}\,.\qquad
\eeq
We note that $E_\smtwo$ \rf{Frosecordope02} is the standard
second-order Fronsdal operator represented in terms of the
oscillators.

\medskip
\noindent{\bf 2nd representation for Lagrangian}. We now discuss the
second representation for Lagrangian. This representation allows us
to introduce various de Donder like gauge conditions for arbitrary
spin conformal field. Namely, Lagrangian \rf{spi2lag01} can be
represented as (up to total derivative)
\beq
\label{manold-17032012-01} && \LL = \frac{1}{2} \phibr \mubf (\Box -
m^2) \phik + \half \langle \Cb \phi|\Cb\phi\rangle \,,\qquad
\\
\label{manold-17032012-02} && \hspace{1cm} \Cb \equiv \Vb_\perp -
\eb_1 \Pi^\smponetwo + \half e_1 \bar\alpha^2\,,\qquad \Vb_\perp
\equiv \albpar - \half \alpar \bar\alpha^2\,,\qquad  m^2 \equiv
\upsilon^\ominussm \bar\upsilon^\ominussm\,,\qquad
\eeq
where operators $\mubf$ and $\Pi^\smponetwo$ are given in
\rf{manold-18032012-06}. To illustrate the structure of Lagrangian
we note that, in terms of tensor fields $\phi_{k'}^{a_1\ldots
a_{s'}}$ \rf{phiset01}, Lagrangian \rf{manold-17032012-01} takes the
form
\beq
\label{manold-04042012-03} \LL & = &\sum_{s'=0}^s \sum_{k'\in
[k_{s'}]_2} \LL_{k'}^{s'}\,,\hspace{2cm} \hbox{for }\ d\geq 6\,,
\\
\label{manold-04042012-03n1} \LL & = &\sum_{s'=1}^s \sum_{k'\in
[k_{s'}]_2} \LL_{k'}^{s'}\,, \hspace{2cm} \hbox{for }\ d = 4\,,
\\
\LL_{k'}^{s'} & \equiv & \frac{1}{2 s'!}\Bigl( \phi_{-k'}^{a_1\ldots
a_{s'}} \Box \phi_{k'}^{a_1\ldots a_{s'}}
- \frac{s'(s'-1)}{4} \phi_{-k'}^{aaa_3\ldots a_{s'}} \Box
\phi_{k'}^{bba_3\ldots a_{s'}}\Bigr)
\nonumber\\
& - & \frac{1}{2 s'!}\Bigl( \phi_{-k'}^{a_1\ldots a_{s'}} \phi_{k'+
2}^{a_1\ldots a_{s'}}
- \frac{s'(s'-1)}{4} \phi_{-k'}^{aaa_3\ldots a_{s'}}
\phi_{k'+2}^{bba_3\ldots a_{s'}}\Bigr)
\nonumber\\
& + &  \frac{1}{2 (s'-1)!} C_{-k'+1}^{a_1\ldots a_{s'-1}}
C_{k'+1}^{a_1\ldots a_{s'-1}}\,,
\eeq

\beq
\label{manold-04042012-04} && C_{k'+1}^{a_1\ldots a_{s'-1}} \equiv
\partial^b \phi_{k'}^{a_1\ldots a_{s'-1} b}
-\frac{s'-1}{2}\partial^{(a_1} \phi_{k'}^{a_2\ldots a_{s'-1}) bb}
\nonumber\\
&& \hspace{1.6cm} + \, f_{s'} \phi_{k'+1}^{\tr\ a_1\ldots a_{s'-1}}
+ \half f_{s'+1} \phi_{k'+1}^{a_1\ldots a_{s'-1}bb}\,,
\\
\label{manold-04042012-05} && \phi_{k'}^{\tr\ a_1\ldots a_{s'}}
\equiv \phi_{k'}^{a_1\ldots a_{s'}} - \frac{s'(s'-1)}{2(2s'+d-4)}
\eta^{(a_1a_2} \phi_{k'}^{a_3\ldots a_{s'}) bb}\,,
\eeq

\beq
&& f_{s'} \equiv
\Bigl(\frac{(s+1-s')(s+s'+d-4)}{2s'+d-4}\Bigr)^{1/2}\,,
\eeq
We note that $\phi_{k'}^{\tr\ a_1\ldots a_{s'}}$
\rf{manold-04042012-05} is a traceless tensor, $\phi_{k'}^{\tr\
aaa_3\ldots a_{s'}}=0$. In \rf{manold-04042012-04},
\rf{manold-04042012-05}, symmetrization of the indices $a_1\ldots
a_n$ is normalized as $(a_1\ldots a_n) = \frac{1}{n!}(a_1\ldots a_n
+ (n!-1)$ terms).

The quantity $\Cb\phik$ with $\Cb$ given in \rf{manold-17032012-02}
will be referred to as modified de Donder divergency. From
\rf{manold-17032012-01}, we see that the use of the modified de
Donder divergency simplifies considerably the presentation of
Lagrangian. It is the modified de Donder divergency that allows us
to introduce various de Donder like gauges for conformal field.
Before presenting these de Donder like gauges we recall that the
commonly used standard de Donder gauge is defined as
\be
\Vb_\perp \phik = 0\,, \hspace{3cm} \hbox{ de Donder gauge},
\ee
where $\Vb_\perp$ is defined in \rf{manold-17032012-02}. Our
representation for Lagrangian of conformal field given in
\rf{manold-17032012-01} suggests new gauge condition which we refer
to as modified de Donder gauge. This to say that, using operator
$\Cb$ \rf{manold-17032012-02}, we define modified de Donder gauge as
\be \label{manold-17032012-03}
\Cb \phik = 0\,, \hspace{3cm} \hbox{ modified de Donder gauge}.
\ee
In the tensorial notation, modified de Donder gauge
\rf{manold-17032012-03} can be represented as
\be \label{manold-17032012-03n1}
C_{k'+1}^{a_1\ldots a_{s'-1}} = 0 \,, \qquad s'=1,\ldots, s\,,
\qquad k'\in [k_{s'}]_2\,,
\ee
where tensorial form of the modified de Donder divergencies
$C_{k'+1}^{a_1\ldots a_{s'-1}}$ is given in \rf{manold-04042012-04}.
Note that, by virtue of double-tracelessness  constraint
\rf{doutracon01}, the modified de Donder divergencies
$C_{k'+1}^{a_1\ldots a_{s'}}$ with $s'\geq 2$ are traceless,
$C_{k'+1}^{aaa_3\ldots a_{s'}}=0$.

Besides the modified de Donder gauge we introduce another new gauge
condition which we refer to as de Donder-Stueckelberg gauge. The de
Donder-Stueckelberg gauge is defined as
\be \label{manold-17032012-04}
\bar\upsilon^\oplussm \Cb \phik = 0\,, \hspace{3cm} \hbox{ de
Donder-Stueckelberg gauge}.
\ee
In the tensorial notation, de Donder-Stueckelberg gauge
\rf{manold-17032012-04} can be represented as
\be \label{manold-17032012-04n1}
C_{k'+1}^{a_1\ldots a_{s'-1}} = 0 \,, \qquad s'=1,\ldots, s\,,
\qquad k'\in -k_{s'}, [k_{s'}-2]_2\,.
\ee
Comparing values of $k'$ in \rf{manold-17032012-03n1} and
\rf{manold-17032012-04n1}, we see that de Donder-Stueckelberg gauge
conditions \rf{manold-17032012-04n1} constitute a subset of the
modified de Donder gauge conditions \rf{manold-17032012-03n1}. Below
we demonstrate that the de Donder-Stueckelberg gauge turns out to be
very convenient for the study of interrelations between the
higher-derivative formulation of conformal fields in section
\ref{subsec-301} and the ordinary-derivative formulation in this
section.

\subsection{Gauge symmetries of conformal field}\label{gaugsym}

We now discuss gauge symmetries of Lagrangian \rf{spi2lag01}. To
this end we introduce the following set of gauge transformation
parameters:
\beq \label{epsilonset01}
&& \xi_{k'-1}^{a_1\ldots a_{s'}}\,,\hspace{1.5cm}
s'=0,1,\ldots,s-1\,,
\hspace{1.5cm} k' \in  [k_{s'}+1]_2\,,
\eeq
where $k_{s'}$ is given in \rf{oldman-21032012-10}. Alternatively,
gauge transformation parameters \rf{epsilonset01} can be represented
as
\beq
\label{epscovspin2DoF01} &&\xi_{k'-1}^{a_1\ldots a_{s-1}}\,,
\hspace{2.5cm} k' \in [k_s]_2\,;
\\
\label{epscovspin2DoF02} &&\xi_{k'-1}^{a_1\ldots a_{s-2}}\,,
\hspace{2.5cm} k' \in [k_s-1]_2\,;
\\
&&\ldots\quad \ldots \hspace{2.5cm} \ldots \quad \ldots \quad \ldots
\nonumber\\
&&\ldots \quad \ldots \hspace{2.5cm}\ldots \quad \ldots \quad \ldots
\nonumber\\
&& \xi_{k'-1}^a\,, \hspace{3cm} k' \in  [k_s-s+2]_2\,;
\\
\label{epscovspin2DoF03} && \xi_{k'-1}\,, \hspace{3cm} k'\in
[k_s-s+1]_2\,.
\eeq
We note that\\
\ibf) In \rf{epsilonset01}, the gauge transformation parameters
$\xi_{k'-1}$ and $\xi_{k'-1}^a$ are the respective scalar and vector
fields of the Lorentz algebra, while the gauge transformation
parameter $\xi_{k'-1}^{a_1\ldots a_{s'}}$, $s'\geq 2$, is rank-$s'$
totally symmetric traceless tensor field of the Lorentz algebra
$so(d-1,1)$,
\be \label{epsdoutracon01} \xi_{k'-1}^{aaa_3\ldots a_{s'}}=0\,,
\qquad s'\geq 2\,. \ee
\iibf) The gauge transformation parameters $\xi_{k'-1}^{a_1\ldots
a_{s'}}$ have the conformal dimensions
\be \label{epscondimarbspi01} \Delta(\xi_{k'-1}^{a_1\ldots a_{s'}})
= \frac{d-2}{2} + k' -1\,.\ee

Now, as usually, we collect the gauge transformation parameters into
ket-vector $\xik$ defined by

\beq
&& \xik \equiv \sum_{s'=0}^{s-1}
\frac{\zeta^{s-1-s'}}{\sqrt{(s-1-s')!}}|\xi^{s'}\rangle \,,
\\
&& |\xi^{s'}\rangle \equiv  \sum_{k'\in
[k_{s'}+1]_2}\frac{1}{s'!(\frac{k_{s'}+1+k'}{2})!}\alpha^{a_1}
\ldots \alpha^{a_{s'}}
(\upsilon^\oplussm)^{^{\frac{k_{s'}+1+k'}{2}}}
(\upsilon^\ominussm)^{^{\frac{k_{s'}+1-k'}{2}}} \,
\xi_{k'-1}^{a_1\ldots a_{s'}} |0\rangle\,.\qquad
\eeq
The ket-vectors $\xik$, $|\xi^{s'}\rangle$ satisfy the algebraic
constraints,
\beq
\label{oldman-05022012-06} && (N_\alpha + N_\zeta - s +1 ) \xik=0
\,, \hspace{1.7cm} (N_\zeta + N_\upsilon - k_s ) \xik=0 \,,
\\
\label{oldman-05022012-08} && (N_\alpha - s') |\xi^{s'}\rangle = 0
\,, \hspace{3cm} (N_\upsilon - k_{s'} - 1) |\xi^{s'}\rangle = 0 \,,
\eeq
where $k_{s'}$ and $k_s$ are defined in \rf{oldman-21032012-10},
\rf{olmman-05022012-01}. Relations \rf{oldman-05022012-06} tell us
that $\xik$ is a degree-$(s-1)$ homogeneous polynomial in the
oscillators $\alpha^a$, $\zeta$ and degree-$k_s$ homogeneous
polynomial in the oscillators $\zeta$, $\upsilon^\oplussm$,
$\upsilon^\ominussm$, while relations \rf{oldman-05022012-08} tell
us that $|\xi^{s'}\rangle$ is degree-$s'$ homogeneous polynomial in
the oscillators $\alpha^a$ and degree-$(k_{s'}+1)$ homogeneous
polynomial in the oscillators $\upsilon^\oplussm$,
$\upsilon^\ominussm$.

In terms of the ket-vector $\xik$, tracelessness constraint
\rf{epsdoutracon01} takes the form

\be \bar\alpha^2 \xik=0 \,.\ee

Gauge transformations can entirely be written in terms of $\phik$
and $\xik$. This is to say that gauge transformations take the form

\be \label{oldman-05022012-11}
\delta \phik = G \xik  \,,
\qquad G \equiv  \alpar - e_1 - \alpha^2 \frac{1}{2N_\alpha
+d-2}\eb_1\,,
\ee
where operators $e_1$, $\eb_1$ are defined in
\rf{oldman-05022012-10}. As a side remark we note that using
operator $G$ \rf{oldman-05022012-11} we can represent Lagrangian as
in \rf{spi2lag01} with new operator $E$  given by
\be \label{oldman-05022012-11n1}
E = \mubf (\Box - m^2 - G\Cb) \,.
\ee
Operator $E$ \rf{oldman-05022012-11n1} is equal to the one in
\rf{Frosecordope01} up to $(\alpha^2)^2$- and
$(\bar\alpha^2)^2$-terms. Obviously, by virtue of
double-tracelessness constraint \rf{man-15022012-04}, such terms do
not contribute to Lagrangian \rf{spi2lag01}.

\subsection{ Realization of conformal boost symmetries}\label{realiz}

To complete the ordinary-derivative formulation of spin-$s$
conformal field we provide realization of the conformal algebra
symmetries on the space of ket-vector $\phik$. All that is required
is to fix the operators $M^{ab}$, $\Delta$, and $R^a$ for the case
of spin-$s$ conformal field and then use these operators in
\rf{conalggenlis01}-\rf{conalggenlis04}. For the case of arbitrary
spin-$s$ conformal field, the realization of the Lorentz algebra
spin operator $M^{ab}$ and the conformal dimension operator $\Delta$
on the space of $\phik$ is given by
\beq
&& M^{ab} = \alpha^a \bar\alpha^b - \alpha^b \bar\alpha^a \,,
\\
\label{manold-11042012-02} && \Delta  =  \frac{d-2}{2}+\Delta'\,,
\qquad \Delta' \equiv N_{\upsilon^\oplussm} - N_{\upsilon^\ominussm}
\,.
\eeq
Note that realization of the conformal dimension operator $\Delta$
\rf{manold-11042012-02} can be read from \rf{condimarbspi01}.
Realization of the operator $R^a$ on the space of $\phik$ is given
by
\beq
&& R^a  = r_\smzero^a + r_\smone^a + R_\smG^a\,,
\\
&& r_\smzero^a  =   r_{0,1} \bar\alpha^a + \rb_{0,1} \Vwt^a \,,
\qquad
r_\smone^a =  r_{1,1} \partial^a\,,
\\
\label{Rgsmspi204} && R_\smG^a  =   G  r_\smG^a\,,
\\
\label{Rgsmspi108} && \hspace{1cm} r_\smG^a = r_{\smG,1} \Vb_\perp^a
+ r_{\smG,2} V^a\Pi^\smponetwo + r_{\smG,3} V^a \bar\alpha^2 +
r_{\smG,4}\bar\alpha^a \bar\alpha^2\,,
\\
\label{oldman-05022012-12} && \hspace{1cm} r_{0,1} = 2 \zeta \ewt_1
\bar\upsilon^\oplussm\,,
\qquad
\rb_{0,1} = - 2 \upsilon^\oplussm \ewt_1 \bar\zeta\,,
\qquad
r_{1,1} = -2\upsilon^\oplussm \bar\upsilon^\oplussm \,,
\\
&& \hspace{1cm} r_{\smG,n} =  \upsilon^\oplussm  \rwt_{\smG,n}
\bar\upsilon^\oplussm \,, \qquad \qquad \ \ \ n = 1,3\,,
\nonumber\\
&& \hspace{1cm} r_{\smG,2} =  \upsilon^\oplussm \upsilon^\oplussm
\rwt_{\smG,2} \bar\zeta^2 \,, \qquad \qquad r_{\smG,4} =  \zeta^2
\rwt_{\smG,4} \bar\upsilon^\oplussm \bar\upsilon^\oplussm \,,
\\
\label{manold-11042012-03} && \hspace{1cm} \rwt_{\smG,n}
=\rwt_{\smG,n} (N_\zeta, \Delta')\,,\qquad \quad \ \ \ n=1,2,3,4\,,
\eeq
where operators $G$ and $\ewt_1$ appearing in \rf{Rgsmspi204} and
\rf{oldman-05022012-12} are defined in \rf{oldman-05022012-11} and
\rf{ewtdef01} respectively. In \rf{manold-11042012-03}, the
quantities $\rwt_{\smG,n}$ are arbitrary functions of the operators
$N_\zeta$, $\Delta'$.

The following remarks are in order.

\noindent {\bf i}) $r_\smzero^a$ and $r_\smone^a$ parts of the
operator $R^a$ are determined uniquely, while $R_\smG^a$ part, in
view of arbitrary $\rwt_{\smG,n}$, $n=1,2,3,4$, is still to be
arbitrary. The reason for arbitrariness in $R_\smG^a$ is obvious.
Global transformations of gauge fields are defined up to gauge
transformations. Since $R_\smG^a$ \rf{Rgsmspi204} is proportional to
$G$, the action of $R_\smG^a$ on $\phik$ takes the form of gauge
transformation governed by the gauge transformation parameter
$r_\smG^a\phik$.

\noindent {\bf ii}) We check that the operator $R^a$ with
$R_\smG^a=0$ satisfies the commutator $[K^a,K^b]=0$.

\noindent {\bf iii}) If we consider the operator $R^a$ with
$R_\smG^a \ne 0 $, then we find that the commutator $[K^a,K^b]$ for
such $R^a$ takes the form
\be \label{man-15022012-05}
[K^a,K^b] = G r_\smG^{ab}\,,
\qquad
r_\smG^{ab} \equiv r^a r_\smG^b + r_\smG^a r^b + r_\smG^a G r_\smG^b
- (a \leftrightarrow b)\,, \qquad  r^a \equiv r_\smzero^a +
r_\smone^a\,.
\ee
From \rf{man-15022012-05}, we see that the commutator $[K^a,K^b]$ is
proportional to the operator of gauge transformations $G$
\rf{oldman-05022012-11}, as it should be in gauge theory.

To summarize, having been introduced the field content, we find that
the Lagrangian, gauge transformations,
and the operator $R^a$ are determined by requiring that%
\footnote{ In the framework of higher-derivative formulation, the
uniqueness of interacting spin-2 conformal field theory was
discussed in Ref.\cite{Boulanger:2001he}.}
\\
\ibf) Lagrangian should not involve higher than second order terms
in derivatives, while gauge transformations should not involve
higher than first order terms in derivatives;
\\
\iibf) the operator $R^a$ should not involve higher than first order
terms
in derivatives;\\
\iiibf) Lagrangian should be invariant with respect to gauge
transformations and conformal algebra transformations.

These requirements allow us to determine the Lagrangian and gauge
transformations uniquely. The operator $R^a$ is determined uniquely
up to the gauge transformation operator $G$ \rf{oldman-05022012-11}
(as it should be in any theory of gauge fields). For the derivation
of Lagrangian, gauge transformations and operator $R^a$, see
Appendix B.

\newsection{ \large Interrelations between ordinary-derivative
and higher-derivative
\\
approaches to conformal fields}\label{sec-05onter}

In this section, we demonstrate that our ordinary-derivative
approach in Sec.\ref{lagran} is equivalent to the higher-derivative
approach in Sec.\ref{subsec-301}. To demonstrate the equivalence it
is convenient to use the de Donder-Stueckelberg gauge frame. Before
going into technical details we discuss the general setup of the
equivalence. This is to say that, in the framework of the de
Donder-Stueckelberg gauge frame, the equivalence of two approaches
is realized as follows.

\medskip
\noindent \ibf) Fields appearing in the ordinary-derivative
formulation \rf{phiset01} are separated into two groups,%
\footnote{ To separate fields \rf{phiset01} into two groups
\rf{manold-22032012-23}, \rf{manold-22032012-24} we note that values
$k'$ in \rf{phiset01}, $k'\in [k_{s'}]_2$, can be represented as
$k'\in -k_{s'},[k_{s'}-2]_2,k_{s'}$. Fields with $k'= -k_{s'}$ are
collected in \rf{manold-22032012-23}, while fields with $k'\in
[k_{s'}-2]_2,k_{s'}$ are collected in \rf{manold-22032012-24}.}
\beq
\label{manold-22032012-23} && \phi_{-k_{s'}}^{a_1\ldots a_{s'}}\,,
\hspace{1cm}
s'= \left\{\begin{array}{l}
0,1,\ldots,s;\quad \quad \hbox{for }\  d \geq 6;
\\[3pt]
1,2,\ldots,s;\qquad  \hbox{for }d = 4;
\end{array}\right.
\\
\label{manold-22032012-24} && \phi_{k'}^{a_1\ldots a_{s'}}\,,
\hspace{1cm}
s'= \left\{\begin{array}{l}
0,1,\ldots,s;\quad \quad \hbox{for }\  d \geq 6;
\\[3pt]
1,2,\ldots,s;\qquad  \hbox{for }d = 4;
\end{array}\right.
\hspace{1.7cm} k' \in  [k_{s'}-2]_2\,, k_{s'}\,.\qquad
\eeq

\noindent \iibf) Fields appearing in our ordinary-derivative
formulation \rf{manold-22032012-23} are identified with fields
$\phi^{a_1\ldots a_{s'}}$, $s'=0,1,\ldots, s$, appearing in our
higher-derivative formulation \rf{manold-06032012-02} in the
following way:
\beq
\label{manold-23032012-01} && \phi_{-k_{s'}}^{a_1\ldots a_{s'}}
\equiv \phi^{a_1\ldots a_{s'}}\,,\qquad \qquad s'=0,1,\ldots, s\,,
\hspace{3.2cm} \hbox{for } \ d\geq 6\,,
\\[10pt]
&& \phi_{-k_{s'}}^{a_1\ldots a_{s'}} \equiv  \phi^{a_1\ldots
a_{s'}}\,,\qquad \qquad s'=1,\ldots, s\,,
\nonumber\\[-10pt]
\label{manold-23032012-02}  && \hspace{11cm} \hbox{for } \ d = 4\,.
\qquad
\\[-10pt]
&& -\sqrt{\frac{2}{s(s+1)}}\Bigl(\partial^a\phi_0^a +
\frac{1}{4}\sqrt{(s-1)(s+2)}\Box\phi_{-1}^{aa}\Bigr)  \equiv \phi\,,
\nonumber
\eeq
Note that, in \rf{manold-23032012-02}, the scalar field $\phi$
appearing in the field content of higher-derivative approach
\rf{manold-06032012-02} is related to the vector field $\phi_0^a$
and tensor field $\phi_{-1}^{ab}$ of ordinary-derivative approach
because, for $d=4$, the field content of ordinary-derivative
approach \rf{phiset01} does not involve a scalar field.

\medskip
\noindent \iiibf) Fields appearing in \rf{manold-22032012-24} are
separated into two groups: Stueckelberg fields and auxiliary fields.
The Stueckelberg fields in \rf{manold-22032012-24} are gauged away
by using Stueckelberg gauge symmetries, while the auxiliary fields
in \rf{manold-22032012-24} are expressed in terms of fields
\rf{manold-22032012-23}. Plugging solution for the auxiliary fields
into ordinary-derivative Lagrangian \rf{manold-17032012-01} and
using identification \rf{manold-23032012-01},
\rf{manold-23032012-02}, we find that ordinary-derivative Lagrangian
\rf{manold-17032012-01} becomes higher-derivative Lagrangian
\rf{manold-06032012-05}.

\medskip
\noindent \ivbf)  To gauge away the Stueckelberg fields appearing in
\rf{manold-22032012-24} we use de Donder-Stueckelberg gauge
condition \rf{manold-17032012-04}. The de Donder-Stueckelberg gauge
condition of ordinary-derivative approach \rf{manold-17032012-04}
leads to the differential constraint of higher-derivative approach
\rf{manold-06032012-04}.

We note that for the study of equivalence of the ordinary-derivative
and higher-derivative approaches we could use a gauge which we refer
to as Stueckelberg gauge (see below Sec. \ref{sub-502}). Our study
of the de Donder-Stueckelberg and Stueckelberg gauges leads us to
the conclusion that the de Donder-Stueckelberg gauge is very useful
for practical computations, while the Stueckelberg gauge seems to be
less useful for practical computations. Therefore, in
Sec.\ref{sub-501} we present the detailed discussion of the de
Donder-Stueckelberg gauge frame, while, in Sec. \ref{sub-502}, the
Stueckelberg gauge frame is discussed without going into details.

\subsection{ \large de Donder-Stueckelberg gauge frame }\label{sub-501}

We now discuss details of equivalence of the ordinary-derivative and
higher-derivative approaches by using the de Donder-Stueckelberg
gauge frame. To elucidate details of the study we consider simple
example of spin-2 conformal field. As we have already noticed,
scalar fields enter the field content of ordinary-derivative
approach when $d\geq 6$, while, for $d=4$, the field content of
ordinary-derivative approach does not involve scalar fields. For
this reason, the treatment of the equivalence of ordinary-derivative
and higher-derivative approaches for $d=4$ is slightly different
from the one for $d\geq 6$. The simplest example of spin-2 conformal
field theory involving scalar field corresponds to $d=6$. Therefore
to illustrate all details of the equivalence we consider separately
$6d$ and $4d$ spin-2 conformal field theories in the respective
sections \ref{subsub-6d} and \ref{subsub-4d}. Arbitrary spin
conformal field in $d$-dimensional space, $d\geq 4$, is considered
in section \ref{subsub-arbd}.

\subsubsection{ 6d conformal gravity in de Donder-Stueckelberg
gauge frame} \label{subsub-6d}

In this section, we illustrate the use of the de Donder-Stueckelberg
gauge frame for matching the ordinary-derivative and higher
derivative approaches to spin-2 conformal theory in six-dimensional
space ($6d$ conformal gravity).%
\footnote{ In the Stueckelberg gauge frame, the detailed discussion
of matching of the ordinary-derivative and higher-derivative
approaches for spin-2 conformal field in arbitrary $d$-dimensional
space may be found in Sec.5.2, Ref.\cite{Metsaev:2007fq}.}
Higher-derivative Lagrangian of $6d$ conformal gravity is obtained
from \rf{oldman-21032012-07} by equating $d=6$. We now briefly
review the ordinary-derivative formulation of $6d$ conformal gravity
developed in Ref.\cite{Metsaev:2007fq} (see also
Ref.\cite{Metsaev:2010kp}).

\noindent {\bf Ordinary-derivative formulation of $6d$ conformal
gravity}. In the framework of ordinary-de\-ri\-vative approach, $6d$
conformal gravity is described by three rank-2 tensor fields
$\phi_{k'}^{ab}$, $k'=0,\pm 2$, two vector fields $\phi_{k'}^a$,
$k'=\pm 1$, and one scalar field $\phi_{0}$:
\beq
& \phi_{-2}^{ab}\qquad \phi_0^{ab}\qquad \phi_2^{ab} &
\nonumber\\
& \phi_{-1}^a\qquad \phi_1^a &
\\
& \phi_0 &
\nonumber
\eeq
Ordinary-derivative Lagrangian of $6d$ conformal gravity is given by
\beq \label{05032012-14}
\LL & = & \frac{1}{2} \phi_2^{ab} \Box \phi_{-2}^{ab} - \frac{1}{4}
\phi_2^{aa} \Box \phi_{-2}^{bb} + \frac{1}{4} \phi_0^{ab} \Box
\phi_0^{ab} - \frac{1}{8} \phi_0^{aa} \Box \phi_0^{bb} + \phi_{1}^a
\Box \phi_{-1}^a + \half \phi_{0} \Box \phi_0 \qquad\quad
\nonumber\\[3pt]
& +  & C_{-1}^a C_3^a + \half C_1^a C_1^a + C_0 C_2 - \half
\phi_2^{ab} \phi_0^{ab} + \frac{1}{4} \phi_2^{aa} \phi_0^{bb} -
\half \phi_1^a \phi_1^a\,,
\eeq
where the modified de Donder divergencies are defined by
\beq
&& \hspace{1cm} C_{-1}^a  = \partial^b \phi_{-2}^{ab} - \half
\partial^a \phi_{-2}^{bb} + \phi_{-1}^a\,,
\nonumber\\
&& \hspace{1cm} C_1^a = \partial^b \phi_0^{ab} - \half \partial^a
\phi_0^{bb} + \phi_1^a\,,
\nonumber\\
\label{05032012-10} && \hspace{1cm} C_3^a = \partial^b \phi_2^{ab} -
\half \partial^a \phi_2^{bb}\,,
\\
&& \hspace{1cm} C_0 =  \partial^a \phi_{-1}^a + \half \phi_0^{aa} +
u\phi_0\,,
\nonumber\\
&& \hspace{1cm} C_2 =  \partial^a \phi_1^a + \half \phi_2^{aa}\,,
\nonumber\\
&& \qquad u\equiv \sqrt{5/2}\,.
\eeq
Lagrangian \rf{05032012-14} is invariant under gauge transformations
given by
\beq
&& \delta \phi_{-2}^{ab} = \partial^a \xi_{-3}^b +
\partial^b \xi_{-3}^a + \half \eta^{ab}\xi_{-2}\,,
\nonumber\\
&& \delta \phi_0^{ab} = \partial^a \xi_{-1}^b +
\partial^b \xi_{-1}^a + \half \eta^{ab}\xi_0\,,
\nonumber\\
&& \delta \phi_2^{ab} = \partial^a \xi_1^b +
\partial^b \xi_1^a\,,
\nonumber\\[-10pt]
\label{05032012-15} &&
\\[-10pt]
&& \delta \phi_{-1}^a  = \partial^a \xi_{-2} - \xi_{-1}^a\,,
\nonumber\\
&& \delta \phi_1^a = \partial^a \xi_0 - \xi_1^a\,,
\nonumber\\
&& \delta \phi_0 = - u \xi_0\,,
\nonumber
\eeq
where $\xi_{-3}^a$, $\xi_{-1}^a$, $\xi_1^a$, $\xi_{-2}$, $\xi_0$ are
gauge transformation parameters.

Now our purpose is to demonstrate how higher-derivative Lagrangian
\rf{oldman-21032012-07}, differential constraints
\rf{oldman-21032012-08}, \rf{oldman-21032012-09} and gauge
transformations \rf{oldman-21032012-15} are obtained from the
ordinary-derivative approach. In the de Donder-Stueckelberg gauge
frame, matching of the higher-derivative and ordinary-derivative
approaches is realized as follows.

\medskip
\noindent \ibf) {\bf de Donder-Stueckelberg gauge frame}. We start
with the detailed description of the de Donder-Stueckelberg gauge
\rf{manold-17032012-04} for the case of $6d$ conformal gravity. To
this end we note that under gauge transformations \rf{05032012-15}
the modified de Donder divergencies \rf{05032012-10} transform  as
\beq
\label{oldman-21032012-01} && \delta C_{-1}^a  = \Box \xi_{-3}^a
-\xi_{-1}^a\,,
\\
\label{oldman-21032012-02} && \delta C_1^a =  \Box \xi_{-1}^a
-\xi_1^a\,,
\\
\label{oldman-21032012-03} && \delta C_3^a =  \Box \xi_1^a\,,
\\
\label{oldman-21032012-04} && \delta C_0 =   \Box \xi_{-2} -\xi_0\,,
\\
\label{oldman-21032012-05} && \delta C_2 =   \Box \xi_0\,.
\eeq
From \rf{oldman-21032012-01},\rf{oldman-21032012-02}, and
\rf{oldman-21032012-04}, we see that under gauge transformations
governed by the gauge transformation parameters $\xi_{-1}^a$,
$\xi_1^a$, and $\xi_0$ the respective modified de Donder
divergencies $C_{-1}^a$, $C_1^a$, and $C_0$ transform as
Stueckelberg fields. It is these transformation rules of the
modified de Donder divergencies $C_{-1}^a$, $C_1^a$, and $C_0$ that
motivate us to introduce de Donder-Stueckelberg gauge
\rf{manold-17032012-04} which, for $6d$ conformal gravity, takes the
form
\be \label{05032012-01}
C_{-1}^a =0 \,, \qquad C_1^a =0 \,, \qquad C_0 =0 \,,\qquad \hbox{
de Donder-Stueckelberg gauge}.
\ee
Obviously, the de Donder-Stueckelberg gauge does not completely fix
gauge.  This is to say that one can still make off-shell left-over
gauge transformations governed by the gauge transformation
parameters $\xi_{-3}^a$, $\xi_{-2}$ which are not subject to any
differential constraints. These left-over gauge symmetries are found
by requiring the modified de Donder divergencies shown in
\rf{05032012-01} to be invariant under gauge transformations,
\be \label{oldman-21032012-17}
\delta C_{-1}^a =0 \,, \qquad \delta C_1^a =0 \,, \qquad \delta C_0
=0.
\ee
Using \rf{oldman-21032012-01},\rf{oldman-21032012-02}, and
\rf{oldman-21032012-04}, we see that solution to
Eqs.\rf{oldman-21032012-17} is given by
\be \label{oldman-21032012-18}
\xi_{-1}^a = \Box\xi_{-3}^a\,, \qquad \xi_1^a = \Box^2\xi_{-3}^a\,,
\qquad \xi_0 = \Box\xi_{-2}\,.
\ee
This implies that left-over gauge transformations take the form as
in \rf{05032012-15}, where the gauge transformation parameters
$\xi_{-1}^a$, $\xi_1^a$, $\xi_0$ are given in
\rf{oldman-21032012-18}, while the gauge transformation parameters
$\xi_{-3}^a$, $\xi_{-2}$ are not subject to differential
constraints. Below we demonstrate that these left-over gauge
transformations are related to the gauge transformations entering
higher-derivative approach \rf{oldman-21032012-15}.

\medskip
\noindent \iibf) {\bf Matching of higher-derivative Lagrangian and
ordinary-derivative Lagrangian}. We now explain how
higher-derivative Lagrangian \rf{oldman-21032012-07} is obtained
from the ordinary derivative Lagrangian \rf{05032012-14}. Using
gauge \rf{05032012-01} in equations of motion for the auxiliary
fields $\phi_0^{ab}$, $\phi_2^{ab}$, $\phi_1^a$, we get the
following gauge-fixed equations of motion:
\beq
\label{05032012-02} \Box \phi_{-2}^{ab} - \phi_0^{ab} = 0\,,
\qquad
\Box \phi_0^{ab} - \phi_2^{ab} - \half \eta^{ab} C_2 = 0 \,,
\qquad
\Box \phi_{-1}^a - \phi_1^a = 0 \,.
\eeq
Solution of Eqs.\rf{05032012-02} is given by
\beq
\label{05032012-05} && \phi_0^{ab} =  \Box \phi_{-2}^{ab} \,,
\\
\label{05032012-06} && \phi_2^{ab} = \Box^2 \phi_{-2}^{ab} -  \half
\eta^{ab} C_2 \,,
\\
\label{05032012-07} && \phi_1^a = \Box \phi_{-1}^a \,.
\eeq
Plugging \rf{05032012-05}-\rf{05032012-07} and \rf{05032012-01} into
Lagrangian \rf{05032012-14}, we get
\be \label{oldman-21032012-06}
\LL =  \frac{1}{4} \phi_{-2}^{ab} \Box^3 \phi_{-2}^{ab} -
\frac{1}{8} \phi_{-2}^{aa} \Box^3 \phi_{-2}^{bb} + \half \phi_{-1}^a
\Box^2 \phi_{-1}^a + \half \phi_0 \Box \phi_0\,.
\ee
Making use of the identification of fields in
\rf{oldman-21032012-06} with the ones in \rf{oldman-21032012-07},
\be \label{oldman-21032012-19}
\phi_{-2}^{ab}\equiv \phi^{ab}\,, \qquad \phi_{-1}^a\equiv \phi^a\,,
\qquad \phi_0 \equiv \phi\,,
\ee
we see that Lagrangian \rf{oldman-21032012-06} coincides with the
one in \rf{oldman-21032012-07} when $d=6$.

\noindent \iiibf) {\bf Matching of de Donder-Stueckelberg gauge
conditions and differential constraints}. We now explain how the
differential constraints appearing in higher-derivative approach
\rf{oldman-21032012-08}, \rf{oldman-21032012-09} are derived from
the de Donder-Stueckelberg gauge conditions of ordinary-derivative
approach \rf{05032012-01}. Firstly, we note that gauge condition
$C_{-1}^a=0$ in \rf{05032012-01} supplemented with identification
\rf{oldman-21032012-19} coincides automatically with differential
constraint \rf{oldman-21032012-08}. Secondly, plugging solution for
field $\phi_0^{ab}$ \rf{05032012-05} into expression for $C_0$
\rf{05032012-10}, we get
\be \label{oldman-21032012-20}
C_0 =  \partial^a \phi_{-1}^a + \half \Box\phi_{-2}^{aa} +
u\phi_0\,.
\ee
From \rf{oldman-21032012-20},  we see that gauge condition $C_0=0$
\rf{05032012-01} supplemented with identification
\rf{oldman-21032012-19} coincides with differential constraint
\rf{oldman-21032012-09}. Thus we see that the de Donder-Stueckelberg
gauge conditions $C_{-1}^a=0$ and $C_0=0$ \rf{05032012-01} lead to
the respective differential constraints \rf{oldman-21032012-08} and
\rf{oldman-21032012-09} when $d=6$. Note that the remaining de
Donder-Stueckelberg gauge condition $C_1^a=0$ \rf{05032012-01} does
not lead to new differential constraint. This is to say that using
\rf{05032012-05},\rf{05032012-07} we find
\be
C_1^a = \Box C_{-1}^a\,,
\ee
i.e., we see that the relation $C_1^a=0$ is automatically satisfied
by virtue of the relation $C_{-1}^a=0$.

As a side remark we note that plugging
\rf{05032012-06},\rf{05032012-07} into expression for $C_2$ in
\rf{05032012-10}, we find
\be
\label{05032012-25} C_2 = - \frac{1}{u}\Box \phi_0\,.
\ee
We now note that using \rf{05032012-25} in
\rf{05032012-05}-\rf{05032012-07} allows us to express all fields in
terms of the ones entering higher-derivative approach
\rf{oldman-21032012-19},
\beq
\phi_0^{ab} =  \Box \phi_{-2}^{ab} \,,
\qquad
\phi_2^{ab} = \Box^2 \phi_{-2}^{ab} +  \frac{1}{2u} \eta^{ab} \Box
\phi_0\,,
\qquad\
\phi_1^a = \Box \phi_{-1}^a \,.
\eeq

\noindent \ivbf) {\bf Matching of gauge symmetries of
ordinary-derivative and higher-derivative approaches}. Finally, we
explain how the gauge transformations of higher-derivative approach
\rf{oldman-21032012-15} are obtained from the gauge transformations
of ordinary-derivative approach \rf{05032012-15}. As we have already
noticed, de Donder-Stueckelberg gauge conditions \rf{05032012-01}
are invariant under left-over gauge symmetries \rf{05032012-15},
where the gauge transformation parameters $\xi_{-1}^a$, $\xi_1^a$,
$\xi_0$ take the form as in \rf{oldman-21032012-18}, while the gauge
transformation parameters $\xi_{-3}^a$, $\xi_{-2}$ are not subject
to differential constraints. Using \rf{oldman-21032012-18} in
\rf{05032012-15}, we find the following gauge transformations of
fields $\phi_{-2}^{ab}$ , $\phi_{-1}^a$, $\phi_0$:
\beq
&& \delta \phi_{-2}^{ab} =\partial^a \xi_{-3}^b +
\partial^b \xi_{-3}^a + \half \eta^{ab} \xi_{-2}\,,
\nonumber\\
\label{oldman-21032012-22} && \delta \phi_{-1}^a = \partial^a \xi_0
- \Box \xi_{-3}^a \,,
\\
&& \delta \phi_0 = - u \Box \xi_{-2}\,.
\nonumber
\eeq
Introducing the identification of gauge transformation parameters in
\rf{oldman-21032012-22} with the ones in \rf{oldman-21032012-15},
\be
\xi_{-3}^a \equiv \xi^a \,, \qquad \xi_{-2} \equiv \xi\,,
\ee
we see that gauge transformations \rf{oldman-21032012-22} coincide
with the ones in \rf{oldman-21032012-15} when $d=6$.

\subsubsection{ 4d conformal gravity in de
Donder-Stueckelberg gauge frame}\label{subsub-4d}

We now proceed to illustrate the use of the de Donder-Stueckelberg
gauge frame for matching the ordinary-derivative and higher
derivative approaches to spin-2 conformal field in four-dimensional
space ($4d$ conformal gravity). Higher-derivative Lagrangian of $4d$
conformal gravity is obtained from \rf{oldman-21032012-07} by
equating $d=4$. We begin by recalling the ordinary-derivative
formulation of $4d$ conformal gravity developed in
Ref.\cite{Metsaev:2007fq}.

\noindent {\bf Ordinary-derivative formulation of $4d$ conformal
gravity}. In the framework of ordinary-deri\-va\-tive approach, $4d$
conformal gravity is described by two rank-2 tensor fields
$\phi_{-1}^{ab}$, $\phi_1^{ab}$ and one vector field $\phi_0^a$:
\beq
& \phi_{-1}^{ab}\qquad \phi_1^{ab} &
\nonumber\\
& \phi_0^a  &
\eeq
Ordinary-derivative Lagrangian of $4d$ conformal gravity is given by
\beq \label{oldman-22032012-01}
&& \hspace{-1.5cm} \LL = \frac{1}{2} \phi_{-1}^{ab} \Box \phi_1^{ab}
- \frac{1}{4} \phi_{-1}^{aa} \Box \phi_1^{bb} + \half \phi_0^a \Box
\phi_0^a +   C_0^a C_2^a + \half C_1 C_1 - \frac{1}{4} \phi_1^{ab}
\phi_1^{ab} + \frac{1}{8} \phi_1^{aa} \phi_1^{bb}\,,
\eeq
where the modified de Donder divergencies are defined by
\beq
\label{oldman-22032012-03n1} && \hspace{1cm} C_0^a  = \partial^b
\phi_{-1}^{ab} - \half
\partial^a\phi_{-1}^{bb} + \phi_0^a\,,
\nonumber\\
\label{oldman-22032012-03} && \hspace{1cm} C_2^a  = \partial^b
\phi_1^{ab} - \half
\partial^a\phi_1^{bb}\,,
\\
&& \hspace{1cm} C_1  = \partial^a \phi_0^a + \half \phi_1^{bb}\,.
\nonumber
\eeq
Lagrangian \rf{oldman-22032012-01} is invariant under gauge
transformations given by
\beq
&& \delta\phi_{-1}^{ab} =\partial^a\xi_{-2}^b +
\partial^b \xi_{-2}^a + \eta^{ab}\xi_{-1}\,,
\nonumber\\
\label{oldman-22032012-05}  && \delta\phi_1^{ab} =\partial^a\xi_0^b
+ \partial^b \xi_0^a\,,
\\
&& \delta\phi_0^a = \partial^a\xi_{-1} - \xi_0^a\,,
\nonumber
\eeq
where $\xi_{-2}^a$, $\xi_0^a$, and $\xi_{-1}$ are gauge
transformation parameters.

Now our purpose is to demonstrate how higher-derivative Lagrangian
\rf{oldman-21032012-07}, differential constraints
\rf{oldman-21032012-08}, \rf{oldman-21032012-09} and gauge
transformations \rf{oldman-21032012-15} are obtained from the
ordinary-derivative approach. In the de Donder-Stueckelberg gauge
frame, matching of the higher-derivative and ordinary derivative
approaches is realized as follows.

\medskip
\noindent \ibf) {\bf de Donder-Stueckelberg gauge}. We start with
the detailed discussion of de Donder-Stueckelberg gauge
\rf{manold-17032012-04} for the case of $4d$ conformal gravity. To
this end we note that under gauge transformations
\rf{oldman-22032012-05} the modified de Donder divergencies
\rf{oldman-22032012-03} transform as
\beq
\label{oldman-22032012-07} && \delta C_0^a = \Box \xi_{-2}^a -
\xi_0^a\,,
\\
\label{oldman-22032012-08} && \delta C_2^a = \Box \xi_0^a\,,
\\
\label{oldman-22032012-09} && \delta C_1 = \Box \xi_{-1}\,.
\eeq
From \rf{oldman-22032012-07}, we see that  under gauge
transformation governed by the gauge transformation parameter
$\xi_0^a$ the modified de Donder divergency $C_0^a$ transforms as
Stueckelberg field. This motivates us to introduce  de
Donder-Stueckelberg  gauge condition \rf{manold-17032012-04} which,
in the case under consideration, takes the form
\be \label{20032012-01}
C_0^a =0 \,, \qquad\qquad \hbox{ de Donder-Stueckelberg gauge}.
\ee
Obviously, the de Donder-Stueckelberg gauge does not completely fix
gauge.  This is to say that one can still make off-shell left-over
gauge transformations governed by the gauge transformation
parameters $\xi_{-2}^a$, $\xi_{-1}$ which are not subject to any
differential constraints. These left-over gauge symmetries are found
by requiring the de Donder-Stueckelberg gauge condition
\rf{20032012-01} to be invariant under gauge transformations,
\be \label{oldman-22032012-10}
\delta C_0^a =0 \,.
\ee
Using \rf{oldman-22032012-07}, we see that solution to
Eq.\rf{oldman-22032012-10} is given by
\be \label{oldman-22032012-18}
\xi_0^a = \Box\xi_{-2}^a\,.
\ee
This implies that left-over gauge transformations take the form
given in \rf{oldman-22032012-05}, where the gauge transformation
parameter $\xi_0^a$ is given in \rf{oldman-22032012-18}, while the
gauge transformation parameters $\xi_{-2}^a$, $\xi_{-1}$ are not
subject to differential constraints. Below we demonstrate that these
off-shell left-over gauge transformations are related to the gauge
transformations in higher-derivative approach
\rf{oldman-21032012-15}.

\medskip
\noindent \iibf) {\bf Matching of higher-derivative Lagrangian and
ordinary-derivative Lagrangian}. We now explain how
higher-derivative Lagrangian \rf{oldman-21032012-07} is obtained
from ordinary derivative Lagrangian \rf{oldman-22032012-01}. Using
gauge \rf{20032012-01} in equations of motion for the auxiliary
field $\phi_1^{ab}$, we get the following gauge-fixed equations of
motion:
\be
\label{20032012-02} \Box \phi_{-1}^{ab} - \phi_1^{ab} -\eta^{ab} C_1
= 0\,.
\ee
Eq.\rf{20032012-02} allows us to solve the auxiliary field
$\phi_1^{ab}$ as
\beq
\label{20032012-04} && \phi_1^{ab} =  \Box \phi_{-1}^{ab} -
\eta^{ab} C_1 \,.
\eeq
Plugging \rf{20032012-04} and \rf{20032012-01} into Lagrangian
\rf{oldman-22032012-01}, we get
\be \label{20032012-11}
\LL =  \frac{1}{4} \phi_{-1}^{ab} \Box^2 \phi_{-1}^{ab} -
\frac{1}{8} \phi_{-1}^{aa} \Box^2 \phi_{-1}^{bb} + \half \phi_0^a
\Box \phi_0^a + \frac{3}{2} C_1 C_1\,.
\ee
We now introduce a scalar field $\phi_1$ by the relation
\be \label{20032012-10}
C_1 + \frac{u}{3}\phi_1 =0 \,,\qquad u \equiv \sqrt{3}\,.
\ee
Using \rf{20032012-10} in \rf{20032012-11}, we get the Lagrangian
\beq \label{20032012-12}
&& \LL =  \frac{1}{4} \phi_{-1}^{ab} \Box^2 \phi_{-1}^{ab} -
\frac{1}{8} \phi_{-1}^{aa} \Box^2 \phi_{-1}^{bb} + \half \phi_0^a
\Box \phi_0^a + \half \phi_1 \phi_1\,.
\eeq
Making use of the identification of fields in \rf{20032012-12} with
the ones in \rf{oldman-21032012-07},
\be \label{oldman-22032012-19}
\phi_{-1}^{ab}\equiv \phi^{ab}\,, \qquad \phi_0^a\equiv \phi^a\,,
\qquad \phi_1 \equiv \phi\,,
\ee
we see that Lagrangian \rf{20032012-12} coincides with the one in
\rf{oldman-21032012-07} when $d=4$.

\noindent \iiibf) {\bf Matching of de Donder-Stueckelberg gauge
conditions and differential constraints}. We now explain how
differential constraints \rf{oldman-21032012-08},
\rf{oldman-21032012-09} are derived from the ordinary-derivative
approach. Firstly, we note that gauge condition $C_0^a=0$
\rf{20032012-01} supplemented with identification
\rf{oldman-22032012-19} coincides automatically with differential
constraint \rf{oldman-21032012-08}. Secondly, plugging $\phi_1^{ab}$
\rf{20032012-04} into expression for $C_1$ in
\rf{oldman-22032012-03}, we get
\be
\label{20032012-09}  \partial^a \phi_0^a + \half \Box \phi_{-1}^{aa}
-3 C_1 = 0 \,.
\ee
Using \rf{20032012-10} in \rf{20032012-09},  we see that constraint
\rf{20032012-09} supplemented with identification
\rf{oldman-22032012-19} coincides with differential constraint
\rf{oldman-21032012-09}. Thus we see that de Donder-Stueckelberg
gauge condition $C_0^a=0$ \rf{20032012-01} and relation
\rf{20032012-10} lead to the respective differential constraints
\rf{oldman-21032012-08} and \rf{oldman-21032012-09} when $d=4$.

As a side remark we note that using \rf{20032012-10} in
\rf{20032012-04} allows us to represent the auxiliary field
$\phi_1^{ab}$ in terms of the fields entering higher-derivative
approach \rf{oldman-22032012-19},
\be
\phi_1^{ab} = \Box \phi_{-1}^{ab} +  \frac{1}{u} \eta^{ab} \phi_1\,.
\ee

\noindent \ivbf) {\bf Matching of gauge symmetries of
ordinary-derivative and higher-derivative approaches}. Finally, we
explain how gauge transformations \rf{oldman-21032012-15} are
obtained from the ones in \rf{oldman-22032012-05}. As we have
already noticed, de Donder-Stueckelberg gauge condition
\rf{20032012-01} is invariant under left-over gauge symmetries
\rf{oldman-22032012-05}, where the gauge transformation parameter
$\xi_0^a$ is given in \rf{oldman-22032012-18}, while the gauge
transformation parameters $\xi_{-2}^a$, $\xi_{-1}$ are not subject
to differential constraints. Using \rf{oldman-22032012-18} in
\rf{oldman-22032012-05}, we find the following gauge transformations
of fields $\phi_{-1}^{ab}$, $\phi_0^a$, $\phi_1$:
\beq
&& \delta \phi_{-1}^{ab} =\partial^a \xi_{-2}^b +
\partial^b \xi_{-2}^a +   \eta^{ab} \xi_{-1}\,,
\nonumber\\
\label{oldman-22032012-21} && \delta \phi_0^a = \partial^a \xi_{-1}
- \Box \xi_{-2}^a \,,
\\
&& \delta \phi_1 = - u \Box \xi_{-1}\,,
\nonumber
\eeq
where for the derivation of gauge transformation $\delta\phi_1$ we
use \rf{oldman-22032012-09} and \rf{20032012-10}. Introducing the
identification of gauge transformation parameters in
\rf{oldman-22032012-21} with the ones in \rf{oldman-21032012-15},
\be
\xi_{-2}^a \equiv \xi^a \,, \qquad \xi_{-1} \equiv \xi\,,
\ee
we see that gauge transformations \rf{oldman-22032012-21} coincide
with the ones in \rf{oldman-21032012-15} when $d=4$.

\subsubsection{ Arbitrary spin conformal field
in de Donder-Stueckelberg gauge frame } \label{subsub-arbd}

In this section, we are going to demonstrate equivalence of
ordinary-derivative and higher-deri\-vative approaches to arbitrary
spin conformal field by using the de Donder-Stueckelberg gauge
condition. Before going into technical details we would like to
formulate our main statements in this section. Let field $\phik$
\rf{phikdef01} entering the ordinary-derivative approach in
Sec.\ref{lagran} satisfies the following two equations:
\beq
\label{10032012-17} && \bar\upsilon^\oplussm \Cb \phik = 0\,,
\\
\label{11032012-01} && \bar\upsilon^\oplussm (\Box - m^2 -
G\Cb)\phik = 0\,,
\eeq
where $m^2$, $\Cb$ and $G$ are given in \rf{manold-17032012-02} and
\rf{oldman-05022012-11} respectively. We recall that
Eq.\rf{10032012-17} is the de Donder-Stueckelberg gauge, while
Eqs.\rf{11032012-01} are equations of motion for auxiliary fields
which are obtained from Lagrangian \rf{manold-17032012-01}. Note
that Eqs.\rf{11032012-01} are equations of motion only for those
auxiliary fields that appear in the list of fields in
\rf{manold-22032012-24}. We now formulate our main statements in
this section.

\noindent {\bf Statement 1}. Solution to
Eqs.\rf{10032012-17},\rf{11032012-01} is given by

\be \label{oldman-17032012-01}
\phik = e^{X \Box} (\upsilon^\ominussm)^\kwh |\Phi'\rangle  +
\alpha^2 \frac{1}{2N_\alpha+d-2} \ewt_1  Z \Phik \,,
\ee
where ket-vector $\Phik$ satisfies the differential constraint
\be \label{16032012-05}
\Cb_\sh \Phik = 0\,, \hspace{3cm} \hbox{for } \ d \geq 4\,,
\ee
and operator $\Cb_\sh$ in \rf{16032012-05} takes the same form as in
\rf{manold-06032012-04n1}. In \rf{oldman-17032012-01}, we use the
following notation (see also notation in
\rf{manold-18032012-02}-\rf{manold-18032012-06}):
\beq
\label{11032012-05} && X \equiv \upsilon^\oplussm
\frac{1}{N_{\upsilon^\ominussm} + 1}\bar\upsilon^\oplussm\,,
\qquad
Y \equiv \frac{1}{(N_\zeta+2)\ewt_1^\smone} V \upsilon^\oplussm
\bar\zeta\,,
\\
\label{11032012-05n1} && Z \equiv e^Y \upsilon^\oplussm
\frac{1}{N_{\upsilon^\oplussm}+1} e^{- Y}
\frac{1}{(N_\zeta+2)\ewt_1^\smone} \bar\zeta^2 \Pi^\smponetwo
\frac{(\upsilon^\oplussm \Box)^{\kwh+1} }{\Gamma(\kwh+2)}\,,
\eeq

\beq
\label{oldman-19032012-01} && \Phik  \equiv \sum_{s'=0}^s
\frac{\zeta^{s-s'}}{s'!\sqrt{(s-s')!}} \alpha^{a_1} \ldots
\alpha^{a_{s'}} \, \phi_{-k_{s'}}^{a_1\ldots a_{s'}}|0\rangle\,,
\qquad \hbox{for } \ d \geq 4 \,,\qquad
\\
\label{oldman-19032012-02} && |\Phi'\rangle \equiv \sum_{s'=1}^s
\frac{\zeta^{s-s'}}{s'!\sqrt{(s-s')!}} \alpha^{a_1} \ldots
\alpha^{a_{s'}} \, \phi_{-k_{s'}}^{a_1\ldots a_{s'}}|0\rangle\,,
\qquad \hbox{for } \ d=4 \,,\qquad
\\
\label{oldman-19032012-03} && |\Phi'\rangle \equiv \Phik\,,
\hspace{7cm} \hbox{for } \ d \geq 6\,.
\eeq
Operator $\ewt_1^\smone$ \rf{11032012-05} is defined in
\rf{ewtdef01}, while $\Gamma$ stands for the Euler gamma function.

Making use of the identification of fields entering the
ordinary-derivative and higher-derivative approaches
\rf{manold-23032012-01}, \rf{manold-23032012-02}, we note that
Statement 1 implies that, in the de Donder-Stueckelberg gauge frame,
all fields of the ordinary-derivative approach can be expressed in
terms of fields entering the higher-derivative approach. For the
case of $d=4$, the identification \rf{manold-23032012-02} and the
appearance of scalar field $\phi_1$ in \rf{oldman-19032012-01}
require some comment. From \rf{oldman-19032012-01},
\rf{oldman-19032012-02}, we find the relation
\be \label{manold-10042012-01}
\Phik = |\Phi'\rangle + |\Phi_0\rangle \,, \qquad |\Phi_0\rangle
\equiv \frac{\zeta^s}{\sqrt{s!}}\phi_1|0\rangle\,, \qquad\quad
\hbox{for }\ d=4\,.
\ee
Plugging \rf{manold-10042012-01} into \rf{16032012-05} we can
represent constraint \rf{16032012-05} in terms of $|\Phi'\rangle$
and $|\Phi_0\rangle$,
\be \label{manold-10042012-02}
\Cb_\sh |\Phi'\rangle + \ewt_1 \bar\zeta |\Phi_0\rangle = 0\,.
\ee
Considering $\zeta^{s-1}$-terms in \rf{manold-10042012-02}, we find
the relation
\be \label{manold-10042012-03}
-\sqrt{\frac{2}{s(s+1)}}\Bigl(\partial^a\phi_0^a +
\frac{1}{4}\sqrt{(s-1)(s+2)}\Box\phi_{-1}^{aa}\Bigr) = \phi_1\,,
\qquad \hbox{ for }\ d=4\,.
\ee
We now note that, because scalar fields do not enter the field
content of $4d$ conformal field theory, relation
\rf{manold-10042012-03} should be considered as the definition of
the scalar field $\phi_1$ when $d=4$. Then, using identification of
the scalar field $\phi_1$ with the scalar field $\phi$ appearing in
the higher-derivative approach, $\phi_1\equiv \phi$, we arrive at
the relation given in \rf{manold-23032012-02}.

\noindent {\bf Statement 2}. Plugging solution
\rf{oldman-17032012-01} into ordinary-derivative Lagrangian
\rf{manold-17032012-01} leads to the following higher-derivative
Lagrangian:
\be \label{oldman-19032012-04}
\LL = \half \langle \Phi|\mubf\Box^{\kwh+1}|\Phi\rangle\,, \qquad
\hbox{for } \ d \geq 4\,.
\ee

Making use of the identification of fields entering the
ordinary-derivative and higher-derivative approaches
\rf{manold-23032012-01}, \rf{manold-23032012-02}, we note that
Statement 2 implies that, in de Donder-Stueckelberg gauge frame,
ordinary-derivative Lagrangian \rf{manold-17032012-01} coincides
with higher-derivative Lagrangian \rf{manold-06032012-05}.

\noindent {\bf Statement 3}. de Donder-Stueckelberg gauge
\rf{10032012-17} is invariant under left-over gauge transformations
which take the form as in \rf{oldman-05022012-11}, where the gauge
transformation parameter $\xik$ satisfies the equation
\be \label{manold-18032012-07}
\bar\upsilon^\oplussm(\Box - m^2)\xik = 0 \,, \qquad m^2\equiv
\upsilon^\ominussm\bar\upsilon^\ominussm\,.
\ee
Solution to Eq.\rf{manold-18032012-07} is given by

\be \label{manold-18032012-08}
\xik = e^{X \Box} (\upsilon^\ominussm)^{\kwh+1} \Xik \,,
\ee
where operator $X$ is given in \rf{11032012-05}, while ket-vector
$\Xik$ is defined as
\be \label{oldman-19032012-06}
\Xik  \equiv \sum_{s'=0}^{s-1}
\frac{\zeta^{s-1-s'}}{s'!\sqrt{(s-1-s')!}} \alpha^{a_1} \ldots
\alpha^{a_{s'}} \, \xi_{-k_{s'}-2}^{a_1\ldots a_{s'}}|0\rangle\,,
\qquad \hbox{for } \ d \geq 4 \,. \qquad
\ee
Gauge transformations \rf{oldman-05022012-11} amount to the
following gauge transformations:
\be \label{oldman-19032012-07}
\delta \Phik = G_\sh \Xik\,,
\ee
where operator $G_\sh$ takes the same form as in
\rf{manold-06032012-04n1}.

Making use of the identification of the gauge transformation
parameters in \rf{oldman-19032012-06} and the ones in
\rf{manold-06032012-06n1},
\be \label{oldman-19032012-08}
\xi_{-k_{s'}-2}^{a_1\ldots a_{s'}} \equiv \xi^{a_1\ldots a_{s'}}\,,
\ee
we note that Statement 3 implies that, in the de Donder-Stueckelberg
gauge frame, gauge transformations of ordinary-derivative approach
\rf{oldman-05022012-11} lead to gauge transformations of
higher-derivative approach \rf{manold-06032012-18}.

Proof of the Statement 1 may be found in Appendix C, while the
Statements 2,3 are proved in Appendix D.

\subsection{ Stueckelberg gauge frame }\label{sub-502}

In this section, we describe a Stueckelberg gauge frame. Although
helpful in the studying general properties of the
ordinary-derivative conformal field theory, this gauge frame seems
to be less useful for practical computation as compared to the de
Donder-Stueckelberg gauge frame. Therefore we outline the
Stueckelberg gauge frame without going into details.

We recall that to develop the ordinary-derivative formulation we use
fields given in \rf{phiset01}. Fields \rf{phiset01} can be separated
into three groups: dynamical fields, auxiliary fields, and
Stueckelberg fields. We now identify fields in \rf {phiset01} that
are realized as dynamical, auxiliary and Stueckelberg fields. To
this end we use the shortcut $\phi_{k'}^{s'}$ for the field
$\phi_{k'}^{a_1\ldots a_{s'}}$ and note that the tensor field
$\phi_{k'}^{s'}$, $s'\geq 2$, can be decomposed into two traceless
tensor fields,
\be \label{manold-09032012-06}
\phi_{k'}^{s'} = \phi_{k'}^{(\Irm)\, s'} \oplus \phi_{k'}^{(\IIrm)\,
s'-2}\,, \qquad s'= 2,3,\ldots,s\,,\qquad k' \in [k_{s'}]_2\,,
\ee
where $\phi_{k'}^{(\Irm)\,s'}$ and $\phi_{k'}^{(\IIrm)\, s'-2}$
stand for the respective rank-$s'$ and rank-$(s'-2)$ traceless
tensor fields of the Lorentz algebra $so(d-1,1)$. Taking into
account decomposition \rf{manold-09032012-06}, we note that fields
entering the ordinary-derivative approach \rf{phiset01} can be
represented as
\beq
\label{manold-09032012-02} && \phi_{k'}^{(\Irm)\, s'}\,,
\hspace{0.8cm} s'=2,3,\ldots,s\,, \hspace{2cm} k'\in [k_{s'}]_2\,,
\\
\label{manold-09032012-03} && \phi_{k'}^{(\IIrm)\, s'}\,,
\hspace{0.8cm} s'=0,1,\ldots,s-2\,,  \hspace{1.2cm} k'\in
[k_{s'}+2]_2\,,\qquad
\\
\label{manold-09032012-04} && \phi_{k'}^1 \,, \hspace{6cm} k'\in
[k_1]_2\,,
\\
\label{manold-09032012-05} && \phi_{k'}^0\,, \hspace{6cm} k' \in
[k_0]_2\,.
\eeq
Now we are going to prove the following two related statements.

\noindent {\bf i}) Fields
\rf{manold-09032012-02}-\rf{manold-09032012-05} are separated into
dynamical, auxiliary and Stueckelberg fields as follows:
\beq
&& \hspace{-2cm} \hbox{dynamical field:}
\nonumber\\
&& \phi_{-k_s}^{(\Irm)\, s};
\\
&& \hspace{-2cm} \hbox{auxiliary fields:}
\nonumber\\
\label{oldman-08032012-10} && \phi_{k'}^{(\Irm)\, s}\,, \qquad \ k'
\in [k_s-2]_2,k_s \,,
\\
\label{oldman-08032012-11} && \phi_{k'}^{(\IIrm) \,s'}\,, \qquad s'=
0,1,\ldots,s-2\,,\qquad k' \in [k_{s'}]_2, k_{s'}+2;
\\
&& \hspace{-2cm} \hbox{Stueckelberg fields:}
\nonumber\\
\label{oldman-10032012-04} && \phi_{k'}^{(\Irm)\, s'}\,,
\hspace{1.2cm} s'=2,3,\ldots,s-1\,, \hspace{1.2cm} k'\in
[k_{s'}]_2\,,
\\
\label{oldman-10032012-05} && \phi_{k'}^1 \,, \hspace{6.4cm} k'\in
[k_1]_2\,,
\\
\label{oldman-10032012-06} && \phi_{k'}^0\,, \hspace{6.4cm} k' \in
[k_0]_2\,,
\\
\label{oldman-10032012-07} && \phi_{-k_{s'}-2}^{(\IIrm)\,s'}\,,
\qquad  s' = 0,1,\ldots, s-2\,.
\eeq

\noindent {\bf ii}) With exception of the gauge transformation
parameter $\xi_{-k_s-1}^{a_1\ldots a_{s-1}}$ \rf{epsilonset01}, all
gauge transformation parameters \rf{epsilonset01} are related to
Stueckelberg gauge transformations which can be used to gauge away
the Stueckelberg fields in
\rf{oldman-10032012-04}-\rf{oldman-10032012-07}.

These two statements can easily be proved by using gauge
transformations \rf{oldman-05022012-11}. To this end let us
introduce the following simplified notation for fields, gauge
transformation parameters, derivatives, and flat metric tensor:
\be
\phi_{k'}^{s'} \sim \phi_{k'}^{a_1\ldots a_{s'}}\,,\qquad
\xi_{k'}^{s'} \sim \xi_{k'}^{a_1\ldots a_{s'}}\,,\quad
\partial \sim \partial^a\,,\quad \eta \sim \eta^{ab}\,.\ee
Using such notation, gauge transformations \rf{oldman-05022012-11}
can schematically be represented as
\beq
\label{gautraspi2def02} && \delta \phi_{k'}^{s'} =
\partial \xi_{k'-1}^{s'-1} + \xi_{k'}^{s'}  +
\eta \xi_{k'}^{s'-2}  \,, \qquad s' = 2,3,\ldots,s\,,\quad k'\in
[k_{s'}]_2\,,\qquad
\\
\label{gautraspi2def03} && \delta \phi_{k'}^1 = \partial
\xi_{k'-1}^0 + \xi_{k'}^1 \,, \hspace{5.7cm} k'\in [k_1]_2\,,
\\
\label{gautraspi2def04} && \delta \phi_{k'}^0 = \xi_{k'}^0 \,,
\hspace{7.3cm} k'\in [k_0]_2\,.
\eeq
Plugging decomposition \rf{manold-09032012-06} into
\rf{gautraspi2def02} gives the following gauge transformations:
\beq
\label{oldman-10032012-01} && \delta \phi_{k'}^{(\Irm)\, s'} =
\partial \xi_{k'-1}^{s'-1} + \xi_{k'}^{s'} \,, \qquad s' =
2,3,\ldots,s\,,\hspace{1.7cm}  k'\in [k_{s'}]_2\,, \qquad
\\
\label{oldman-10032012-02} && \delta \phi_{k'}^{(\IIrm)\, s'} =
\partial \xi_{k'-1}^{s'+1}  +
\xi_{k'}^{s'}  \,, \qquad s' = 0,1,\ldots,s-2\,, \qquad k'\in
[k_{s'}+2]_2\,.\qquad
\eeq
We now note that gauge transformation parameters \rf{epsilonset01}
can be represented as
\be \label{manold-09032012-01}
\xi_{k'}^{s'}\,,\hspace{1.5cm} s'=0,1,\ldots,s-1;
\hspace{1.5cm} k' \in -k_{s'}-2, [k_{s'}]_2\,.
\ee

From \rf{gautraspi2def03}-\rf{oldman-10032012-01}, we see that all
scalar fields $\phi_{k'}^0$, all vector fields $\phi_{k'}^1$, and
the tensor fields $\phi_{k'}^{(\Irm)\, s'}$, $s=2,3,\ldots, s-1$,
transform as Stueckelberg fields. In other words, using the
Stueckelberg gauge symmetries governed by the gauge transformation
parameters
\be \xi_{k'}^{s'}\,, \qquad s'= 0,1,\ldots,s-1\,, \qquad \ \ k' \in
[k_{s'}]_2 \,,\ee
we can gauge away the following fields:
\beq
\label{manold-09032012-07} && \phi_{k'}^{(\Irm)\, s'}\,,
\hspace{0.8cm} s'=2,3,\ldots,s-1\,, \hspace{1.2cm} k'\in
[k_{s'}]_2\,,
\\
\label{manold-09032012-08} && \phi_{k'}^1 \,, \hspace{6cm} k'\in
[k_1]_2\,,
\\
\label{manold-09032012-09} && \phi_{k'}^0\,, \hspace{6cm} k' \in
[k_0]_2\,.
\eeq
After gauging away fields
\rf{manold-09032012-07}-\rf{manold-09032012-09}, we are left with
the fields
\beq \label{phiintset01}
&& \phi_{k'}^{(\Irm)\, s}\,, \qquad \quad \ k' \in [k_s]_2\,,
\\
\label{phiintset02} &&\phi_{k'}^{(\IIrm)\, s'}\,, \qquad \ \ \ s'=
0,1,\ldots,s-2\,, \qquad k' \in  [k_{s'}+2]_2\,,
\eeq
and gauge symmetries governed by the following gauge transformation
parameters:
\be
\label{epsintgau01} \xi_{-k_{s'}-2}^{s'}\,,\qquad s'=0,1,\ldots,
s-1\,.
\ee
From \rf{oldman-10032012-02}, we see that gauge transformations
governed by gauge transformation parameters \rf{epsintgau01} with
$s'=0,1,\ldots,s-2$ are realized as Stueckelberg gauge
transformations. This is to say that, using gauge transformations
\rf{oldman-10032012-02} governed by the gauge transformation
parameters
\be
\xi_{-k_{s'}-2}^{s'}\,, \qquad  s' = 0,1,\ldots, s-2\,,
\ee
we can gauge away the following set of fields in \rf{phiintset02}:

\be \label{oldman-10032012-08}
\phi_{-k_{s'}-2}^{(\IIrm)\,s'}\,, \qquad  s' = 0,1,\ldots, s-2\,.
\ee
Thus, gauging away all Stueckelberg fields given in
\rf{manold-09032012-07}-\rf{manold-09032012-09} and
\rf{oldman-10032012-08}, we are left with the fields
\beq
\label{phiintset04} && \phi_{k'}^{(\Irm)\, s}\,, \qquad \ k' \in
[k_s]_2\,,
\\
\label{phiintset05} && \phi_{k'}^{(\IIrm)\,s'}\,, \qquad s'=
0,1,\ldots,s-2\,, \qquad k' \in [k_{s'}]_2, k_{s'}+2\,,
\eeq
and gauge symmetry governed by the gauge transformation parameter
\be \label{epsintgau03}
\xi_{-k_s-1}^{s-1}\,.
\ee
We now note that the following fields in \rf{phiintset04}
\be \label{oldman-10032012-03}
\phi_{k'}^{(\Irm)\, s}\,, \qquad \ k' \in [k_s-2]_2, k_s\,,
\ee
and fields in \rf{phiintset05} turn out to be auxiliary fields,
while the field $\phi_{-k_s}^{(\Irm), s}$ in \rf{phiintset04} is
realized as dynamical field. This is to say that, using equations of
motion for fields in \rf{phiintset05},\rf{oldman-10032012-03} one
can make sure that fields in
\rf{phiintset05},\rf{oldman-10032012-03} can be expressed in terms
of the field $\phi_{-k_s}^{(\Irm), s}$. This leads to the
higher-derivative description of conformal field which is formulated
in terms of the traceless field $\phi_{-k_s}^{(\Irm), s}$ in
Ref.\cite{Fradkin:1985am,Segal:2002gd}. Gauge symmetries of this
traceless field are governed by gauge transformation parameter
\rf{epsintgau03}.

\newsection{ \large Interrelation between ordinary-derivative
description of conformal field and gauge invariant description of
massive field}\label{interel}

The ordinary-derivative formulation of conformal field developed in
this paper involves Stueckelberg fields. As is well known, the gauge
invariant description of massive field is also formulated by using
Stueckelberg fields. Obviously, field contents in the
ordinary-derivative conformal field theory and the massive gauge
field theory are different. However, it turns out that there are
interesting interrelations between the ordinary-derivative
description of conformal field and the gauge invariant description
of massive field. These interrelations can straightforwardly be
illustrated by using the oscillator formulation we use in this
paper. To do this, we need the oscillator form of Lagrangian and
gauge transformations for massive field. In the tensorial notation,
Lagrangian and gauge transformations for arbitrary spin massive
field were obtained in Ref.\cite{Zinoviev:2001dt}. Using the
oscillators and suitable modified de Donder divergencies, we find
alternative simple representation for the gauge invariant Lagrangian
of massive field. This is to say that it is the use of the modified
de Donder divergencies that simplifies significantly our
representation for gauge invariant Lagrangian of massive field. We
discuss our representation for the gauge invariant Lagrangian of
massive field in Sec. \ref{sec-06}.

\subsection{ Gauge invariant Lagrangian of massive field via modified
de Donder divergency}\label{sec-06}

Gauge invariant formulation of totally symmetric spin-$s$ massive
field in flat space of dimension $d\geq 4$ involves the following
set of scalar, vector, and tensor fields of the Lorentz algebra
$so(d-1,1)$ (see Ref.\cite{Zinoviev:2001dt}):
\beq \label{masphiset01}
&& \phi^{a_1\ldots a_{s'}}\,, \hspace{2cm} s'=0,1,\ldots,s\,.
\eeq

We note that, in \rf{masphiset01}, the fields $\phi$ and $\phi^a$
are the respective scalar and vector fields of the Lorentz algebra,
while the field $\phi^{a_1\ldots a_{s'}}$, $s'>1$, is rank-$s'$
totally symmetric tensor field of the Lorentz algebra $so(d-1,1)$.
The tensor fields $\phi^{a_1\ldots a_{s'}}$ with $s'\geq 4 $ satisfy
the double-tracelessness constraint,
\be
\label{masdoutracon01} \phi^{aabba_5\ldots a_{s'}}=0\,, \qquad
s'\geq 4\,.
\ee

In order to obtain the gauge invariant description in an
easy--to--use form we use the oscillators $\alpha^a$, $\zeta$. The
fields \rf{masphiset01} can then be collected into ket-vector
$\phik$ defined by
\be
\label{masphikdef01} \phik \equiv \sum_{s'=0}^s
\frac{\zeta^{s-s'}}{\sqrt{(s-s')!}} |\phi^{s'}\rangle \,,
\qquad \quad  |\phi^{s'}\rangle \equiv  \frac{1}{s'!} \alpha^{a_1}
\ldots \alpha^{a_{s'}}\, \phi^{a_1\ldots a_{s'}} |0\rangle\,.
\ee
From \rf{masphikdef01}, we see that the ket-vectors $\phik$,
$|\phi^{s'}\rangle$ satisfy the relations
\be \label{oldman-05022012-14}
(N_\alpha + N_\zeta - s)\phik = 0 \,, \qquad  (N_\alpha - s')
|\phi^{s'}\rangle =0\,.
\ee
Relations \rf{oldman-05022012-14} tell us that the ket-vector
$\phik$ is degree-$s$ homogeneous polynomial in the oscillators
$\alpha^a$, $\zeta$, while the ket-vector $|\phi^{s'}\rangle$ is
degree-$s'$ homogeneous polynomial in the oscillators $\alpha^a$. In
terms of ket-vector $\phik$ \rf{masphikdef01}, double-tracelessness
constraint \rf{masdoutracon01} takes the form $(\bar{\alpha}^2)^2
\phik = 0$.

In terms of ket-vector \rf{masphikdef01}, the new representation for
Lagrangian we find takes the form
\beq
\label{masspi2lag01} && \LL = \frac{1}{2} \phibr \mubf(\Box - m^2)
\phik + \half \langle \Cb \phi|\Cb\phi\rangle \,,
\\
\label{oldman-07042012-01} && \hspace{1cm} \Cb = \albpar - \half
\alpar \bar\alpha^2 - \eb_1 \Pi^\smponetwo + \half e_1
\bar\alpha^2\,,
\\
\label{oldman-05122012-17}  && \hspace{1cm} e_1 = \zeta \ewt_1 m
\,,\qquad \qquad \eb_1 = - m \ewt_1 \bar\zeta\,, \qquad \ewt_1 =
\Bigl(\frac{2s+d-4-N_\zeta}{2s+d-4-2N_\zeta}\Bigr)^{1/2}\,,\qquad
\eeq
where the operators $\mubf$ and $\Pi^\smponetwo$ are given in
\rf{manold-18032012-06}. We recall that the quantity $\Cb\phik$ is
referred to as modified de Donder divergency in this paper. From
\rf{masspi2lag01}, we see that it is the use of the modified de
Donder divergency that simplifies considerably the presentation of
the Lagrangian.

To illustrate the structure of the Lagrangian we note that, in terms
of fields $\phi^{a_1\ldots a_{s'}}$ \rf{masphiset01}, Lagrangian
\rf{masspi2lag01} takes the form
\beq
\LL & = & \sum_{s'=0}^s   \LL^{s'}\,,
\\
\LL^{s'} & \equiv & \frac{1}{2 s'!}\Bigl( \phi^{a_1\ldots a_{s'}}
(\Box - m^2)\phi^{a_1\ldots a_{s'}}
- \frac{s'(s'-1)}{4} \phi^{aaa_3\ldots a_{s'}} (\Box-m^2)
\phi^{bba_3\ldots a_{s'}}\Bigr)\qquad
\nonumber\\
& + &  \frac{1}{2 (s'-1)!} C^{a_1\ldots a_{s'-1}} C^{a_1\ldots
a_{s'-1}}\,,
\\
&& C^{a_1\ldots a_{s'-1}} \equiv
\partial^b \phi^{a_1\ldots a_{s'-1} b}
-\frac{s'-1}{2}\partial^{(a_1} \phi^{a_2\ldots a_{s'-1}) bb}
\nonumber\\
&& \hspace{1.6cm} + \, f_{s'} m\phi^{\tr\ a_1\ldots a_{s'-1}} +
\half f_{s'+1} m \phi^{a_1\ldots a_{s'-1}bb}\,,
\\
&& \phi^{\tr\ a_1\ldots a_{s'}} \equiv \phi^{a_1\ldots a_{s'}} -
\frac{s'(s'-1)}{2(2s'+d-4)} \eta^{(a_1a_2} \phi^{a_3\ldots a_{s'})
bb}\,,
\eeq

\beq
&& f_{s'} \equiv
\Bigl(\frac{(s+1-s')(s+s'+d-4)}{2s'+d-4}\Bigr)^{1/2}\,.
\eeq

We now discuss gauge symmetries of Lagrangian \rf{masspi2lag01}. As
in Ref.\cite{Zinoviev:2001dt}, we introduce the following set of
gauge transformation parameters:
\be \label{masepsilonset01}
\xi^{a_1\ldots a_{s'}}\,,\qquad\qquad s'=0,1,\ldots,s-1\,.
\ee
We note that, in \rf{masepsilonset01}, the gauge transformation
parameters $\xi$ and $\xi^a$ are the respective scalar and vector
fields of the Lorentz algebra, while the gauge transformation
parameter $\xi^{a_1\ldots a_{s'}}$, $s'\geq 2$, is rank-$s'$ totally
symmetric traceless tensor field of the Lorentz algebra $so(d-1,1)$,
\be \label{masepsdoutracon01} \xi^{aaa_3\ldots a_{s'}}=0\,, \qquad
s'\geq 2\,. \ee
Now, as usually, we collect gauge transformation parameters in
ket-vector $\xik$ defined by
\be
\xik \equiv \sum_{s'=0}^{s-1}
\frac{\zeta^{s-1-s'}}{\sqrt{(s-1-s')!}} |\xi^{s'}\rangle \,, \qquad
|\xi^{s'}\rangle \equiv  \frac{1}{s'!} \alpha^{a_1} \ldots
\alpha^{a_{s'}} \xi^{a_1\ldots a_{s'}} |0\rangle\,.
\ee
The ket-vectors $\xik$, $|\xi^{s'}\rangle$ satisfy the relations
\be
(N_\alpha + N_\zeta - s +1 ) \xik=0 \,,
\qquad  (N_\alpha -s')|\xi^{s'}\rangle = 0 \,,
\ee
which tell us that the ket-vector $\xik$ is degree-$(s-1)$
homogeneous polynomial in the oscillators $\alpha^a$, $\zeta$, while
the ket-vector $|\xi^{s'}\rangle$ is degree-$s'$ homogeneous
polynomial in the oscillators $\alpha^a$. In terms of the ket-vector
$\xik$, tracelessness constraint \rf{masepsdoutracon01} takes the
form $\bar\alpha^2 \xik=0$.

Gauge transformations can entirely be written in terms of the
ket-vectors $\phik$ and $\xik$. This is to say that gauge
transformations take the form
\be \label{masgautraarbspi01}
\delta \phik =  G \xik  \,, \qquad G\equiv \alpar - e_1 -
\alpha^2\frac{1}{2N_\alpha+d-2}\eb_1\,,
\ee
where operators $e_1$, $\eb_1$ are defined in
\rf{oldman-05122012-17}.

\subsection{ Comparison of ordinary-derivative description of
conformal field and gauge invariant description of massive field}

We are now ready to make comparison of the ordinary-derivative
description of the conformal field and the gauge invariant
description of the massive field. As we have already said, the field
contents of conformal field theory and massive field theory are
different (see \rf{phiset01} and \rf{masphiset01}). We note however
that the formulation of conformal and massive field theories in
terms of the respective ket-vectors \rf{phikdef01} and
\rf{masphikdef01} leads to remarkable interrelations between the
structures of Lagrangians and gauge transformations. Namely,
comparing representation for Lagrangian and gauge transformations of
conformal field \rf{manold-17032012-01}, \rf{oldman-05022012-11}
with the ones for the massive field \rf {masspi2lag01}, \rf
{masgautraarbspi01}, we see that Lagrangians and gauge
transformations are constructed out of the operators $\Cb$, $m^2$,
$G$. Moreover, comparing the operators $\Cb$, $m^2$, $G$ for the
conformal field with the ones for massive field we conclude that:

\medskip
\noindent \ibf) Operators $\Cb$ \rf{manold-17032012-02}, $G$
\rf{oldman-05022012-11} of conformal field depend on the oscillators
$\alpha^a$ and the derivatives $\partial^a$ in the same way as
operators $\Cb$ \rf{oldman-07042012-01}, $G$ \rf{masgautraarbspi01}
of massive field. Also note that the operators $\Cb$, $G$ for
conformal field and the ones for massive field are distinguished
only by the operators $e_1$, $\eb_1$.

\medskip
\noindent \iibf) Operators $e_1$, $\eb_1$, \rf{oldman-05022012-10},
$m^2$ \rf{manold-17032012-01} for conformal field
 can be obtained from the ones for massive
field \rf{oldman-05122012-17} by the substitutions
\be
\zeta m \rightarrow  \zeta\bar\upsilon^\ominussm \,, \qquad
\qquad
m \bar\zeta \rightarrow \upsilon^\ominussm \bar\zeta\,,\qquad
\qquad
 m^2 \rightarrow \upsilon^\ominussm \bar\upsilon^\ominussm\,.
\ee
We note that, to some extent, these substitutions are realized
unambiguously.

At present time, many interesting approaches to gauge invariant
dynamics of massive fields are available in the literature.
Obviously, use of just mentioned interrelations between conformal
and massive fields theories might be helpful for the straightforward
generalization of those approaches to the case of conformal fields.
Note however that it is important to keep in mind the following
difference between the conformal and massive fields theories.
Lagrangian and gauge transformations of massive fields theory are
uniquely determined by requiring the Lagrangian to be invariant (up
to total derivative) under gauge transformations. For the case of
conformal fields theory such requirement does not allow to determine
the ordinary-derivative Lagrangian uniquely. One needs to impose the
additional requirement. Namely, one needs to require that the
Lagrangian be invariant under conformal algebra transformations.

\newsection{Conclusions}\label{conlcus}

In this paper, we applied ordinary-derivative approach developed in
Ref.\cite{Metsaev:2007fq} to the study of totally symmetric
arbitrary spin conformal bosonic fields in flat space of even
dimension greater than or equal to four. In our approach the gauge
symmetries are realized, among other things, by involving the
Stueckelberg fields. As is well known, the Stueckelberg approach
turned out be successful for the study of theories involving massive
fields. This is to say that all Lorentz covariant formulations of
string theory are realized by using Stueckelberg gauge symmetries.
Therefore we think that use of the Stueckelberg fields might be
useful for the study of the conformal fields. The results presented
here should have a number of interesting applications and
generalizations, some of which are:

\noindent \ibf) application of our approach to theory of interacting
higher-spin conformal fields. Various approaches to theory of
interacting higher-spin conformal fields were discussed in
Refs.\cite{Fradkin:1990ps,Fradkin:1989md,Segal:2002gd}. As we have
illustrated, in our approach, use of Stueckelberg fields is very
similar to the one in gauge invariant formulation of massive fields.
Stueckelberg fields provide interesting possibilities for the study
of interacting massive gauge fields theory (see, e.g.,
Refs.\cite{Zinoviev:2006im,Metsaev:2006ui}). We expect therefore
that application of our approach to theory of interacting conformal
fields might lead to new interesting development.

\noindent \iibf) application of BRST approach to the study of
ordinary-derivative conformal fields theory. In recent time, BRST
approach was extensively used for the study of gauge invariant
formulation of massive fields (see, e.g.,
Ref.\cite{Buchbinder:2007vq}). Because gauge invariant formulation
of massive fields and ordinary-derivative formulation of conformal
fields share many common features, application of BRST approach to
the study of ordinary-derivative conformal fields theories should be
straightforward.

\noindent \iiibf) ordinary-derivative formulation of conformal
fields theory in terms of unconstrained fields. In
Ref.\cite{Francia:2002aa}, various formulations of higher-spin
dynamics in terms of unconstrained gauge fields were developed.
Application of those formulations to the study of
ordinary-derivative conformal fields theory could be of some
interest.

\noindent \ivbf) extension of our approach to mixed-symmetry fields.
Mixed-symmetry fields \cite{Aulakh:1986cb} have extensively been
studied in recent time (see, e.g., Ref.\cite{Burdik:2001hj}).
Higher-derivative Lagrangian of mixed-symmetry conformal fields was
obtained in Ref.\cite{Vasiliev:2009ck}.%
\footnote{ Discussion of equations and constraints for mixed
symmetry conformal fields may be found in
Ref.\cite{Shaynkman:2004vu}.}
We think that ordinary-derivative formulation of mixed-symmetry
conformal fields could be of some interest (see, e.g.,
Ref.\cite{Metsaev:2008ba}). In this respect, it would be interesting
to understand the phenomenon discovered for AdS mixed-symmetry
fields in Ref.\cite{Brink:2000ag} from the point of view of the
mixed-symmetry conformal fields.

\noindent \vbf) light-cone gauge formulation of the
ordinary-derivative conformal fields theory. In
Refs.\cite{Metsaev:2005ar,Metsaev:2007rn}, we developed general
methods for the building interaction vertices of light-cone gauge
fields. We note that these methods can straightforwardly be
generalized to the case of conformal fields. This will provides the
possibility for the studying interaction vertices for conformal
fields and currents constructed out of fields of the supersymmetric
Yang-Mills theory. We note also that action of Green-Schwarz
superstring in AdS/Ramond-Ramond background simplifies considerably
in the light-cone gauge \cite{Metsaev:2000yf,Metsaev:2000yu}.
Therefore we expect that from the stringy perspective of AdS/CFT
correspondence the light-cone approach to ordinary-derivative
conformal fields theory might be helpful for the study of various
aspects of the string/gauge theory duality.

\bigskip

\bigskip
{\bf Acknowledgments}. This work was supported by the INTAS project
03-51-6346, by the RFBR Grant No.05-02-17217, RFBR Grant for Leading
Scientific Schools, Grant No. 4401-2006-2 and Russian Science
Support Foundation.

\setcounter{section}{0}\setcounter{subsection}{0}
\appendix{ \large Counting of on-shell D.o.F for conformal spin-s field }\label{app-A}

In this Appendix, we analyze on-shell D.o.F for arbitrary spin-$s$
conformal field by using the higher-derivative approach in
Ref.\cite{Segal:2002gd}. Analysis of on-shell D.o.F for the
arbitrary spin-$s$ conformal field follows the procedure we used for
the spin-2 conformal field in Appendix C, in
Ref.\cite{Metsaev:2007fq}. Therefore to avoid the discussion of
unnecessary technical details we outline procedure of the derivation
and present the result.

In Ref.\cite{Segal:2002gd}, the higher-derivative Lagrangian of
spin-$s$ conformal field is constructed out of field
$\phi^{a_1\ldots a_s}$ which is totally symmetric traceless rank-$s$
tensor field of the Lorentz algebra $so(d-1,1)$. The Lagrangian is
invariant under gauge transformations described by gauge
transformation parameter which is totally symmetric traceless
rank-$(s-1)$ tensor field of the Lorentz algebra $so(d-1,1)$. We now
impose the light-cone gauge condition to fix the gauge symmetries
\be \label{man-15022012-06}
\phi^{+a_2\ldots a_s} = 0 \,.
\ee
Note that because the tensor field $\phi^{a_1\ldots a_s}$ is
traceless the number of gauge conditions we impose in
\rf{man-15022012-06} is equal to the number of gauge
transformations. Now using gauge condition \rf{man-15022012-06} in
equations of motion for higher-derivative Lagrangian in
Ref.\cite{Segal:2002gd}, we make sure that these equations of motion
amount to the following equations of motion and constraints:
\beq
\label{man-15022012-07} && \hspace{-1cm} \Box^{k_s + 1 - n}
\partial^{a_1} \ldots
\partial^{a_n} \phi^{a_1 \ldots a_n a_{n+1} \ldots a_s
} = 0 \,, \qquad \quad n = 0,1,\ldots , s \,, \qquad k_s\equiv s +
\frac{d-6}{2}\,,\qquad
\\
\label{man-15022012-08} &&  \hspace{-1cm} \phi^{aaa_3\ldots a_s} =
0\,,
\eeq
where, in \rf{man-15022012-08}, we recall that the field
$\phi^{a_1\ldots a_s}$ is traceless.

We now cast equations \rf{man-15022012-07} into ordinary-derivative
form. To this end we introduce the following fields:
\be
\label{man-15022012-10} \phi_{k'}^{a_1\ldots a_{s'}}\,, \hspace{1cm}
s'=0,1,\ldots,s-1,s; \qquad  k' \in  [k_{s'}]_2\,, \qquad k_{s'}
\equiv s'+ \frac{d-6}{2}\,.
\ee
Identifying the generic conformal field $\phi^{a_1\ldots a_s}$ in
\rf{man-15022012-07} with field $\phi_{-k_s}^{a_1\ldots a_s}$ in
\rf{man-15022012-10}, $\phi_{-k_{s}}^{a_1\ldots a_s} \equiv
\phi^{a_1\ldots a_s}$, we cast equations \rf{man-15022012-07} into
the following ordinary-derivative form:
\beq
\label{manold-10042012-04} && \Box \phi_{k'}^{a_1\ldots a_s} -
\phi_{k'+2}^{a_1\ldots a_s} = 0\,,
\hspace{3.2cm} k' \in [k_s]_2; \qquad
\\
&& \Box \phi_{k'}^{a_1\ldots a_{s-1}} - \phi_{k'+2}^{a_1\ldots
a_{s-1}} = 0\,,
\hspace{2.5cm} k' \in [k_s-1]_2;
\\
&& \Box \phi_{k'}^{a_1\ldots a_{s-2}} - \phi_{k'+2}^{a_1\ldots
a_{s-2}} = 0\,,
\hspace{2.5cm} k' \in [k_s-2];
\\
&& \ldots \quad \ldots \quad \ldots \quad  \ldots \quad \ldots
\hspace{3cm} \ldots \quad \ldots
\\
&& \ldots \quad \ldots \quad \ldots \quad  \ldots \quad \ldots
\hspace{3cm} \ldots \quad \ldots
\\
\label{man-15022012-15} && \Box \phi_{k'}^a - \phi_{k'+2}^a = 0\,,
\hspace{4cm} k' \in [k_s-s+1]_2;
\\
&& \Box \phi_{k'} - \phi_{k'+2} = 0\,,
\hspace{4cm} k' \in [k_s-s];
\\[20pt]
\label{man-15022012-17} && \partial^a \phi_{k'}^{aa_2\ldots a_s}  -
\phi_{k'+1}^{a_2\ldots a_s} = 0\,,
\hspace{3cm} k' \in [k_s]_2;
\\
&& \partial^a \phi_{k'}^{aa_2\ldots a_{s-1}}  -
\phi_{k'+1}^{a_2\ldots a_{s-1}} = 0\,,
\hspace{2.3cm} k' \in [k_s-1]_2;
\\
&& \ldots \quad \ldots \quad \ldots \quad  \ldots \quad \ldots
\hspace{3cm} \ldots \quad \ldots
\\
&& \ldots \quad \ldots \quad \ldots \quad  \ldots \quad \ldots
\hspace{3cm} \ldots \quad \ldots
\\
&& \partial^a \phi_{k'}^{a a_2} - \phi_{k'+1}^{a_2} = 0\,,
\hspace{3.6cm} k' \in [k_s-s+2]_2;
\\
\label{man-15022012-20} && \partial^a\phi_{k'}^a - \phi_{k'+1} =
0\,,
\hspace{3.8cm} k' \in [k_s-s+1]_2\,.\quad
\eeq
In \rf{manold-10042012-04}-\rf{man-15022012-20}, fields with $k'$
not appearing in \rf{phiset01} should be equated to zero. Note that
normalization of fields in this Appendix differs from the one in
section \ref{lagran}.

Using relations \rf{manold-10042012-04}-\rf{man-15022012-15}, we
note that gauge condition \rf{man-15022012-06} and
Eqs.\rf{manold-10042012-04}-\rf{man-15022012-15} lead to gauge
conditions for fields entering the ordinary-derivative formulation,
\beq
\label{man-15022012-21} && \phi_{k'}^{+a_2 \ldots a_{s'}} =0\,,
\hspace{2cm} s'=1,2,\ldots, s; \qquad k' \in  [k_{s'}]_2\,;
\\
\label{man-15022012-22} && \phi_{k'}^{aaa_3 \ldots a_{s'}} =0\,,
\hspace{2cm} s'=2,3,\ldots, s; \qquad k' \in  [k_{s'}]_2\,.
\eeq

Before proceeding we recall some terminology. In the light-cone
gauge frame, all fields $\phi_{k'}^{a_1\ldots a_{s'}}$ that do not
appear in \rf{man-15022012-21} are separated into the following two
groups of the $so(d-2)$ algebra fields:
\beq
\label{man-15022012-23} && \phi_{k'}^{i_1\ldots i_{s'}}\,,
\hspace{2cm} s'=0,1,\ldots, s; \qquad  k' \in  [k_{s'}]_2\,;
\\
\label{man-15022012-24} && \phi_{k',n}^{-\ldots - i_{n+1} \ldots
i_{s'}}\,, \hspace{1cm} n= 1,\ldots, s'\,,\qquad \ \ \
s'=0,1,\ldots,s; \qquad  k' \in  [k_{s'}]_2\,.\qquad
\eeq
Subscript $n$ in field $\phi_{k',n}^{-\ldots - i_{n+1} \ldots
i_{s'}}$ \rf{man-15022012-24} indicates that the field
$\phi_{k',n}^{-\ldots - i_{n+1} \ldots i_{s'}}$ involves $n$ indices
in `minus' light-cone direction. In the light-cone gauge frame,
fields \rf{man-15022012-23} are referred to as dynamical fields,
while fields \rf{man-15022012-24} are referred to as non-dynamical
fields.

Using light-cone gauge \rf{man-15022012-21} and constraints
\rf{man-15022012-17}-\rf{man-15022012-20}, we make sure that all
non-dynamical fields can be expressed in terms of the dynamical
fields. Also, by virtue of \rf{man-15022012-21},
\rf{man-15022012-22}, all dynamical fields with $s'>2$ turn out to
be traceless.

To summarize, we are left with on-shell D.o.F given by the fields
$\phi_{k'}^{i_1\ldots i_{s'}}$ \rf{man-15022012-23} which, for
$s'=0$, $s'=1$, and $s'\geq 2$, are the respective scalar, vector,
and rank-$s'$ traceless tensor fields of the $so(d-2)$ algebra.

\appendix{ \large Derivation of
Lagrangian, gauge transformations, and operator $R^a$
}\label{appspin2}

We begin with the study of restrictions imposed by gauge symmetries.
For arbitrary spin conformal field, general ordinary-derivative
Lagrangian and gauge transformations are given by
\beq
\label{20072011-08} \LL &  = & \frac{1}{2} \phibr E \phik\,,
\\
\label{20072011-09} E  & = & E_\smtwo + E_\smone + E_\smzero\,,
\\
\label{21072011-05}  && E_\smtwo \equiv \Box - \alpar \albpar +
\frac{1}{2}(\alpar)^2\bar\alpha^2 + \frac{1}{2} \alpha^2 (\albpar)^2
- \frac{1}{2}\alpha^2 \Box \bar\alpha^2 -\frac{1}{4}\alpha^2\alpar\,
\albpar \bar\alpha^2\,,\qquad
\\
\label{21072011-05a1} && E_\smone =  \alpar \eb_1 + \albpar e_1 +
\alpha^2 \albpar \eb_2 + \alpar \bar\alpha^2 e_2
\nonumber\\
&& \hspace{1.2cm} + \, \alpha^2 \alpar \bar\alpha^2 \eb_3 + \alpha^2
\albpar \bar\alpha^2 e_3 + \alpha^2 \alpar \eb_4 +
\bar\alpha\partial \bar\alpha^2 e_4\,,\qquad \ \ \
%
\\
&& E_\smzero  \equiv   m_1 + \alpha^2\bar\alpha^2m_2 + \mb_3
\alpha^2 + m_3 \bar\alpha^2\,,
\\
\label{21072011-01} \delta \phik & = &  (G_\smone+ G_\smzero)
\xik\,,\qquad G_\smone \equiv \alpar\,, \qquad G_\smzero \equiv  b_1
+ \alpha^2 b_2\,,
\\
\label{21072011-06} && e_n = \zeta \ewt_1 \bar\upsilon^\ominussm\,,
\qquad\qquad \ \
\eb_n =  \upsilon^\ominussm \ebwt_1 \bar\zeta\,,
\qquad
n=1,2,3\,,
\qquad
\nonumber\\
&& e_4 = \zeta^3 \ewt_4 \bar\upsilon^\oplussm \bar\upsilon^\ominussm
\bar\upsilon^\ominussm \,,
\qquad
\eb_4 = \upsilon^\oplussm \upsilon^\ominussm\upsilon^\ominussm
\ebwt_4 \bar\zeta^3\,,
\\
\label{21072011-07} && m_1 = \upsilon^\ominussm \mwt_1
\bar\upsilon^\ominussm \,,\quad
m_2 = \upsilon^\ominussm \mwt_2 \bar\upsilon^\ominussm \,,\quad
\nonumber\\
\label{21072011-08} && m_3 = \zeta^2 \mwt_3\bar\upsilon^\ominussm
\bar\upsilon^\ominussm\,, \quad
\mb_3 = \upsilon^\ominussm \upsilon^\ominussm \mbwt_3 \bar\zeta^2\,,
\\
\label{21072011-09} && b_1 = \zeta \bwt_1
\bar\upsilon^\ominussm\,,\qquad b_2 = \upsilon^\ominussm \bwt_2
\bar\zeta\,,
\\
\label{21072011-02} && \ewt_n=\ewt_n(N_\zeta,\Delta')\,,\quad
\ebwt_n=\ebwt_n(N_\zeta,\Delta')\,,\quad n=1,2,3,4\,,
\\
\label{21072011-03} && \mwt_n=\mwt_n(N_\zeta,\Delta')\,,\qquad
n=1,2,3\,,\qquad \mbwt_3=\mbwt_3(N_\zeta,\Delta')\,,\qquad
\\
\label{21072011-04} && \bwt_1 =  \bwt_1(N_\zeta,\Delta')\,, \qquad
\bwt_2 = \bwt_2(N_\zeta,\Delta')\,,
\\
\label{21072011-04x1} && \mwt_1^\dagger = \mwt_1\,, \quad
\mwt_2^\dagger = \mwt_2\,, \quad  \mwt_3^\dagger = \mbwt_3\,, \qquad
\ewt_n^\dagger = -\ebwt_n\,,\quad n=1,2,3,4\,.\qquad
\eeq
We note that:\\
\ibf) The Fronsdal operator $E_\smtwo$ \rf{21072011-05} is fixed by
requiring the $\phibr E_\smtwo\phik$ part of the Lagrangian to be
invariant under the standard gradient $\alpar\xik$ part of gauge
transformations \rf{21072011-01};\\
\iibf) Dependence of operators $E$ \rf{20072011-09}, $G_\smone$,
$G_\smzero$ \rf{21072011-01} on the oscillators in
\rf{21072011-06}-\rf{21072011-04} is fixed by requiring the
ket-vectors $E\phik$, $(G_\smone+G_\smzero)\xik$ to satisfy
constraints given in \rf{olmman-05022012-02},
\rf{olmman-05022012-04} and the dilatation symmetry. Also, the
ket-vector $G_\smzero \xik$ should respect
constraint \rf{man-15022012-04}.\\
\iiibf) Quantities $\ewt$, $\mwt$, $\bwt$ in
\rf{21072011-02}-\rf{21072011-04} depend on the operators $N_\zeta$
and $\Delta'$.

Thus all that is required is to find the dependence of the
quantities $\ewt$, $\mwt$, $\bwt$ on the operators $N_\zeta$ and
$\Delta'$. To this end we study restrictions imposed by the gauge
symmetries.

Variation of Lagrangian \rf{20072011-08} under gauge transformations
\rf{21072011-01} takes the form (up to total derivative)
\be
\delta \label{oldman-14072011-01} \LL  =  \phibr (\VV_\smtwo +
\VV_\smone + \VV_\smzero)\xik\,,
\ee
where $\VV_{(n)}$ stands for contribution involving $n$ derivatives.
Obviously, gauge invariance of the Lagrangian requires the equations
$\phibr\VV_{(n)}\xik = 0$, $n=2,1,0$, which we now analyze in turn.

Firstly, we consider the equation $\phibr\VV_\smtwo\xik = 0$. To
this end we compute $\VV_\smtwo$,
\beq
\VV_\smtwo
& =&  \Box X_{\smtwo,1}
+ \alpha\partial\bar\alpha\partial X_{\smtwo,2}
+ \alpha^2(\bar\alpha\partial)^2 X_{\smtwo,3}
+ \alpha^2\Box X_{\smtwo,4}
\nonumber\\
& + &  (\alpha\partial)^2 X_{\smtwo,5}
+ \alpha^2\alpha\partial\bar\alpha\partial X_{\smtwo,6}
+ (\bar\alpha\partial)^2 X_{\smtwo,7}
+ \alpha^2(\alpha\partial)^2 X_{\smtwo,8}\,,
\\
&& X_{\smtwo,1} \equiv b_1 +  e_1\,,
\\
&& X_{\smtwo,2} \equiv - b_1 +  e_1 + 2 e_2\,,
\\
&& X_{\smtwo,3} \equiv \frac{1}{2} b_1 +  2e_3\,,
\\
&& X_{\smtwo,4} \equiv -(2N_\alpha+d-2) b_2 +  \eb_2\,,
\\
&& X_{\smtwo,5} \equiv (2N_\alpha+d-2) b_2 +  \eb_1\,,
\\
&& X_{\smtwo,6} \equiv -\frac{1}{2}(2N_\alpha+d-2) b_2 + \eb_2 + 2
\eb_3\,,
\\
&& X_{\smtwo,7} \equiv 2e_4\,, \qquad X_{\smtwo,8} = \eb_4\,,
\eeq
and note that solution to the equation $ \phibr\VV_\smtwo \xik = 0$
is given by
\beq
\label{man-130220120-15} && e_1 = - b_1\,, \qquad e_2 =  b_1\,,
\qquad\quad e_3 = - \frac{1}{4} b_1\,,
\\
\label{man-130220120-18} && \eb_1 = -(2N_\alpha+d-2) b_2\,, \qquad \
\ \ \eb_2 = (2N_\alpha+d-2) b_2\,,
\\
\label{man-130220120-21} && \eb_3 = -\frac{1}{4}(2N_\alpha+d-2)
b_2\,, \qquad \ e_4 = 0\,,\qquad \eb_4 = 0\,.
\eeq
Note that relations \rf{man-130220120-15}-\rf{man-130220120-21}
should be considered on space of the ket-vector $\xik$. Also note
that relations \rf{man-130220120-15}-\rf{man-130220120-21} imply the
following representation for $E_\smone$ \rf{21072011-05a1}:
\be \label{E1sol01}
E_\smone  \equiv  e_1\bar\AA  + \eb_1 \AA \,,
\ee
where the operators $\AA$, $\bar\AA$ are given in
\rf{man-13022012-22}, \rf{man-13022012-23}.

Secondly, we consider the equation $\phibr\VV_\smone\xik=0$. Using
$E_\smone$ \rf{E1sol01}, we compute the gauge variation of
Lagrangian \rf{20072011-08} to obtain the following expression for
$\VV_\smone$ \rf{oldman-14072011-01}:
\beq \label{E1sol01a1}
\VV_\smone
& = &  \alpha\partial X_{\smone,1}
+ \alpha^2\bar\alpha\partial X_{\smone,2}
+ \alpha^2 \alpha\partial X_{\smone,3}
+ \bar\alpha\partial X_{\smone,4}\,,
\\
&& X_{\smone,1} \equiv m_1 + \eb_1 b_1 - 2 (2N_\alpha + d - 1 )e_1
b_2\,,
\\
&& X_{\smone,2} \equiv 2 m_2 - \eb_1 b_1 + \half ( 2N_\alpha + d +2
)e_1 b_2\,,
\\
&& X_{\smone,3} \equiv \mb_3 + \half (2N_\alpha + d -2 )\eb_1 b_2
\,,
\\
&& X_{\smone,4} \equiv 2m_3 + e_1 b_1 \,.
\eeq
Using \rf{man-130220120-15}-\rf{man-130220120-21}, we make sure that
the
equation $\phibr\VV_\smone \xik = 0$ amounts to the equations%
\beq
\label{man-13022012-24} m_1 &  = & \eb_1 e_1 - 2\frac{2s + d-3
-2N_\zeta}{2s + d-4 - 2N_\zeta} e_1 \eb_1\,,
\\
\label{man-13022012-25} m_2 &  =  &  - \half \eb_1 e_1 + \frac{1}{4}
\frac{2s + d -2N_\zeta}{2s + d-4 - 2N_\zeta} e_1 \eb_1\,,
\\
\label{man-13022012-26} m_3 & = & \frac{1}{2}e_1 e_1 \,, \qquad
\mb_3  =  \frac{1}{2} \eb_1 \eb_1 \,,
\eeq
which should be considered on space of the ket-vector $\xik$.

Finally, we consider the equation $\phibr\VV_\smzero\xik=0$.
Expression for $\VV_\smzero$ \rf{oldman-14072011-01} can be
presented as
\beq
\VV_\smzero & = & X_{\smzero,1} +\alpha^2 X_{\smzero,2}\,,
\\
&& X_{\smzero,1} \equiv  m_1 b_1 + 2(2N_\alpha+d)m_3 b_2\,,
\\
&& X_{\smzero,2}  \equiv m_1 b_2 + 2(2N_\alpha+d)m_2 b_2 + \mb_3
b_1\,.
\eeq
Using \rf{man-130220120-15}-\rf{man-130220120-21}, we find that the
equation $\phibr\VV_\smzero \xik = 0$ amounts to the equations
\beq
\label{man-13022012-28} &&  m_1 e_1 + 2
\frac{2s+d-2N_\zeta}{2s+d-2-2N_\zeta}m_3 \eb_1 = 0 \,,
\\
\label{man-13022012-29} &&  \frac{1}{2s+d-6-2N_\zeta} m_1 \eb_1 + 2
\frac{2s+d-4-2N_\zeta}{2s+d-6-2N_\zeta}m_2 \eb_1 + m_3 e_1 = 0 \,,
\eeq
which should be considered on space of the ket-vector $\xik$.

Equations \rf{man-130220120-15}-\rf{man-130220120-21},
\rf{man-13022012-24}-\rf{man-13022012-26}, and \rf{man-13022012-28},
\rf{man-13022012-29} provide the complete list of restrictions
imposed by gauge symmetries. These restrictions by themselves are
not sufficient to determine quantities $\ewt$, $\mwt$, $\bwt$
\rf{21072011-02}-\rf{21072011-04} uniquely. In order to determine
quantities $\ewt$, $\mwt$, $\bwt$ \rf{21072011-02}-\rf{21072011-04}
uniquely we proceed with analysis of restrictions imposed by the
conformal boost symmetries.

{\bf Analysis of restrictions imposed by conformal boost
symmetries}. Requiring Lagrangian \rf{20072011-08} to be invariant
(up to total derivative) under conformal boost transformations given
in \rf{man-14022012-01}, \rf{conalggenlis04}, we find the equations
\be
\label{app3-15082009-09}  R^{a \dagger}E + E R^a + E_\smzero^a +
E_\smone^a \approx 0\,,
\ee
where operators $E_\smone^a$, $E_\smzero^a$ are defined as
\beq
\label{man-15022012-03} E_\smone^a  & = & \Delta' ( 2
-\alpha^2\bar\alpha^2)\partial^a
\nonumber\\
& - &   (\Delta' +N_\alpha + \frac{d-6}{2}) \alpha^a \bar\AA +
 ( - \Delta' + N_\alpha + \frac{d-6}{2})  \AA \bar\alpha^a
\nonumber\\
& - & \frac{1}{4} \alpha^2  \alpar \alpha^a (\bar\alpha^2)^2 +
\frac{1}{4} (\alpha^2)^2 \bar\alpha^a \albpar\bar\alpha^2\,.
\\
\label{man-15022012-03n1} E_\smzero^a &= &
e_1(\Delta' - N_\alpha - \frac{d}{2})\bar\alpha^a + \alpha^a
(\Delta' + N_\alpha + \frac{d}{2})\eb_1
\nonumber\\
& - & \alpha^a e_1 \Bigl( \Delta' + N_\alpha + \frac{d-4}{2} \Bigr)
\bar\alpha^2 - \alpha^2  \Bigl( \Delta' - N_\alpha - \frac{d-4}{2}
\Bigr)\eb_1 \bar\alpha^a
\nonumber\\
& + &  \frac{1}{4} \alpha^2 e_1 \Bigl(\Delta' - N_\alpha -
\frac{d-8}{2} \Bigr) \bar\alpha^2 \bar\alpha^a + \frac{1}{4}
\alpha^2 \alpha^a \Bigl(\Delta' + N_\alpha + \frac{d-8}{2}\Bigr)
\eb_1 \bar\alpha^2  \ \
\nonumber\\
& + & \frac{1}{4} e_1 \alpha^2 \alpha^a (\bar\alpha^2)^2 -
\frac{1}{4} \eb_1 (\alpha^2)^2 \bar\alpha^a \bar\alpha^2\,.
\eeq
In \rf{app3-15082009-09} and below, we simplify our formulas as
follows. Let $A$ be some operator. We use the relation $A \approx 0$
in place of $\phibr A\phik=0$, where $\phik$ is defined in
\rf{phikdef01}. Note that, by virtue of double-tracelessness
constraint \rf{man-15022012-04}, the last terms in
\rf{man-15022012-03}, \rf{man-15022012-03n1}, being proportional to
$(\alpha^2)^2$ and $(\bar\alpha^2)^2$, are not relevant when we
consider Eqs.\rf{app3-15082009-09}.

Because the operator $R^a$ turns out to be degree-1 polynomial in
the derivative we represent the operator $R^a$ as power series in
the derivative,
\be \label{2kapkap00kpa01}
R^a =  R_\smzero^a+ R_\smone^a\,,
\ee
where $R_\smn^a$ stands for the contribution involving $n$
derivatives. Using power series expansions of operators $E$
\rf{20072011-09} and $R^a$ \rf{2kapkap00kpa01}, we see that
Eqs.\rf{app3-15082009-09} amount to the following equations:
\beq
\label{2E2kap1eq01}
&& E_\smtwo R_\smone^a + h.c. \approx 0 \,,
\\
\label{2E2kap1eq02}
&& E_\smtwo R_\smzero^a +E_\smone R_\smone^a + h.c. \approx 0 \,,
\\
\label{2E2kap1eq03}
&& ( E_\smone R_\smzero^a + E_\smzero R_\smone^a + h.c. ) +
E_\smone^a \approx 0 \,,
\\
\label{2E2kap1eq04}
&& ( E_\smzero R_\smzero^a + h.c. )  + E_\smzero^a \approx 0 \,.
\eeq
Most general operator $R^a$ that respects constraints
\rf{olmman-05022012-02}-\rf{man-15022012-04} and the dilatation
symmetry  is given by
\beq \label{manold-11112011-05}
R^a & = & r_\smzero^a + r_\smone^a + R_\smG^a \,,
\\
r_\smzero^a & = & \Bigl( r_{0,1} + r_{0,2}\alpha^2 \bar\alpha^2 +
\rb_{0,3}\alpha^2 \Pi^\smponetwo +
r_{0,4}\bar\alpha^2\Bigr)\bar\alpha^a
\nonumber\\
& + & \Vwt^a \Bigl(\rb_{0,1} + \rb_{0,2} \alpha^2 \bar\alpha^2 +
r_{0,3}\bar\alpha^2 + \rb_{0,4}\alpha^2\Pi^\smponetwo\Bigr)\,,
\\
r_\smone^a & = & \Bigl( r_{1,1} + r_{1,2}\bar\alpha^2 +
r_{1,3}\alpha^2 \Pi^\smponetwo + r_{1,4}\alpha^2\bar\alpha^2 \Bigr)
\partial^a
\nonumber\\
& + & \Bigl( r_{1,5}\Vwt^a +  r_{1,6}\Vwt^a \bar\alpha^2 +
r_{1,7}\alpha^2 V^a\Pi^\smponetwo + r_{1,8}\alpha^2 V^a \bar\alpha^2
\Bigr) \albpar
\nonumber\\
& + & \Bigl( r_{1,9} + r_{1,10}\bar\alpha^2 + r_{1,11}\alpha^2
\Pi^\smponetwo + r_{1,12}\alpha^2\bar\alpha^2 \Bigr) \bar\alpha^a
\albpar
\nonumber\\
& + & \alpha^2 V \Bigl( r_{1,13}\Pi^\smponetwo\bar\alpha^a +
r_{1,14} \bar\alpha^2\bar\alpha^a
+ r_{1,15}V^a \Pi^\smponetwo +   r_{1,16}V^a \bar\alpha^2\Bigr)\,,
\\
R_\smG^a  & = &  G r_\smG^a\,,\qquad G = G_\smone + G_\smzero\,,
\\
&& r_\smG^a = r_{\smG,1} \Vb_\perp^a + r_{\smG,2} V^a\Pi^\smponetwo
+ r_{\smG,3} V^a \bar\alpha^2 + r_{\smG,4}\bar\alpha^a
\bar\alpha^2\,,
\\
\label{man-14022012-02} &&  r_{0,n} =\zeta
\rwt_{0,n}\bar\upsilon^\oplussm\,, \qquad  \rb_{0,n} =
\upsilon^\oplussm \rbwt_{0,n}\bar\zeta\,,\qquad  n=1,2,3\,,
\nonumber\\
&& r_{0,4} =\zeta^3 \rwt_{0,4}\bar\upsilon^\oplussm
\bar\upsilon^\oplussm\bar\upsilon^\ominussm\,, \qquad \rb_{0,4} =
\upsilon^\oplussm \upsilon^\oplussm \upsilon^\ominussm
\rbwt_{0,4}\bar\zeta^3\,,
\\
&& r_{\smG,n} =  \upsilon^\oplussm  \rwt_{\smG,a}
\bar\upsilon^\oplussm \,, \qquad n= 1,3\,;
\nonumber\\
&& r_{\smG,2} =  \upsilon^\oplussm \upsilon^\oplussm \rwt_{\smG,2}
\bar\zeta^2 \,, \qquad r_{\smG,4} =  \zeta^2 \rwt_{\smG,4}
\bar\upsilon^\oplussm \bar\upsilon^\oplussm \,,
\\
&& r_{1,n} = \upsilon^\oplussm \rwt_{1,n}\bar\upsilon^\oplussm\,,
\hspace{2cm} n=1,4,5,8,11,14\,,
\nonumber\\
&& r_{1,n} = \zeta^2 \rwt_{1,n}\bar\upsilon^\oplussm
\bar\upsilon^\oplussm\,, \hspace{1.6cm} n=2,6,9,12\,,
\nonumber\\
&& r_{1,n} = \upsilon^\oplussm \upsilon^\oplussm
\rwt_{1,n}\bar\zeta^2\,, \hspace{1.6cm}  n=3,7,13,16\,,
\nonumber\\
&& r_{1,10} = \zeta^4
\rwt_{1,10}\bar\upsilon^\oplussm\bar\upsilon^\oplussm\bar\upsilon^\oplussm
\bar\upsilon^\ominussm\,,
\nonumber\\
&& r_{1,15} = \upsilon^\oplussm \upsilon^\oplussm \upsilon^\oplussm
\upsilon^\ominussm \rwt_{1,15}\bar\zeta^4\,.
\\
&& \rwt_{0,n} = \rwt_{0,n}(N_\zeta,\Delta')\,, \qquad
\rbwt_{0,n}=\rbwt_{0,n}(N_\zeta,\Delta')\,, \qquad n=1,2,3,4\,,
\\
&& \rwt_{1,n} = \rwt_{0,n}(N_\zeta,\Delta')\,,  \qquad n=1,2,\ldots
16\,,
\\
\label{man-14022012-20} && \rwt_{\smG,n} =
\rwt_{\smG,n}(N_\zeta,\Delta')\,, \qquad n=1,2,3,4\,,
\eeq
where $G_\smone$, $G_\smzero$ are given in \rf{21072011-01}. We see
that in order to fix the operator $R^a$ we have to find quantities
$\rwt$ in \rf{man-14022012-02}-\rf{man-14022012-20}. Note that
because operator $R^a$ \rf{manold-11112011-05} can be presented as
in \rf{2kapkap00kpa01} with
\be
R_\smzero^a = r_\smzero^a + G_\smzero r_\smG^a \,,
\qquad\quad R_\smone^a = r_\smone^a + G_\smone r_\smG^a\,,
\ee
and by virtue of the relation $\phibr E Gr_\smG^a\phik = 0$, all
terms proportional to $r_\smG^a$ cancel automatically in
Eqs.\rf{2E2kap1eq01}-\rf{2E2kap1eq04}. This implies that, in
\rf{2kapkap00kpa01}, we can replace $r_\smzero^a$ for $R_\smzero^a$
and $r_\smone^a$ for $R_\smone^a$ in analysis of
Eqs.\rf{2E2kap1eq01}-\rf{2E2kap1eq04}. We now analyze
Eqs.\rf{2E2kap1eq01}-\rf{2E2kap1eq04} in turn.

\noindent {\bf i)} Because analysis of Eqs.\rf{2E2kap1eq01} is
straightforward we just present our result. This is to say that
Eqs.\rf{2E2kap1eq01} lead to the following constraints:
\be r_{1,1}^\dagger = r_{1,1}\,,\quad r_{1,n} =  0\,,\qquad
n=2,3,\ldots, 16\,. \ee
{\bf ii)} Analysis of Eqs.\rf{2E2kap1eq02} is also straightforward.
Therefore we just present result of our analysis. Namely,
Eqs.\rf{2E2kap1eq02} amount to the following relations:
\be
\label{k01l1k11} \rb_{0,1} = [\eb_1,r_{1,1}]\,,
\qquad r_{0,1} = - \rb_{0,1}^\dagger\,,\qquad
r_{0,n} = 0\,, \qquad \rb_{0,n} = 0 \,, \qquad n =2,3,4\,.
\ee
To summarize, Eqs.\rf{2E2kap1eq01},\rf{2E2kap1eq02} and hermicity
conditions \rf{21072011-04x1} lead to the following expressions for
$r_\smzero^a$, $r_\smone^a$:
\beq
&& r_\smzero^a = r_{0,1} \bar\alpha^a + \rb_{0,1}\Vwt^a \,,
\qquad
r_\smone^a = r_{1,1}\partial^a \,, \qquad r_{1,1}^\dagger =
r_{1,1}\,,
\\
\label{man-14022012-26} && r_{0,1} = [r_{1,1},e_1]\,,
\qquad\rb_{0,1} = [\eb_1,r_{1,1}]\,.
\eeq
\noindent \iiibf) We proceed with the detailed analysis of
Eqs.\rf{2E2kap1eq03}. To this end we note the relations
\beq \label{man-15022012-01}
E_\smone r_\smzero^a + h.c. & = & [e_1, r_{0,1}] \bar\AA\bar\alpha^a
+ [\eb_1, \rb_{0,1}] \alpha^a \AA
\nonumber\\
& + & \Bigl( [\eb_1, r_{0,1}] +
\frac{2}{2N_\alpha+d-4}e_1\rb_{0,1}\Bigr) \AA \bar\alpha^a
\nonumber\\
& + & \Bigl( [e_1, \rb_{0,1}] - \frac{2}{2N_\alpha+d-4}r_{0,1}
\eb_1\Bigr) \alpha^a \bar\AA
\nonumber\\
&+& (e_1\rb_{0,1} - r_{0,1}\eb_1) ( 1 +
\frac{1}{4}\alpha^2\bar\alpha^2)\partial^a
\nonumber\\
& + & r_{0,1} e_1 \bar\alpha^2\partial^a -  \eb_1\rb_{0,1}
\alpha^2\partial^a\,,
\\
\label{man-15022012-02} E_\smzero r_\smone^a + h.c. & = & \Bigl(
[m_1,r_{1,1}] + [m_2,r_{1,1}]\alpha^2\bar\alpha^2 +
[m_3,r_{1,1}]\bar\alpha^2 +
[\mb_3,r_{1,1}]\alpha^2\Bigr)\partial^a\,,\qquad
\eeq
where in \rf{man-15022012-01} we drop terms proportional to
$(\alpha^2)^2$ and $(\bar\alpha^2)^2$. By virtue of
double-tracelessness constraint \rf{man-15022012-04}, such terms do
not contribute to Eqs.\rf{2E2kap1eq03}. Using
\rf{man-15022012-01},\rf{man-15022012-02}, and \rf{man-15022012-03}
we find that Eqs.\rf{2E2kap1eq03} amount to the following equations:
\beq
\label{09022012-01}  && [e_1, r_{0,1}]\Bigr|_{N_\zeta = 0,1,\ldots,
s-2} = 0\,,
\\
\label{09022012-04}  && [e_1, \rb_{0,1}] - \frac{2}{2k_s + 2 -
2N_\zeta}r_{0,1} \eb_1 - \Delta' + N_\zeta  - k_s\Bigr|_{N_\zeta =
0,1,\ldots, s-1} = 0 \,,
\\
\label{09022012-05}  && e_1\rb_{0,1} - r_{0,1}\eb_1 + [m_1,r_{1,1}]
+2\Delta'\Bigr|_{N_\zeta = 0,1,\ldots, s} = 0 \,,
\\
\label{09022012-06} && \frac{1}{4}(e_1\rb_{0,1} - r_{0,1}\eb_1) +
[m_2,r_{1,1}] -\Delta'\Bigr|_{N_\zeta = 0,1,\ldots, s-2} = 0 \,,
\\
\label{09022012-07} && r_{0,1} e_1  + [m_3,r_{1,1}]\Bigr|_{N_\zeta =
0,1,\ldots, s-2} = 0 \,,
\eeq
where $k_s$ appearing in \rf{09022012-04} is given in
\rf{olmman-05022012-01}. In \rf{09022012-01}-\rf{09022012-07}, we
use shortcut $A|_{N_\zeta=0,1,\ldots N}=0$ for the respective $N+1$
equations $A\zeta^n|\phi^{s-N}\rangle=0$, $n=0,1,\ldots,N$, where
the ket-vectors $|\phi^{s'}\rangle$ are defined in \rf{phikdef02}.
Note that, for the derivation of \rf{09022012-04}, we use the first
constraint in \rf{olmman-05022012-02}.

We proceed with analysis of Eqs.\rf{09022012-04}. For $N_\zeta=0$,
Eqs.\rf{09022012-04} lead to
\be \label{12022012-02}
\rbwt_{0,1}(0,\Delta')\ewt_1(0,\Delta') =-2\,.
\ee
Eq.\rf{12022012-02} and hermicity conditions \rf{21072011-04x1}
imply
\be \label{12022012-17}
\rwt_{0,1}(0,\Delta')\ebwt_1(0,\Delta') =-2\,.
\ee
Using Eqs.\rf{09022012-04} for $N_\zeta=1$ and relations
\rf{12022012-02}, \rf{12022012-17}, we find
\be \label{12022012-03}
\rbwt_{0,1}(1,\Delta')\ewt_1(1,\Delta') =-\frac{2k_s+1}{k_s}\,.
\ee
Eq.\rf{12022012-03} and hermicity conditions \rf{21072011-04x1}
imply
\be \label{12022012-05}
\rwt_{0,1}(1,\Delta')\ebwt_1(1,\Delta') =-\frac{2k_s+1}{k_s}\,.
\ee
Repeating these steps for $N_\zeta=2,3,\ldots, s-1$, we get
\be \label{12022012-06}
\rbwt_{0,1}\ewt_1 =-2\frac{2k_s+2-N_\zeta}{2k_s+2-2N_\zeta}\,.
\ee

We now proceed with analysis of Eq.\rf{09022012-01}. From
\rf{12022012-02}, we see that $\ewt_1(0,\Delta)\ne 0$. Using field
redefinitions we can set
\be\label{12022012-11}
\ewt_1(0,\Delta) = 1 \,.
\ee
Hermicity conditions \rf{21072011-04x1} and relations
\rf{12022012-02}, \rf{12022012-11} imply
\be \label{12022012-12}
\ebwt_1(0,\Delta) = -1\,, \qquad \rwt_{0,1}(0,\Delta) = 2\,,\qquad
\rbwt_{0,1}(0,\Delta) = - 2\,.
\ee
Using Eq.\rf{09022012-01} for $N_\zeta=0$ and relations
\rf{12022012-12}, we find

\be  \label{12022012-14}
\rwt_{0,1}(1,\Delta') = 2 \ewt_1(1,\Delta')\,, \qquad
\rbwt_{0,1}(1,\Delta') = 2 \ebwt_1(1,\Delta')\,.
\ee
Using \rf{12022012-14}, \rf{12022012-03}, we find
\be \label{12022012-15}
\ewt_1(1,\Delta')\ebwt_1(1,\Delta') = - \frac{2k_s+1}{2k_s}\,.
\ee
Using field redefinitions, the phase of $\ewt_1(1,\Delta')$ can be
set equal to zero. Relation \rf{12022012-15} and hermicity
conditions \rf{21072011-04x1} imply then
\be \label{12022012-16}
\ewt_1(1,\Delta') = \Bigl(\frac{2k_s+1}{2k_s}\Bigr)^{1/2}\,, \qquad
\ebwt_1(1,\Delta') = - \Bigl(\frac{2k_s+1}{2k_s}\Bigr)^{1/2}\,.
\ee
Repeating the above-described procedure for $N_\zeta =1,2,\ldots,
s-2$, we get
\be \label{man-14022012-21}
\ewt_1 = \Bigl(\frac{2k_s + 2 - N_\zeta}{2k_s + 2 -
2N_\zeta}\Bigr)^{1/2}\,, \qquad \ebwt_1 = - \ewt_1\,,
\qquad \rwt_{0,1} = 2 \ewt_1 \,,  \qquad \rbwt_{0,1} = - 2 \ewt_1
\,.
\ee
Using \rf{man-14022012-21} and
\rf{man-13022012-24}-\rf{man-13022012-26},
\rf{man-13022012-28},\rf{man-13022012-29}, allows us to determine
the operators $m_1$, $m_2$, $m_3$, $\mb_3$. Our result for these
operators is given in \rf{man-14022012-22}.

Using above-obtained expression for the operators $e_1$, $m_1$, we
make sure that solution of Eq.\rf{09022012-05} is given by
$\rwt_{1,1} =-2$.

Using above-obtained expression for operators $e_1$, $r_{0,1}$
$r_{1,1}$, $m_2$, $m_3$, we make sure that Eqs.\rf{09022012-06},
\rf{09022012-07} are automatically satisfied.

To summarize, we finished our study of Eqs.\rf{2E2kap1eq03}.

\noindent \ivbf) We now consider Eqs.\rf{2E2kap1eq04}. Using results
above obtained, we make sure that Eqs.\rf{2E2kap1eq04} are
automatically satisfied.

Finally, we check that Eqs.\rf{man-14022012-26} are also satisfied.

\appendix{ \large Derivation of relations
\rf{oldman-17032012-01} and \rf{16032012-05}}

{\bf Derivation of \rf{oldman-17032012-01}}. We note that, in
general, ket-vector $\Cb\phik$ depends on the oscillators
$\upsilon^\oplussm$, $\upsilon^\ominussm$, $\alpha^a$, $\zeta$.
Obviously, the ket-vector $\Cb\phik$, which satisfies
Eq.\rf{10032012-17}, does not depend on the oscillator
$\upsilon^\ominussm$. In other words, the ket-vector $\Cb\phik$,
which satisfies  Eq.\rf{10032012-17}, can be represented as
\be \label{11032012-06}
\Cb\phik = |\Cb(\upsilon^\oplussm)\rangle\,,
\ee
where $|\Cb(\upsilon^\oplussm)\rangle$ stands for ket-vector which
depends on the oscillator $\upsilon^\oplussm$, $\alpha^a$, $\zeta$.
Thus our problem is to find ket-vectors $\phik$ and
$|\Cb(\upsilon^\oplussm)\rangle$ where the ket-vector $\phik$ should
satisfy Eq.\rf{11032012-01}, while the ket-vector
$|\Cb(\upsilon^\oplussm)\rangle$ is defined by \rf{11032012-06}.

Taking relation \rf{11032012-06} into account, we note that
Eq.\rf{11032012-01} can be represented as nonhomogeneous
differential equation for the ket-vector $\phik$,
\be \label{11032012-07}
\Bigl(\bar\upsilon^\oplussm \Box - (N_{\upsilon^\ominussm}+1)
\bar\upsilon^\ominussm \Bigr)\phik - \alpha^2
\frac{1}{2N_\alpha+d-2} \ewt_1 \bar\zeta
|\Cb(\upsilon^\oplussm)\rangle = 0 \,,
\ee
where $\ewt_1$ is given in \rf{ewtdef01}. General solution of
Eq.\rf{11032012-07} can be presented as
\be \label{11032012-08}
\phik = e^{X\Box} |\phi(\upsilon^\ominussm)\rangle +
|\tau(\upsilon^\oplussm)\rangle\,,
\qquad
|\tau(\upsilon^\oplussm)\rangle \equiv  \alpha^2
\frac{1}{2N_\alpha+d-2} \ewt_1 |\Cwt(\upsilon^\oplussm)\rangle\,,
\ee
where operator $X$ is given in \rf{11032012-05}, while the
ket-vector $ |\Cwt(\upsilon^\oplussm)\rangle$ should satisfy the
equation
\be \label{11032012-09}
\bar\upsilon^\ominussm |\Cwt(\upsilon^\oplussm)\rangle = -\bar\zeta
| \Cb(\upsilon^\oplussm)\rangle\,.
\ee
In \rf{11032012-08}, $|\phi(\upsilon^\ominussm)\rangle$ stands for
ket-vector that depends on the oscillators $\upsilon^\ominussm$,
$\alpha^a$, $\zeta$. We note that
\\
\ibf) In \rf{11032012-08}, the ket-vector
$e^{X\Box}|\phi(\upsilon^\ominussm)\rangle$ is a general solution of
homogeneous equation obtained from Eq.\rf{11032012-07} by equating
$|\Cb(\upsilon^\oplussm)\rangle=0$;
\\
\iibf) The ket-vector $|\tau(\upsilon^\oplussm)\rangle$ is a
particular solution of nonhomogeneous equation \rf{11032012-07}
provided the ket-vector $|\Cwt(\upsilon^\oplussm)\rangle$ satisfies
Eq.\rf{11032012-09}. In order to fix the particular solution
uniquely we should supplement Eq.\rf{11032012-09} by initial
condition. We assume the following initial condition for the
ket-vector $|\Cwt(\upsilon^\oplussm)\rangle$:
\be \label{oldman-30032012-01}
|\Cwt(\upsilon^\oplussm)\rangle\bigr|_{\upsilon^\oplussm=0} = 0 \,.
\ee
Note that \rf{11032012-09}, \rf{oldman-30032012-01} imply that the
ket-vector $|\Cwt(\upsilon^\oplussm)\rangle$ depends on the
oscillators $\upsilon^\oplussm$, $\alpha^a$, $\zeta$. Considering
relation \rf{11032012-08} for $\upsilon^\oplussm=0$ and taking into
account \rf{oldman-30032012-01}, \rf{phikdef01}, we find the
relations
\beq
&& |\phi(\upsilon^\ominussm)\rangle = (\upsilon^\ominussm)^\kwh
\Phik\,,\hspace{2cm} \hbox{for } \ d \geq 6\,,
\nonumber\\[-10pt]
\label{manold-02042012-01} &&
\\[-10pt]
&& |\phi(\upsilon^\ominussm)\rangle = (\upsilon^\ominussm)^\kwh
|\Phi'\rangle\,, \hspace{2cm} \hbox{for } \ d =4\,,
\nonumber
\eeq
where ket-vectors $\Phik$, $|\Phi'\rangle$ are defined in
\rf{oldman-19032012-01}, \rf{oldman-19032012-02}, while the operator
$\kwh$ is given in \rf{manold-31102011-05}.

We now use representation for the ket-vector $\phik$ given in
\rf{11032012-08} and find the following representation for the
ket-vector $\Cb\phik$:
\beq
\label{11032012-11} \Cb\phik & = & e^{X\Box} \Cb'
|\phi(\upsilon^\ominussm)\rangle - V \ewt_1
|\Cwt(\upsilon^\oplussm)\rangle  - \frac{2N_\alpha+d}{2N_\alpha+d-2}
N_\zeta \ewt_1^\smminone \ewt_1^\smminone
|\Cb(\upsilon^\oplussm)\rangle
\nonumber\\
& - &  \ewt_1 \bar\zeta \Pi^\smponetwo \Wk\,,
\\
\label{manold-02042012-02} && \Cb' \equiv \Vb_\perp +
\upsilon^\ominussm \ewt_1 \bar\zeta \Pi^\smponetwo +
\frac{1}{2(N_{\upsilon^\ominussm}+1)} \zeta \ewt_1
\bar\upsilon^\oplussm \bar\alpha^2 \Box\,,
\\
\label{manold-02042012-02n1} && \Wk \equiv
\frac{\Box^{N_{\upsilon^\oplussm}}}{\Gamma(N_{\upsilon^\oplussm}+1)}
\upsilon^\oplussm |\phi(\upsilon^\oplussm)\rangle\,,
\eeq
where $\Gamma$ in \rf{manold-02042012-02n1} stands for the Euler
gamma function. Plugging \rf{11032012-11} into \rf{11032012-06}, we
find the equation
\beq \label{11032012-12n1}
e^{X\Box} \Cb' |\phi(\upsilon^\ominussm)\rangle - V \ewt_1
|\Cwt(\upsilon^\oplussm)\rangle  - (N_\zeta +1)\ewt_1\ewt_1
|\Cb(\upsilon^\oplussm)\rangle - \ewt_1 \bar\zeta \Pi^\smponetwo \Wk
= 0 \,.
\eeq
Multiplying Eq.\rf{11032012-12n1} by $e^{-X\Box}$ and taking into
account the relations
\be
X |\Cwt(\upsilon^\oplussm)\rangle =0 \,, \qquad
X|\Cb(\upsilon^\oplussm)\rangle = 0 \,, \qquad X \Wk =0 \,,
\ee
we see that Eq.\rf{11032012-12n1} amounts to the following equation:
\beq \label{11032012-12}
\Cb' |\phi(\upsilon^\ominussm)\rangle - V \ewt_1
|\Cwt(\upsilon^\oplussm)\rangle  - (N_\zeta +1)\ewt_1\ewt_1
|\Cb(\upsilon^\oplussm)\rangle - \ewt_1 \bar\zeta \Pi^\smponetwo \Wk
= 0 \,.
\eeq
The procedure for solving Eq.\rf{11032012-12} is slightly different
for the cases $d\geq 6$ and $d=4$. We now consider these two cases
in turn.

\noindent {\bf Solving Eq.\rf{11032012-12} for  $d\geq 6$}. We note
that, for $d\geq 6$, the ket-vector
$\Cb'|\phi(\upsilon^\ominussm)\rangle$ in \rf{11032012-12} is
independent of $\upsilon^\oplussm$ and polynomial in
$\upsilon^\ominussm$, while the ket-vectors
$|\Cwt(\upsilon^\oplussm)\rangle$, $|\Cb(\upsilon^\oplussm)\rangle$,
and $\Wk$ in \rf{11032012-12} are polynomial in $\upsilon^\oplussm$
and independent of $\upsilon^\ominussm$. Also we note that
$\Cb'|\phi(\upsilon^\ominussm)\rangle|_{\upsilon^\ominussm=0}=0$.
This implies that Eq.\rf{11032012-12} amounts to the following two
equations:
\beq
\label{11032012-14} && \Cb' |\phi(\upsilon^\ominussm)\rangle = 0\,,
\\
\label{11032012-15} && V |\Cwt(\upsilon^\oplussm)\rangle  + (N_\zeta
+1) \ewt_1 |\Cb(\upsilon^\oplussm)\rangle + \bar\zeta \Pi^\smponetwo
\Wk  = 0 \,.
\eeq
Below, we demonstrate that Eq.\rf{11032012-14} amounts to constraint
\rf{16032012-05}, while Eq.\rf{11032012-15} allows us to determine
$|\Cwt(\upsilon^\oplussm)\rangle$. We now analyze
Eq.\rf{11032012-15}.

From Eq.\rf{11032012-15}, we find
\beq \label{11032012-16}
-|\Cb(\upsilon^\oplussm)\rangle  = \frac{1}{(N_\zeta +1)\ewt_1}
\Bigl( V |\Cwt(\upsilon^\oplussm)\rangle +  \bar\zeta \Pi^\smponetwo
\Wk \Bigr) \,.
\eeq
Plugging \rf{11032012-16} into \rf{11032012-09}, we find the
equation
\beq \label{11032012-17}
\bar\upsilon^\ominussm |\Cwt(\upsilon^\oplussm)\rangle =
\frac{1}{(N_\zeta +2)\ewt_1^\smone}\Bigl( V
\bar\zeta|\Cwt(\upsilon^\oplussm)\rangle + \bar\zeta^2
\Pi^\smponetwo \Wk \Bigr) \,.
\eeq
Solution to Eq.\rf{11032012-17}  with initial condition
\rf{oldman-30032012-01} is given by
\be
\label{11032012-03} |\Cwt(\upsilon^\oplussm)\rangle = e^Y
\upsilon^\oplussm \frac{1}{N_{\upsilon^\oplussm}+1} e^{- Y}
\frac{1}{(N_\zeta+2)\ewt_1^\smone} \bar\zeta^2 \Pi^\smponetwo \Wk\,,
\ee
where $Y$ is given in \rf{11032012-05}. Plugging \rf{11032012-03}
into \rf{11032012-08}, we get $\phik$ given in
\rf{oldman-17032012-01}.

\noindent {\bf Solving Eq.\rf{11032012-12} for  $d=4$}. We note
that, for $d=4$, the ket-vector $|\phi(\upsilon^\ominussm)\rangle$
\rf{manold-02042012-01} can be represented as
\beq
\label{15032012-01} && |\phi(\upsilon^\ominussm)\rangle =
|\phi_{\geq 3}\rangle + |\phi_2\rangle + |\phi_1\rangle\,,
\\
\label{manold-02042012-06}  && |\phi_{\geq 3}\rangle  \equiv
(\upsilon^\ominussm)^\kwh \sum_{s'=3}^s |\Phi_{s'}\rangle\,, \qquad
|\phi_2\rangle  \equiv (\upsilon^\ominussm)^\kwh |\Phi_2\rangle \,,
\qquad |\phi_1\rangle \equiv (\upsilon^\ominussm)^\kwh
|\Phi_1\rangle \,, \qquad
\\
&& |\Phi_{s'}\rangle  \equiv \frac{\zeta^{s-s'}}{s'!\sqrt{(s-s')!}}
\alpha^{a_1} \ldots \alpha^{a_{s'}} \, \phi_{-k_{s'}}^{a_1\ldots
a_{s'}}|0\rangle\,.
\eeq
Using \rf{15032012-01}, we represent the ket-vector
$\Cb'|\phi(\upsilon^\ominussm)\rangle$ as
\beq
\label{15032012-02} \Cb'|\phi(\upsilon^\ominussm)\rangle & = & \Cb'
|\phi_{\geq 3} \rangle + (\Vb_\perp + \upsilon^\ominussm \ewt_1
\bar\zeta \Pi^\smponetwo )|\phi_2\rangle + \upsilon^\ominussm \ewt_1
\bar\zeta \Pi^\smponetwo |\phi_1\rangle
\nonumber\\
& + & \Vb_\perp |\phi_1\rangle + \half\zeta \ewt_1
\bar\upsilon^\oplussm \bar\alpha^2 \Box |\phi_2\rangle\,.
\eeq
We now introduce scalar field $\phi_1$ by the relation
\be \label{15032012-03}
\Vb_\perp |\phi_1\rangle  + \half\zeta \ewt_1 \bar\upsilon^\oplussm
\bar\alpha^2 \Box |\phi_2\rangle + \ewt_1 \bar\zeta |\phi_0\rangle =
0\,, \qquad |\phi_0\rangle  \equiv  \frac{\zeta^s}{\sqrt{s!}} \phi_1
|0\rangle
\ee
and note that ket-vector $|\phi_0\rangle$ does not depend on the
oscillators $\upsilon^\oplussm$, $\upsilon^\ominussm$, $\alpha^a$.
Using \rf{15032012-02}, we note that Eq.\rf{11032012-12} amounts to
the following equations:
\beq
\label{15032012-04} && \Cb' |\phi_{\geq 3} \rangle + (\Vb_\perp +
\upsilon^\ominussm \ewt_1 \bar\zeta \Pi^\smponetwo )|\phi_2\rangle +
\upsilon^\ominussm \ewt_1 \bar\zeta \Pi^\smponetwo |\phi_1\rangle =
0\,,
\\
\label{15032012-05} && V |\Cwt(\upsilon^\oplussm)\rangle  + (N_\zeta
+1) \ewt_1 |\Cb(\upsilon^\oplussm)\rangle + \bar\zeta \Pi^\smponetwo
(\Wk + |\phi_0\rangle)  = 0 \,.
\eeq
The fact that Eq.\rf{11032012-12} amounts to Eqs.\rf{15032012-04},
\rf{15032012-05} can easily be understood by noticing that  left
hand side of Eq.\rf{15032012-04} is independent of
$\upsilon^\oplussm$ and polynomial in $\upsilon^\ominussm$, while
left hand side of Eq.\rf{15032012-05} is polynomial in
$\upsilon^\oplussm$ and independent of $\upsilon^\ominussm$. Note
also also left hand side of Eq.\rf{15032012-04} is automatically
equal to zero when $\upsilon^\ominussm=0$. Below, we demonstrate
that Eqs.\rf{15032012-03},\rf{15032012-04} amount to constraint
\rf{16032012-05}, while Eq.\rf{15032012-05} allows us to determine
$|\Cwt(\upsilon^\oplussm)\rangle$. We now analyze
Eq.\rf{15032012-05}.

From Eq.\rf{15032012-05}, we find
\be \label{15032012-06}
-|\Cb(\upsilon^\oplussm)\rangle  = \frac{1}{(N_\zeta +1)\ewt_1}
\Bigl( V |\Cwt(\upsilon^\oplussm)\rangle +  \bar\zeta \Pi^\smponetwo
(\Wk  + |\phi_0\rangle) \Bigr) \,.
\ee
Plugging \rf{15032012-06} into \rf{11032012-09}, we find the
equation
\beq \label{15032012-07}
\bar\upsilon^\ominussm |\Cwt(\upsilon^\oplussm)\rangle =
\frac{1}{(N_\zeta +2)\ewt_1^\smone}\Bigl( V
\bar\zeta|\Cwt(\upsilon^\oplussm)\rangle + \bar\zeta^2
\Pi^\smponetwo (\Wk  + |\phi_0\rangle) \Bigr) \,.
\eeq
Solution to Eq.\rf{15032012-07} with initial condition
\rf{oldman-30032012-01} is given by
\be
\label{11032012-03n1} |\Cwt(\upsilon^\oplussm)\rangle = e^Y
\upsilon^\oplussm \frac{1}{N_{\upsilon^\oplussm}+1} e^{- Y}
\frac{1}{(N_\zeta+2)\ewt_1^\smone} \bar\zeta^2 \Pi^\smponetwo (\Wk
+|\phi_0\rangle)\,.
\ee
Plugging \rf{11032012-03n1} into \rf{11032012-08}, we get $\phik$
given in \rf{oldman-17032012-01}.

{\bf Derivation of \rf{16032012-05}}. For $d\geq 6$, derivation of
\rf{16032012-05} is straightforward. Namely, using
\rf{manold-02042012-01}, \rf{manold-02042012-02}, we find the
relation
\be \label{manold-02042012-03}
\Cb' |\phi(\upsilon^\ominussm)\rangle = (\upsilon^\ominussm)^{\kwh
+1} \Cb_\sh \Phik\,.
\ee
From \rf{manold-02042012-03}, we see that constraint
\rf{11032012-14} amounts to \rf{16032012-05}.

For $d = 4$, the derivation of \rf{16032012-05} is as follows. Using
\rf{manold-02042012-06}, we note that constraints \rf{15032012-04},
\rf{15032012-03} amount to the following respective constraints:
\beq
\label{manold-02042012-04} && \Cb_\sh |\Phi_{\geq 3} \rangle +
(\Vb_\perp + \ewt_1 \bar\zeta \Pi^\smponetwo )|\Phi_2\rangle +
\ewt_1 \bar\zeta \Pi^\smponetwo |\Phi_1\rangle = 0\,,
\\
\label{manold-02042012-05} && \Vb_\perp |\Phi_1\rangle  + \half\zeta
\ewt_1 \bar\alpha^2 \Box |\Phi_2\rangle + \ewt_1 \bar\zeta
|\Phi_0\rangle = 0\,, \qquad |\Phi_0\rangle\equiv |\phi_0\rangle\,.
\eeq
Taking into account the relation
\be
\Phik = |\Phi_{\geq 3}\rangle + |\Phi_2\rangle + |\Phi_1\rangle +
|\Phi_0\rangle\,,
\ee
we see that constraints \rf{manold-02042012-04},
\rf{manold-02042012-05} amount to constraint \rf{16032012-05}.

\appendix{ \large Derivation of Lagrangian \rf{oldman-19032012-04}
and gauge transformations \rf{oldman-19032012-07}}

{\bf Derivation of Lagrangian \rf{oldman-19032012-04}}. We now show
that plugging solution \rf{oldman-17032012-01} into
ordinary-derivative Lagrangian \rf{manold-17032012-01} leads to
higher-derivative Lagrangian \rf{oldman-19032012-04}. To this end we
represent $\phik$ \rf{oldman-17032012-01} as in \rf{11032012-08},
where $|\Cwt(\upsilon^\oplussm)\rangle$ for $d\geq 6$ and $d=4$ is
given in \rf{11032012-03} and \rf{11032012-03n1} respectively, while
$|\phi(\upsilon^\ominussm)\rangle$ is defined in
\rf{manold-02042012-01}. Using $\phik$ \rf{11032012-08}, we note the
helpful formula
\beq \label{manold-01042012-02}
&& \langle \phi |\Box - \upsilon^\ominussm
\bar\upsilon^\ominussm|\phi\rangle = \langle
\phi(\upsilon^\ominussm)|
\frac{\Box^{N_{\upsilon^\oplussm}+1}}{\Gamma(N_{\upsilon^\oplussm}+1)}
|\phi(\upsilon^\oplussm)\rangle
 + 2\langle \tau(\upsilon^\oplussm)|
\Box^{N_{\upsilon^\oplussm}+1} |\phi(\upsilon^\oplussm)\rangle
\nonumber\\
&& \hspace{3.1cm} + \, \langle \tau(\upsilon^\oplussm)| \Box -
\upsilon^\ominussm \bar\upsilon^\ominussm
|\tau(\upsilon^\oplussm)\rangle\,,
\eeq
where $|\phi(\upsilon^\oplussm)\rangle$ is obtained from
\rf{manold-02042012-01} by the substitution $\upsilon^\ominussm
\rightarrow \upsilon^\oplussm $. Using \rf{11032012-06} and
\rf{manold-01042012-02}, we see that Lagrangian
\rf{manold-17032012-01} can be represented as
\beq \label{15032012-08}
\LL & = &  \half \langle \phi(\upsilon^\ominussm)|
\mubf\frac{\Box^{N_{\upsilon^\oplussm}+1}}{\Gamma(N_{\upsilon^\oplussm}+1)}
|\phi(\upsilon^\oplussm)\rangle  + \langle
\tau(\upsilon^\oplussm)|\mubf \Box^{N_{\upsilon^\oplussm}+1}
|\phi(\upsilon^\oplussm)\rangle
\nonumber\\
& + & \half \langle \tau(\upsilon^\oplussm)| \mubf(\Box -
\upsilon^\ominussm \bar\upsilon^\ominussm)
|\tau(\upsilon^\oplussm)\rangle +  \half \langle
\Cb(\upsilon^\oplussm)|\Cb(\upsilon^\oplussm)\rangle\,,
\eeq
where $|\Cb(\upsilon^\oplussm)\rangle$ for $d\geq 6$ and $d=4$ is
given in \rf{11032012-16} and \rf{15032012-06} respectively. For
$d\geq 6$, contributions of
$\langle\tau(\upsilon^\oplussm)|\phi(\upsilon^\oplussm)\rangle$-,
$\langle\tau(\upsilon^\oplussm)|\tau(\upsilon^\oplussm)\rangle$-,
and $\langle\Cb(\upsilon^\oplussm)|\Cb(\upsilon^\oplussm)\rangle$
-terms in \rf{15032012-08} are equal to zero. For $d =4$,
contribution of
$\langle\tau(\upsilon^\oplussm)|\phi(\upsilon^\oplussm)\rangle$-term
in \rf{15032012-08} is equal to zero. This is to say that Lagrangian
\rf{15032012-08} takes the form
\beq
\label{manold-15042012-01} \LL & = & \half \langle
\phi(\upsilon^\ominussm)|\mubf
\frac{\Box^{N_{\upsilon^\oplussm}+1}}{\Gamma(N_{\upsilon^\oplussm}+1)}
|\phi(\upsilon^\oplussm)\rangle\,, \hspace{4.1cm} \hbox{for } \ d
\geq 6,
\\
\label{manold-15042012-02} \LL & = & \half \langle
\phi(\upsilon^\ominussm)|\mubf
\frac{\Box^{N_{\upsilon^\oplussm}+1}}{\Gamma(N_{\upsilon^\oplussm}+1)}
|\phi(\upsilon^\oplussm)\rangle
\nonumber\\
& + & \half \langle \tau(\upsilon^\oplussm)|\mubf( \Box -
\upsilon^\ominussm \bar\upsilon^\ominussm)
|\tau(\upsilon^\oplussm)\rangle  + \half \langle
\Cb(\upsilon^\oplussm)|\Cb(\upsilon^\oplussm)\rangle\,, \qquad
\hbox{for } \ d = 4\,.
\eeq
Using the relations
\beq
&& \langle \phi(\upsilon^\ominussm)|\mubf
\frac{\Box^{N_{\upsilon^\oplussm}+1}}{\Gamma(N_{\upsilon^\oplussm}+1)}
|\phi(\upsilon^\oplussm)\rangle = \langle
\Phi|\mubf\Box^{\kwh+1}|\Phi\rangle\,, \hspace{3cm} \hbox{for } \ d
\geq 6\,,
\\
&& \langle \phi(\upsilon^\ominussm)|\mubf
\frac{\Box^{N_{\upsilon^\oplussm}+1}}{\Gamma(N_{\upsilon^\oplussm}+1)}
|\phi(\upsilon^\oplussm)\rangle = \langle
\Phi'|\mubf\Box^{\kwh+1}|\Phi'\rangle\,, \hspace{2.8cm} \hbox{for }
\ d = 4\,,
\\
\label{15032012-09}  && \langle
\Cb(\upsilon^\oplussm)|\Cb(\upsilon^\oplussm)\rangle + \langle
\tau(\upsilon^\oplussm)|\mubf(\Box -
\upsilon^\ominussm\bar\upsilon^\ominussm )|
\tau(\upsilon^\oplussm)\rangle = \langle \Phi_0 |\Phi_0\rangle \,,
\qquad \hbox{for } \ d = 4\,,\qquad
\eeq
we see that Lagrangian \rf{manold-15042012-01},
\rf{manold-15042012-02} takes the form given in
\rf{oldman-19032012-04}. We note that, for the derivation of
relation \rf{15032012-09}, we use \rf{15032012-06} and the relations
\beq
&& \langle \Cb(\upsilon^\oplussm)|\Cb(\upsilon^\oplussm)\rangle  =
\langle \Phi_0 |\frac{2}{s(s+1)}  |\Phi_0\rangle \,,
\\
&& \langle \tau(\upsilon^\oplussm)|\mubf(\Box -
\upsilon^\ominussm\bar\upsilon^\ominussm )|
\tau(\upsilon^\oplussm)\rangle  = \langle \Phi_0
|\frac{(s-1)(s+2)}{s(s+1)}  |\Phi_0\rangle \,.
\eeq

\noindent {\bf  Derivation of gauge transformation
\rf{oldman-19032012-07}}. By definition, the left-over gauge
symmetries of de Donder-Stueckelberg gauge \rf{10032012-17} are
governed by gauge transformation parameter which satisfies the
equation $\bar\upsilon^\oplussm\Cb\delta \phik =0$, where $\delta
\phik$ stands for gauge transformations \rf{oldman-05022012-11}.
Using the relation
\be \label{manold-03042012-01}
\Cb\delta \phik = (\Box - m^2) \xik \,,
\ee
we see that gauge transformation parameter $\xik$ corresponding to
the left-over gauge symmetries  of de Donder-Stueckelberg gauge
\rf{10032012-17} should satisfy Eq.\rf{manold-18032012-07}. It is
easy to check that general solution to Eq.\rf{manold-18032012-07} is
given by \rf{manold-18032012-08}.

Now our purpose is to demonstrate that gauge transformations
\rf{oldman-05022012-11}, where $\phik$ and $\xik$ take the form as
in \rf{oldman-17032012-01} and \rf{manold-18032012-08} respectively,
lead to gauge transformations \rf{oldman-19032012-07}. To this end
we note that plugging $\phik$ \rf{oldman-17032012-01} and $\xik$
\rf{manold-18032012-08} into gauge transformations
\rf{oldman-05022012-11} leads to the following relation:
\be
\label{19032012-01} e^{X\Box} (\upsilon^\ominussm)^\kwh \delta
|\Phi'\rangle + \alpha^2 \frac{1}{2N_\alpha+d-2} \ewt_1 Z \delta
\Phik = G  e^{X\Box} (\upsilon^\ominussm)^{\kwh+1}\Xik \,.
\ee
All that remains is to demonstrate that relation \rf{19032012-01}
holds true provided the gauge transformations take the form as in
\rf{oldman-19032012-07}. We now consider relation \rf{19032012-01}
for $d\geq 6$ and $d=4$ in turn.

\noindent {\bf Proof of relation \rf{19032012-01} for $d\geq 6$}.
Using \rf{oldman-19032012-07}, \rf{oldman-19032012-03}, we note
that, for $d\geq 6$, relation \rf{19032012-01} amounts to
\be \label{19032012-02}
\Bigl( e^{X\Box} (\upsilon^\ominussm)^\kwh  + \alpha^2
\frac{1}{2N_\alpha+d-2} \ewt_1 Z \Bigr)G_\sh \Xik  = G e^{X\Box}
(\upsilon^\ominussm)^{\kwh+1} \Xik\,.
\ee
Relation \rf{19032012-02} can be checked by using the following
relations:
\beq
\label{19032012-03} && e^{X\Box} (\upsilon^\ominussm)^\kwh G_\sh
\approx (\alpar - \zeta \ewt_1 \bar\upsilon^\ominussm) e^{X\Box}
(\upsilon^\ominussm)^{\kwh +1} + \alpha^2 \frac{1}{2N_\alpha+d-2}
\ewt_1 \bar\zeta e^{X\Box} (\upsilon^\ominussm)^{\kwh+2}\,,\qquad
\\
\label{19032012-04} && Z G_\sh \approx - \bar\zeta \frac{(\Box
\upsilon^\oplussm)^{\kwh+2}}{\Gamma(\kwh+3)}\,,
\\
\label{manold-03042012-09} && \upsilon^\ominussm e^{ X \Box}
(\upsilon^\ominussm)^N \approx e^{X \Box} (\upsilon^\ominussm)^{N+1}
- \frac{(\Box\upsilon^\oplussm)^{N+1}}{(N+1)!}\,,
\eeq
where notation $\approx$ is used to indicate the fact that relations
\rf{19032012-03}-\rf{manold-03042012-09} are valid on space of
ket-vector $\Xik$ \rf{oldman-19032012-06}. Proof of relations
\rf{19032012-03}-\rf{manold-03042012-09} is straightforward.

\noindent {\bf Proof of relation \rf{19032012-01} for $d = 4$}. We
represent ket-vectors $\Phik$ \rf{oldman-19032012-01} and $\Xik$
\rf{oldman-19032012-06} as
\be \label{manold-04042012-01}
\Phik = |\Phi'\rangle + |\Phi_0\rangle\,, \qquad \Xik = |\Xi'\rangle
+ |\Xi_0\rangle\,,
\ee
where $|\Phi_0\rangle$ stands for $s'=0$ contribution of the scalar
field $\phi_1$ in \rf{oldman-19032012-01}, while $ |\Xi_0\rangle$
stands for $s'=0$ contribution of the scalar gauge transformation
parameter $\xi_{-1}$ in \rf{oldman-19032012-06}. Plugging
\rf{manold-04042012-01} into \rf{oldman-19032012-07} we represent
gauge transformations \rf{oldman-19032012-07} as
\beq
\label{manold-04042012-02}  && \delta |\Phi'\rangle  = G_\sh
|\Xi'\rangle + (\alpar - \alpha^2 \frac{1}{2N_\alpha+d-2}
\eb_{1,\sh} )|\Xi_0\rangle\,,
\\
&& \delta |\Phi_0\rangle  = - e_{1,\sh} |\Xi_0\rangle\,.
\eeq
For gauge transformations \rf{manold-04042012-02}, we prove the
following relation:
\beq
\label{19032012-05} && \hspace{-1.5cm} e^{X\Box}
(\upsilon^\ominussm)^\kwh \delta |\Phi'\rangle  =  (\alpar - \zeta
\ewt_1 \bar\upsilon^\ominussm) e^{X\Box}
(\upsilon^\ominussm)^{\kwh+1} \Xik +  \alpha^2
\frac{1}{2N_\alpha+d-2} \ewt_1 \bar\zeta e^{X\Box}
(\upsilon^\ominussm)^{\kwh+2}\Xik\,.
\eeq
All that we need to prove \rf{19032012-05} is to note the relation
\be
\bar\upsilon^\ominussm e^{X \Box} (\upsilon^\ominussm)^{\kwh +1}
|\Xi_0\rangle = 0 \,.
\ee
Finally, plugging \rf{19032012-05} and $\delta\Phik$
\rf{oldman-19032012-07} into \rf{19032012-01} and using
\rf{19032012-04}, \rf{manold-03042012-09}, we make sure that
relation \rf{19032012-01} holds true.

\end{document}